%% file: main.tex
\documentclass[10pt,a4paper]{article}
\usepackage[a4paper, left=2cm, right=2cm, top=2cm]{geometry}

\input{./tex/settings.tex}
\input{./tex/plotting.tex}

\usetikzlibrary{external}
\title{Differential Equation Based Wall Distance Approaches for Maritime Engineering Flows}
\author[1]{Niklas K\"uhl\thanks{kuehl@hsva.de}}

\affil[1]{Hamburg Ship Model Basin, Bramfelder Strasse 164, D-22305 Hamburg, Germany}

\begin{document}

\providetoggle{tikzExternal}
\settoggle{tikzExternal}{true}
\settoggle{tikzExternal}{false}

\maketitle

\begin{abstract}

The paper is concerned with modeling and simulating approaches of wall distance functions based on Partial Differential Equations (PDE). The distance to the nearest wall is required for many industrial problems in Computational Fluid Dynamics (CFD).

The first part of the manuscript addresses fundamental aspects of wall distance modeling and simulation. The following approaches are considered: Nonlinear and linear p-Poisson and Screened-Poisson methods, Eikonal and regularized Eikonal or Hamilton-Jacobi methods, and alternatives using Laplace equations. Following the definition of boundary and initial conditions, the discrete approximation and relevant measures to increase its numerical robustness are described, such as the utilization of convective instead of diffusive approximation strategies in the case of the Eikonal equation combined with deferred correction approaches. The implemented methods are verified and validated against the exact but expensive geometric benchmark accelerated using a KD-tree considering an academic and an industrial example of a ship with a rotating propeller.

In the second part, the different methods are applied to hydrodynamic and aerodynamic flow applications from maritime engineering, each relying on Shear Stress Transport (SST) strategies for turbulence modeling that require the distance to the nearest wall at different procedural points. The hydrodynamic behavior of a model scale bulk carrier cruising at $\mathrm{Re}_\mathrm{L} = \SI{7.246}{} \cdot 10^6$ and $\mathrm{Fn}=0.142$ is investigated on the influence of the wall distance formulation for predicting resistance and propulsion behavior in conjunction with statistical turbulence modeling based on a classical Reynolds-Averaged Navier-Stokes (RANS) method. It is shown that the different wall distance modeling barely influences relevant integral hydrodynamic quantities such as drag, trim, and sinkage, and related errors are in the range of $\mathcal{O}(10^{-1}\%)$ and, therefore, significantly below typical modeling, discretization, and approximation errors. Additionally, it is shown that the assumption of a temporally constant wall distance by calculating the distance function once at the beginning of the simulation is sufficient for an adequately accurate propulsion calculation. Subsequently, the wall distance methods were investigated for the aerodynamic analysis of a full-scale feeder ship at $\mathrm{Re}_\mathrm{L} = 5.0 \cdot 10^8$. A hybrid RANS Large-Eddy-Simulation (LES) approach, in line with the Improved Delayed Detached Eddy Simulation (IDDES) model, is utilized, and the results indicate an improved sensitivity to the choice of the wall distance model. 

It is found that the Eikonal approach based on a convective approximation --e.g., with an upwind-biased approximation and thus considerable numerical viscosity or a Hamilton-Jacobi method including a suitable, minor diffusivity-- with deferred correction provides a solid method with low numerical effort and, at the same time, fair approximation quality. The considered p-Poisson methods are --based on the utilized numerical treatment-- numerically too expensive, especially for p-values greater than two, the Laplace method is prone to singularities, and the Screened-Poisson method is generally sensitive with respect to its model parameters. 

\end{abstract}

\begin{flushleft}
\end{flushleft}


\section{Introduction}

Within industrial Computational Fluid Dynamics (CFD) investigations, the distance of a wetted, e.g., node / volume / element to the nearest wall is often necessary at various procedural points. Examples are (a) modern strategies for modeling flow turbulence (\cite{menter2003ten, gritskevich2017comprehensive}) and its transition (\cite{menter2006transition, menter2015one}), (b) the numerical stability of the overall numerical process, for example, by fading out higher approximation orders or damping free surface waves (\cite{woeckner2010efficient}) in regions far from the wall, and (c) robust grid deformation processes, often controlled by inverse wall distance (\cite{haubner2021continuous, mueller2021novel, kuhl2022adjoint}).

Methods for determining the distance to the nearest wall can be roughly divided into methods based on (a) intelligent search/march methods, (b) integral, and (c) differential equation methods. Strategies belonging to type (a) provide the wall distance geometrically precisely based on the underlying discretization. However, the effort scales quadratically with the number of discrete degrees of freedom, e.g., $\mathcal{O}(\mathrm{N^V} \cdot \mathrm{N^S})$, where $\mathrm{N^V}$ and $\mathrm{N^S}$ refer to the number of discrete volume and surface points, respectively, resulting in a comparatively large computational effort. This effort may be justifiable on static grids, where the wall distance is computed once at the simulation's start. Industrial applications often feature dynamic grid geometries and topologies, i.e., those that change at runtime, for example, by sliding, overlapping, or adaptively refined grids. In this case, a new determination of the wall distance is necessary for each grid adjustment, for example, at each discrete time step, which is often more expensive than computing the current flow state when utilizing approaches of type (a). Based on efficient, e.g., Nearest Neighbor Search algorithms, their effort can be reduced to, e.g., $\mathcal{O}(\mathrm{N^V} \cdot \mathrm{log}\mathrm{N^S})$ or $\mathcal{O}(\mathrm{N^V} \cdot \sqrt{\mathrm{N^S}})$, cf. \cite{boger2001efficient, roget2013wall}, that is still too costly in the above mentioned scenarios. Methods based on (b) integral equations are presented in \cite{dyken2009transfinite, belyaev2013signed}, among others. However, due to their nature, they require particular solution strategies, e.g., based on the Boundary-Element-Method. The latter is rarely used in viscous industrial CFD methods, cf. \cite{altair2021theory, ansys2021theory, siemens2021user}, where volume-based procedures are employed. Hence, type (c) methods are beneficial, as they approximate the wall distance by specific model equations formulated by means of classical transport equations. Therein, the wall distance is interpreted as a variable to be solved, which allows the use of often generalized (code) infrastructures of general purpose CFD codes and, thus, efficient solution strategies and the possibility to restart. The latter is especially beneficial for dynamic grids of an industrial simulation process since already invested simulation efforts can be reused, and the new distance to the nearest wall is calculated with only a few iterations compared to the initial calculation from scratch.


Various methods for determining the distance function have been presented in the literature, and they address all three types mentioned above with different numerical strategies. A reader interested in different applications might therefore consult the following exemplary contributions: \cite{fares2002differential}, \cite{xia2010finite}, \cite{zhou2015cpr}, or \cite{kakumani2022use}. For an in-depth introduction, please refer to the publication record of P.G. Tucker, especially \cite{tucker2003differential, tucker2005computations, tucker2011hybrid}.

This paper's contribution is divided into two parts. On the one hand, fundamental CFD-side aspects of various distance function approximation strategies will be analyzed and compared. In addition to accurately specifying the model equations and the associated boundary and initial conditions, particular attention is paid in this context to economic aspects of the discrete approximation, i.e., the trade-off between approximation effort and accuracy, with a focus on their practical implications and applicability. The second part of the paper catches up on this later aspect and deals with the influence of the different methods on technical flows at high Reynolds numbers in a maritime context. 
In addition to the continuous modeling, the numerical approximation and measures promoting the approximation stability are discussed. Without exception, equation-based wall distances are compared with the exact geometric benchmark, and parameter settings for maritime simulations are provided. Several prominent equation-based wall distance determination methods are considered, from well-known, simple linear up to modern, highly non-linear approaches. These are (1) Eikonal and Hamilton-Jacobi, (2) Laplacian, as well as (3) screened and p-Poisson equation based approaches. The latter, in particular, is comparatively new for maritime engineering flows.


The manuscript is structured as follows. Section \ref{sec:equation_models} presents the equations investigated in this paper, including boundary conditions and the approximation strategy based on a state-of-the art second-order cell-centered Finite-Volume-Method. Section \ref{sec:verification_validation} presents the verification of the implementation as well as validation based on academic and industrial test cases. Finally, the presented methods are analyzed in Sec. \ref{sec:applicaiton} based on application cases of industrial character --represented by the hydrodynamic analysis of a bulk carrier in model scale at$\mathrm{Re}_\mathrm{L} = \SI{7.246}{} \cdot 10^6$ and $\mathrm{Fn}=0.142$ as well as the aerodynamic analysis of a feeder ship in full scale at $\mathrm{Re} = 5 \cdot 10^8$ -- for approximation accuracy and effort as well as the influence on local (e.g., velocity prope or turbulence model zones) and global (e.g., force coefficients) flow data. The manuscript concludes in Sec. \ref{sec:conclusion_outlook} with a summary and an outlook for further work. The publication uses Einstein's summation convention for lower-case Latin subscripts and vectors as well as tensors are defined with respect to Cartesian coordinates.


\section{Wall Distance Modeling Procedures}
\label{sec:equation_models}

\subsection{Continuous Modeling}
\label{subsec:continuous_models}
Throughout the document, $w$ denotes the distance to the nearest wall. Regardless of whether it is determined geometrically or approximated by differential equations, the corresponding field solution follows the so-called Eikonal equation, viz., 
\begin{alignat}{2}
    \bigg| \frac{\partial w}{\partial x_k} \bigg| &= 1 \qquad &&\text{in} \qquad \Omega \label{equ:eikonal} \, , \\
    w &= 0 \qquad &&\text{on} \qquad \Gamma^\text{wall} \, , \\
    \frac{\partial w}{\partial n} &= 0 \qquad &&\text{on} \qquad \Gamma^\text{symm} \, , \\
    \frac{\partial w}{\partial n} &= 1 \qquad &&\text{on} \qquad \Gamma^\text{rest} \, .  \label{equ:eikonal_far}
\end{alignat}
Therein, $n$ denotes the local surface normal direction, and $\Gamma^\text{wall}$, $\Gamma^\text{symm}$, as well as $\Gamma^\text{rest}$ refer to wall, symmetry, and all remaining, e.g., velocity inlet or pressure outlet boundaries. In the following, the manuscript's utilized strategies for the equation-based approximation of Eqn. \eqref{equ:eikonal} are introduced. 

\subsubsection{Regularized Eikonal aka Hamilton-Jacobi}
The consistently positive, truncated and thus tricky to differentiate, gradient of the wall distance field from Eqn. \eqref{equ:eikonal} can be converted via simple squaring into a companion that is more straightforward to handle for general-purpose CFD codes, i.e., $|\nabla_k w| = 1 \to \nabla_k w \, \nabla_k w = 1$, where $\nabla_k$ refers to the gradient operator. In order to favor the numerical approximation and to dampen potential (gradient) jumps in the solution, some authors add an additional diffusion term with in-homogeneous viscosity $\mu$. The resulting equation features similarities with a Hamilton-Jacobi formalism and is typically referred to as the Hamilton-Jacobi approach in the context of wall distance determinations, cf. \cite{tucker2003differential, tucker2011hybrid}. The field equation as well as its boundary conditions read
\begin{alignat}{2}
    \frac{\partial w}{\partial x_k} \frac{\partial w}{\partial x_k} - \frac{\partial}{\partial x_k} \left[ \mu \frac{\partial w}{\partial x_k} \right] &= 1 \qquad &&\text{in} \qquad \Omega \label{equ:hamilton-jacobi} \\
    w &= 0 \qquad &&\text{on} \qquad \Gamma^\text{wall} \\
    \frac{\partial w}{\partial n} &= 0 \qquad &&\text{on} \qquad \Gamma^\text{symm} \\
    \frac{\partial w}{\partial n} &= 1 \qquad &&\text{on} \qquad \Gamma^\text{rest} \, .
\end{alignat}
In order to reproduce the actual Eikonal Eqn. \eqref{equ:eikonal}, the viscosity should generally be small and tend towards zero in areas of required high accuracy, i.e., the wall proximity. \cite{tucker2005computations} suggest $\Gamma = \varepsilon \, d \to 0$, which vanishes near the wall, where the parameter $\varepsilon$ is introduced as a free parameter, whereby values between $0 \leq \varepsilon \leq 0.5$ [$0.1 \leq \varepsilon \leq 1.0$] are suggested in \cite{tucker2011hybrid} [\cite{kakumani2022use}]. Note that the Eikonal Eqn. \eqref{equ:eikonal} is reproduced for $\varepsilon = 0$.
Some authors reformulate the squared Eikonal equation, i.e.,
\begin{align}
    \frac{\partial w}{\partial x_k} \frac{\partial w}{\partial x_k} = \frac{\partial}{\partial x_k} \left[ d \frac{\partial w}{\partial x_k} \right] - d \frac{\partial^2 w}{\partial x_k^2} \label{equ:squared_eikonal_rewritten} \, ,
\end{align}
resulting in a more convenient equation form for specific, usually conservative, approximation strategies. The Eikonal equation's initial definition ($|\nabla_k w| = 1 = \mathrm{const.}$) hints at the minor importance of the second term on the right-hand side of Eqn. \eqref{equ:squared_eikonal_rewritten}, which in turn might motivate its neglect during the discrete approximation. The main contribution follows from the first term, which can, e.g., be interpreted as a classical convection (velocity $\nabla_k d$) or diffusion (viscosity $d$) expression. Whereas the former corresponds to a parabolic transport process with constant unit velocity, the latter features an elliptic behavior with large/low diffusivity in the far/near field, respectively. Depending on the discrete approximation strategy, the latter approach might be prone to numerical difficulties when approaching the wall. This paper's implementation relies on the velocity approach. Further details on the discrete approximation and relevant stability measures are provided in the upcoming Sec. \ref{subsec:discrete_approximation}.

\subsubsection{Screened-Poisson}
Following the recognition of \cite{varadhan1967behavior} in the recent overview of \cite{belyaev2015variational}, another way to estimate the distance function is based on a Screeed-Poisson equation. For this purpose, the following field problem is first solved:
\begin{alignat}{2}
    \tilde{w} - t \frac{\partial^2 \tilde{w}}{\partial x_k^2} &= 0 \qquad &&\text{in} \qquad \Omega \label{equ:screened_poisson} \, , \\
    \tilde{w} &= 1 \qquad &&\text{on} \qquad \Gamma^\text{wall} \, , \\
    \frac{\partial \tilde{w}}{\partial n} &= 0 \qquad &&\text{on} \qquad \Gamma^\text{symm} \, , \\
    \frac{\partial \tilde{w}}{\partial n} + \frac{\tilde{w}}{\sqrt{t}} &= 0 \qquad &&\text{on} \qquad \Gamma^\text{rest} \, , \label{equ:screened_poisson_bc}
\end{alignat}
and the solution is then transformed to
\begin{align}
    w = - \sqrt{t} \mathrm{ln}(\tilde{w}) \, . \label{equ:screened_poisson_normalization}
\end{align}
Here, $t \ge 0$ is a small, positive parameter of squared length dimension, i.e., $[t] = \mathrm{L}^2$. The rationale behind the approach follows from inverting Eqn. \eqref{equ:screened_poisson_normalization}, i.e., $\tilde{w} = \mathrm{exp}(-w/\sqrt{t})$, and a subsequent insertion into Eqns. \eqref{equ:screened_poisson}-\eqref{equ:screened_poisson_bc}, viz.
\begin{alignat}{2}
    1 - \frac{\partial w}{\partial x_k} \frac{\partial w}{\partial x_k} + \sqrt{t} \frac{\partial^2 w}{\partial x_k^2} &= 0 \qquad &&\text{in} \qquad \Omega \, , \\
    w &= 0 \qquad &&\text{on} \qquad \Gamma^\text{wall} \, , \\
    \frac{\partial w}{\partial n} &= 0 \qquad &&\text{on} \qquad \Gamma^\text{symm} \, , \\
    \frac{\partial w}{\partial n} &= 1 \qquad &&\text{on} \qquad \Gamma^\text{rest} \, .
\end{alignat}
The process recovers a Cole–Hopf transformation (\cite{evans2022partial}) and converges to the Eikonal Eqn. \eqref{equ:eikonal} for $\sqrt{t} \to 0$. The non-linear transformation leads to a mixed Dirichlet-Neumann, i.e., a Robin boundary condition. For external flows, this condition is relevant only to far-field boundaries and might allow a compromise of implementation consistency, e.g., by specifying only one, i.e., the Dirichlet or Neumann condition. However, the situation might change for internal flows, where walled regions neighbor inlet/outlet conditions. Hence, this aspect is not the subject of this manuscript.
The parameter $t$ is a model parameter, and the choice depends on the underlying problem. A practical aid for selection support might be the relation to a global reference length of the underlying flow situation, e.g., $\sqrt{t}/L$, where $L$ refers to a ship's length between its perpendiculars or a propeller diameter, used to compile further fluid mechanical parameters such as Reynolds and Froude numbers.
The strategy shares ideas of the so-called heat method proposed by \cite{crane2017heat}, used to compute distance functions on curved surfaces by considering a heat kernel similar to Eqn. \ref{equ:screened_poisson_normalization} for a short amount of time, among other aspects.

\subsubsection{p-Poisson}
Standard linear Poisson approaches are well known for computing distance functions, cf. \cite{spalding2013trends, wukie2017p, nishikawa2021hyperbolic, loft2023two}, that provide satisfactory results in the wall proximity but deviate significantly from the correct value in the far field. A natural extension follows the generalization of the p-Poisson equation, viz.
\begin{alignat}{2}
    -\frac{\partial}{\partial x_k} \left[ \left( \frac{\partial \tilde{w}}{\partial x_k} \right)^{(p-2)} \frac{\partial \tilde{w}}{\partial x_k} \right] &= 1 \qquad &&\text{in} \qquad \Omega \label{equ:p_poisson} \\
    \tilde{w} &=0 \qquad &&\text{on} \qquad \Gamma^\text{wall} \\
    \frac{\partial \tilde{w}}{\partial n} &= 0 \qquad &&\text{on} \qquad \Gamma^\text{symm} \\
    \frac{\partial \tilde{w}}{\partial n} &= 1 \qquad &&\text{on} \qquad \Gamma^\text{rest} \, .
\end{alignat}
A subsequent normalization is required
\begin{align}
    w = \left[ \frac{p}{p-1} \tilde{w} + \bigg|\frac{\partial \tilde{w}}{\partial x_k}\bigg|^p \right]^{\frac{p-1}{p}} - \bigg|\frac{\partial \tilde{w}}{\partial x_k}\bigg|^{(p-1)}
\end{align}
that finally ensures dimensional consistency since $[\tilde{w}] = \mathrm{L}^{p/(p-1)}$ but $[w] = L$. Several authors report a theoretical convergence of $\tilde{w} \to d$ if $p \to \infty$, cf. \cite{ting1971elastic, kawohl1990familiy, belyaev2015variational}. In practice, however, large values of $p$ are numerically challenging as, next to typical solver convergence issues for non-linear problems, issues with respect to digit/decimal precision might arise. Hence, realistic values for $p$ are between $2 \leq p \leq 8$ (\cite{belyaev2015variational, wukie2017p, mueller2021novel}), where $p=2$ recovers the classical linear Poisson equation.
Recently, the $p$-Poisson approach gained attention in the shape optimization community when computing parameter-free descent directions within gradient-based optimization strategies, cf. \cite{mueller2021novel, deckelnick2022novel, radtke2023parameter}. The relation between $p$-Poisson approaches and distance functions is further underlined by the fact that alternative descent direction approaches employ scalings of the (inverse) distance to the nearest wall, cf. \cite{kuhl2021phd, kuhl2022adjoint, radtke2023parameter}.

\subsubsection{Laplacian}
Another relatively intuitive strategy is based on the illustrative interpretation of potential fields. These can be extracted from the following linear Laplace equation, i.e.,
\begin{alignat}{2}
    \frac{\partial^2 \tilde{w}}{\partial x_k^2} &= 0 \qquad &&\text{in} \qquad \Omega \label{equ:laplacian} \, , \\
    \tilde{w} &= 0 \qquad &&\text{on} \qquad \Gamma^\text{wall} \, , \\
    \frac{\partial \tilde{w}}{\partial n} &= 0 \qquad &&\text{on} \qquad \Gamma^\text{symm} \, , \\
    \frac{\partial \tilde{w}}{\partial n} &= 1 \qquad &&\text{on} \qquad \Gamma^\text{rest} \, ,
\end{alignat}
by performing the following normalization to extract distance fields
\begin{align}
    d = \frac{\tilde{w}}{\left| \frac{\partial \tilde{w}}{\partial x_k} \right| } \, . \label{equ:laplacian_normalization}
\end{align}
While the standardized approximation of Eqn. \eqref{equ:laplacian} is possible with almost any CFD solver, executing the normalization by Eqn. \eqref{equ:laplacian_normalization} requires special attention at regions of vanishing distance gradients and, therefore, singularity-sensitive areas. It will be shown that the Laplacian approach provides satisfactory results for situations with isolated, central, and gap-free walls, such as resistance studies of ship hulls without appendages. However, the approach features issues with walled boundaries opposite each other, e.g., between a propeller blade and the hull.

\subsubsection{Initial Conditions}
The approximation of all presented equations uses the norm of the spatial position vector as an initial condition, i.e., $w = |x_k|$ or $\tilde{w} = |x_k|$.

\subsection{Discrete Approximation}
\label{subsec:discrete_approximation}

\begin{figure}[!ht]
\centering
\subfigure[]{
\iftoggle{tikzExternal}{
\input{./tikz/01__discretization/finite_volume_approximation_field.tikz}}{
\includegraphics{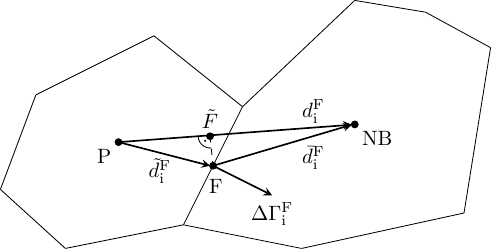}}
}
\hspace{2cm}
\subfigure[]{
\iftoggle{tikzExternal}{
\input{./tikz/01__discretization/finite_volume_approximation_boundary.tikz}}{
\includegraphics{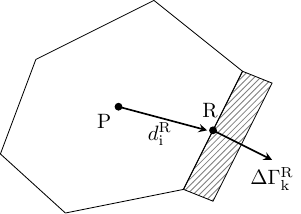}}
}
\caption{Schematic representation of a Finite-Volume arrangement (a) in the field and (b) along the boundary.}
\label{fig:finite_volume_approximation}
\end{figure}

In line with a state-of-the-art Finite-Volume-Method (FVM), the spatial domain is discretized by NP contiguous control volumes of individual size. Local approximations are constructed from values extracted at the cell center $P \in [1, \mathrm{NP}]$, the face centers F separating the control volume around P from the neighboring control volumes, and the centers of the boundary control volume denoted by $R \in [1, \mathrm{NR}]$. The number of faces and neighbors per cell is denoted by $\mathrm{F}(\mathrm{P}) = \left[1, \mathrm{NF} \right]$ as well as $\mathrm{NB}(\mathrm{P}) = \left[1, \mathrm{NB} \right]$, respectively. Figure \ref{fig:finite_volume_approximation} (a) shows an arbitrary arrangement of two adjacent polyhedral control volumes that share the connecting vector between P and NB(P) through F(P). On the right, (b) shows a boundary companion. Control volumes of arbitrary polyhedral shapes that form an unstructured numerical grid are applicable. Temporal movement or deformation is possible, cf. (\cite{volkner2017analysis, schubert2019relation, luo2017computation, luo2019numerical}), but not relevant for the approximation of the paper's steady wall distance describing differential equations. Further information on the employed FVM code FreSCo$^+$ can be found, e.g., in \cite{rung2009challenges, stuck2012adjoint, manzke2018development, schubert2019analysis, kuhl2021adjoint, kuhl2021adjoint_2, kuhl2021continuous}.
A collocated variable arrangement is employed. Hence, quantities of all approximated transport equations are stored in the cell centers. Values at different locations must be reconstructed, e.g., via an appropriate interpolation. The numerical process utilizes the cell size $\Delta \Omega^P$ as well as the area $\Delta \Gamma^F$ and the respective area vector $\Delta \Gamma_i^F = \left[ \Delta \Gamma \, n_i \right]^F$ of all cell adjacent faces where $n_i$ represents a face's normal vector entries. The definition of vectors connecting the relevant positions around the face F reads
\begin{align}
    d_i^F = \tilde{d}_i^F + \bar{d}_i^F
    \qquad \qquad \mathrm{with} \qquad \qquad
    \tilde{d}_i^F = x_i^F - x_i^P
    \qquad \qquad \mathrm{and} \qquad \qquad
    \bar{d}_i^F = x_i^{NB} - x_i^F \, .
\end{align}
The NR degenerated boundary cells have no volume and a midpoint coordinate that coincides with the center of the adjacent face, viz. $\bar{d}_i^F = 0$ and thus $d_i^R = \tilde{d}_i^F$. Typical crucial aspects for the FV approximations apply, e.g., (a) the connecting vector $d_i^F$ is not necessarily co-linear to the face normal $n_i$, (b) faces are not necessarily centered between their neighboring cell's center, and thus (c) $d_i^F$ does not pierce through the face center (or the face at all). These aspects will be further taken up while approximating this paper's balance equations in Sec. \ref{sec:approximation}.

\subsubsection{Iterative Procedure}
A discrete system of equations of size $NP \times NP$ is constructed in order to solve for the cell-centered wall distance approximation via an outer fix-point-iteration of counter M. Each line corresponds to a particular balance of the control volume P to compute the cell-centered wall distance $w^{P, M}$, viz. 
\begin{align}
    \underline{\underline{A}}^M \underline{w}^M = \underline{S}^M
    \qquad \qquad \text{or} \qquad \qquad
    A^\mathrm{P, M} w^\mathrm{P, M} + \sum_\mathrm{NB(P)} A^\mathrm{NB, M} w^\mathrm{NB, M} = S^\mathrm{P, M} \label{equ:discrete_system_of_equations} \, .
\end{align}
The variable changes from $w \to \tilde{w}$ if the Laplacian and Poisson type Eqns. \eqref{equ:screened_poisson} - \eqref{equ:laplacian} are solved. Once the discrete relation is assembled, an under-relaxation supports the diagonal dominance of the matrix and thus the iterative fix point procedure, viz.
\begin{align}
A^\mathrm{P, M} \to \frac{A^\mathrm{P, M}}{\omega}
\qquad \mathrm{and} \qquad
S^\mathrm{P, M} \gets \frac{\left( 1 - \omega \right) A^\mathrm{P, M}}{\omega} w^\mathrm{P, M-1} \, , \label{equ:relaxation}
\end{align}
where $w^\mathrm{P, M-1}$ follows from the previous outer iteration. The case-specific relaxation parameter is around $0 \leq \omega \leq 1$. Linear approaches allow $\omega \to 1$, whereas non-linear problems such as the $p$-Poisson approaches require lower values, usually between $0.1 \leq \omega \leq 0.5$.
The system of equations for the outer iteration M is solved using the Portable, Extensible Toolkit for Scientific Computation (PetSC, cf. \cite{petsc-web-page}), whereby a BI-Conjugate Gradient (BICG) method with Jacobi preconditioning is utilized. The PetSC solver is employed with relatively small convergence criteria around $\mathcal{O}(10^{-2})$ and therefore requires --compared to the outer iteration process-- fewer inner iterations. Convergence of the outer fixed point method is monitored by the following residual $R^M$ expression, viz.
\begin{align}
    R^M = \frac{1}{N} \sum_{P=1}^{NP} |\underline{r}^{M}|
    \qquad \qquad \text{with} \qquad \qquad
    \underline{r}^{M} = \underline{\underline{A}}^{M} \underline{w}^{M-1} - \underline{S}^M \, ,
\end{align}
whereby residual normalization is omitted. The academic studies in the paper stop the outer iteration process once $R^M \le 10^{-14}$. The industrial applications are not as strict and increase the bound to $R^M \le 10^{-5}$.

The approximation of the individual terms of all transport equations and the contribution to the system of equations \eqref{equ:discrete_system_of_equations} are discussed in more detail below.

\subsubsection{Assembling Procedure}
\label{sec:approximation}
The fixed-point iteration requires the gradient of the wall distance field at various procedural positions, wherefore Gaussian's theorem is used, viz.
\begin{align}
    \frac{\partial w}{\partial x_k} \bigg|^{P, M} \approx \frac{1}{\Delta \Omega^P} \sum_{F(P)} \left[ w \, \Delta \Gamma_k \right]^{F, M} \, .
\end{align}
Required face values are interpolated linearly from the neighboring cell-centered values, whereby further extrapolation accounts for potential non-orthogonality, i.e.,
\begin{align}
    w^{F,M} \approx \left( 1 - \lambda^F \right) w^{P, M} + \lambda^F w^{NB, M}
    + \left( x_k^F - \tilde{x}_k^F \right) \frac{\partial w}{\partial x_k} \bigg|^{\tilde{F}, M-1}
    \quad \quad \text{with} \quad \quad
    \lambda^F = \frac{\Delta \Gamma_i^F \left( x_i^P - x_i^F\right)}{\Delta \Gamma_k^F d_k^F} \, . \label{equ:field_interpolation}
\end{align}
Note that this contribution becomes smaller for grids of increased orthogonality, i.e., $\tilde{x}_i^F \to x_i^F$. The required face gradient follows from another linear interpolation based on the value of the previous outer iteration
\begin{align}
    \frac{\partial w}{\partial x_k} \bigg|^{\tilde{F}, M-1} \approx \left( 1 - \lambda^F \right) \frac{\partial w}{\partial x_k}  \bigg|^{P, M-1} + \lambda^F \frac{\partial w}{\partial x_k}  \bigg|^{NB, M-1} \, . \label{equ:gradient_interpolation}
\end{align}
Either Dirichlet $\Gamma^\text{Dir}$ or Neumann $\Gamma^\text{Neu}$ boundaries are employed. In the former case, the wall distance is prescribed and considered implicitly during the assembly process. Unknown gradients follow from a zero-order extrapolation. Along Neumann walls, the wall distance is extrapolated based on the given prescribed normal gradient, e.g., $[ \nabla_k d ]^B = 1$, viz.
\begin{alignat}{4}
    w^{B, M} &= \mathcal{D}^B
    \qquad \qquad
    &&\frac{\partial w}{\partial n} \bigg|^{B,M} &&\approx \frac{\partial w}{\partial n} \bigg|^{P,M}
    \qquad \qquad &&\text{on} \qquad \Gamma^\text{Dir} \\
    w^{B, M} &\approx w^{P, M} + d_i^{B,M} \frac{\partial w}{\partial x_i} \bigg|^{B,M}
    \qquad \qquad
    &&\frac{\partial w}{\partial n} \bigg|^{B,M} &&= \mathcal{N}^B
    \qquad \qquad &&\text{on} \qquad \Gamma^\text{Neu} \, .
\end{alignat}

\paragraph{Convective Fluxes} of the Eikonal or Hamilton-Jacobi approach follow a first-order Upwind Differencing Scheme (UDS) approximation. The gradient field of the previous iteration (cf. Eqn. \eqref{equ:gradient_interpolation}) projected into the face's normal direction serves as flux, viz.
\begin{align}
    \int \frac{\partial}{\partial x_k} \left[ \frac{\partial w}{\partial x_k} d \right] \mathrm{d} \Omega
    \approx
    \sum_{F(P)} &\left[ \frac{\partial w}{\partial x_k} \Delta \Gamma_k w \right]^F \\
    \text{with} \qquad &\left[ \frac{\partial w}{\partial x_k} \Delta \Gamma_k w \right]^F 
    \approx
    \mathrm{max} \left( \left(\frac{\partial w}{\partial x_k} \Delta \Gamma_k \right)^\mathrm{F} , 0 \right) \, w^\mathrm{P} + \mathrm{min} \left( \left( \frac{\partial w}{\partial x_k} \Delta \Gamma_k \right)^\mathrm{F} , 0 \right) \, w^\mathrm{NB} \, .
\end{align}
The formulation allows an implicit consideration that accelerates the solution process and enters the discrete system of equations as follows:
\begin{align}
    A^{P, M} \gets \sum_{F(P)} - \mathrm{min} \left( \left( \frac{\partial w}{\partial x_k} \Delta \Gamma_k \right)^\mathrm{F, M-1} , 0 \right)
    \qquad \text{and} \qquad
    A^{NB, M} \gets \mathrm{min} \left( \left( \frac{\partial w}{\partial x_k} \Delta \Gamma_k \right)^\mathrm{F, M-1} , 0 \right) \, .
\end{align}
The main diagonal is strengthened, but may get no contribution for (iterative) situations with vanishing gradient, cf. further information in Sec. \ref{subsec:stabilization}

\paragraph{Diffusive Fluxes} occur in all of the considered model equations. The implicit approximation supports the diagonal dominance, maintains a positive coefficient, and follows a face-based Central Difference Scheme (CDS) that accounts for the unbiased information transport, where the (possibly variable) diffusivity is interpolated in line with Eqn. \eqref{equ:field_interpolation}, viz.
\begin{align}
    \int \frac{\partial}{\partial x_k} \left[ \mu \frac{\partial w}{\partial x_k} \right] \mathrm{d} \Omega
    \approx
    \sum_{F(P)} &\left[ \mu \frac{\partial w}{\partial x_k} \Delta \Gamma_k \right]^F \nonumber \\
    \qquad \text{with} \qquad &\left[ \mu \frac{\partial w}{\partial x_k} \Delta \Gamma_k \right]^F
    \approx
    \left[ \frac{\mu \, \Delta \Gamma}{| d_\mathrm{i} | } \right]^\mathrm{F} \left[ w^\mathrm{NB} - w^\mathrm{P} \right] + \left[ \frac{\mu \, \Delta \Gamma}{| d_\mathrm{k} |} \frac{\partial \, w}{\partial \, x_\mathrm{k}} \bigg( d_k - n_\mathrm{k} \, |d_\mathrm{i}| \bigg) \right]^F \, .
\end{align}
%
Another explicitly evaluated non-orthogonality correction is introduced, which tends to zero if the computational grid is perfectly orthogonal ($[d_\mathrm{k} / |d_\mathrm{}| - n_\mathrm{k}]^\mathrm{F} \to 0$). The following contributions to the discrete system of equations are derived:
\begin{align}
    A^{P, M} \gets \sum_{F(P)} \left[ \frac{\mu \, \Delta \Gamma}{| l_\mathrm{k} |} \right]^\mathrm{F, M-1} \, ,
    \qquad
    A^{NB, M} &\gets - \left[ \frac{\mu \, \Delta \Gamma}{| d_\mathrm{k} |} \right]^\mathrm{F, M-1} \nonumber \\
    \qquad \text{and} \qquad
    S^\mathrm{P, M} &\gets - \sum_\mathrm{F(P)} \left[ \frac{\mu \, \Delta \Gamma}{| d_\mathrm{i} |} \frac{\partial \, w}{\partial \, x_\mathrm{k}} \bigg( d_k - n_\mathrm{k} \, |d_\mathrm{i}| \bigg) \right]^\mathrm{F, M-1} \, . \label{equ:implicit_laplacian}
\end{align}
Note that the diffusivity $\mu$ can vary between the iterations, e.g., if the non-linear $p$-Poisson approach is considered, where $\mu^M= |(\nabla_k w)^{M-1}|^{p-2}$.

\paragraph{Further Explicit Sources} may arise that are simply added to the right-hand side, viz.
\begin{align}
    \int 1 \mathrm{d} \Omega
    \approx
    \sum_P \Delta \Omega^P
    \qquad \qquad \text{and thus} \qquad \qquad
    S^\mathrm{P, M} \gets \Delta \Omega^P \, .
\end{align}
According to the Eikonal Eqn. \ref{equ:eikonal}, the Laplacian of the reformulated Hamilton-Jacobi procedure usually vanishes, except for regions with insufficient wall-distance accuracy, e.g., regions with high curvature. Hence, the term is also considered explicitly
\begin{align}
    \int \mu \frac{\partial^2 w}{\partial x_k^2} \mathrm{d} \Omega
    \approx
    \sum_{F(P)} \left[ \mu \frac{\partial w}{\partial x_k} \Delta \Gamma_k \right]^F
    \qquad \qquad \text{and thus} \qquad \qquad
    S^\mathrm{P, M} \gets \sum_{F(P)} \left[ \mu \frac{\partial w}{\partial x_k} \Delta \Gamma_k \right]^{F, M-1} \, . \label{equ:explicit_laplacian}
\end{align}
Gradient and diffusivity follow from the previous iteration.

\paragraph{Implicit Sources} of the Screened-Poisson Eqn. \ref{equ:screened_poisson} are considered implicitly, which further strengthens the main diagonal, viz.
\begin{align}
    \int w \mathrm{d} \Omega
    \approx
    \sum_P \left[ w \, \Delta \Omega \right]^P
    \qquad \qquad \text{and thus} \qquad \qquad
    A^\mathrm{P, M} \gets \Delta \Omega^P \, .
\end{align}

\subsubsection{Stability Measures}
\label{subsec:stabilization}
Stabilization measures, crucial for the success of the computational methods, are briefly outlined below.
\begin{itemize}
    \item Negative values in the wall distance field occasionally might occur during the iterative process. At the end of each outer iteration, they are rigorously set to zero, ensuring the process's robustness.
    \item Additionally, situations may arise in which the gradient of the wall distance field disappears. In this case, there are no main diagonal contributions for the Eikonal, Hamilton-Jacobi (for $\varepsilon = 0$), or the p-Poisson (for $p>2$) method. Hence, a comparatively small, positive diffusivity is added in all cases both (implicitly, cf. Eqn. \eqref{equ:implicit_laplacian}) on the left and (explicitly, cf. Eqn. \eqref{equ:explicit_laplacian}) on the right side in the sense of a deferred correction approach. This approach ensures $A^{P,M} > 0$, thus favoring the numerical process, allowing higher relaxation values $\omega$, and eliminating itself (except for non-orthogonality related issues) in a converged scenario due to the same contribution on the right-hand side. The additional viscosity is set to $\mu^\text{dc} = 10^{-5}\SI{}{m}$.
    \item The additional Laplace term resulting from reformulating the Eikonal system disappears in many cases and may only differ from zero in localized areas of specific model or approximation inaccuracies. It has been shown that truncating the contribution by $\mu \nabla_k (\nabla_k \, w) \to \mathrm{max}(\mathrm{min}(\mu \nabla_k (\nabla_k \, w), 1 ),-1)$ hardly influences the solutions but leads to a significantly more stable numerical process. A similar strategy has also been reported by \cite{tucker2005computations}, and the clipping can be justified by raising curvature arguments (cf. \cite{belyaev2015variational}).
    \item The viscosity of the $p$-Poisson approach tends to amplify the employed gradient for higher $p$. The exponent is gradually increased or linearly ramped over the first (typically 1000) outer iterations to favor the numerical process. Especially for high exponents of  $4 \le p \le 6$, the numerical behavior is more stable, cf. \cite{wukie2017p}.
    \item Normalizing the Laplace-based distance function requires a certain regularization in regions of vanishing gradient field to avoid singularities. For this reason, the Eqn. \eqref{equ:laplacian_normalization} is rewritten to $w = \tilde{w}/\mathrm{max}(\nabla_k \tilde{w}, 10^{-15})$.
\end{itemize}


\section{Verification \& Validation}
\label{sec:verification_validation}

\subsection{Reference Solution by Geometric Wall Distance Determination}
\label{subsec:goemetric_reference}
\begin{figure}[!htb]
\centering
\iftoggle{tikzExternal}{
\input{./tikz/02__scaling/scaling.tikz}}{
\includegraphics{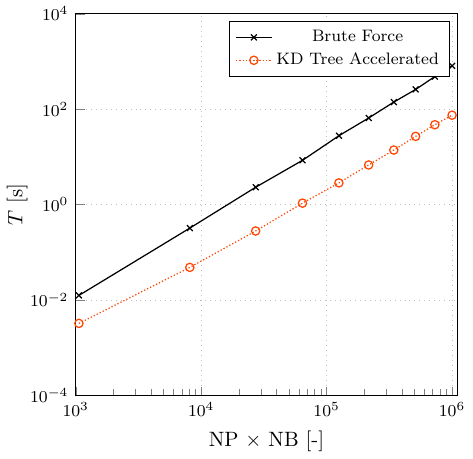}}
\caption{Measured wall clock time $T$ [s] over the number of discrete volume and surface degrees of freedom NP $\times$ NR for a simple search-find method (black), as well as an accelerated KD-Tree-based approach (orange) for the geometrically exact wall distance determination inside a unit box.}
\label{fig:geometric_scaling}
\end{figure}

Throughout the document, the quality of the modeled wall distance is judged by comparing against the exact distance function based on a straightforward geometric method. The latter is exact except for errors induced by the spatial discretization but scales with the number of control volumes (NP) and boundary elements (NR), i.e., the effort is proportional to NP $\times$ NR, cf. Fig. \ref{fig:finite_volume_approximation}. An entirely walled unit cube is cartesian and homogeneously gridded, where the number of discrete degrees of freedom varies for scaling measure purposes. The time to estimate the wall distance field is measured using a brute-force Search \& Find method and an alternative approach accelerated via a KD tree. Each experiment is performed three times, and the mean values of the measured calculation times T [s] are plotted against the number of discrete degrees of freedom NP $\times$ NR in Fig. \ref{fig:geometric_scaling}. The KD-tree method is approximately one order of magnitude faster than the brute-force approach, provides the same wall distance field, and is therefore used as a reference method in the remainder of the paper.

\subsection{1D Academic Case}
\label{subsec:academic_case}
\begin{figure}[!htb]
\centering
\iftoggle{tikzExternal}{
\input{./tikz/00__verification/one_wall_1.tikz}}{
\includegraphics{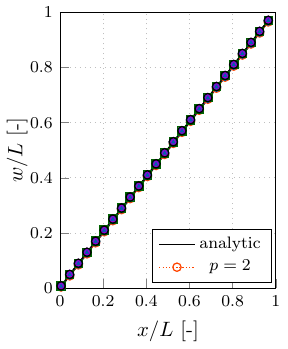}
\includegraphics{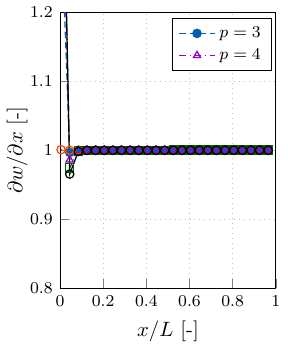}
\includegraphics{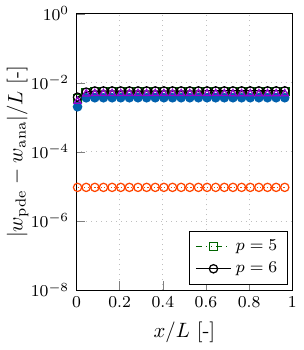}}
\caption{Generic 1D single wall case -- wall left ($x/L=0$), open end right ($x/L=1$): Verification and validation of the predicted, non-dimensional wall distance (left), its gradient (center), and its error against the analytic solution (right) for five considered p-Poisson ($p=[2,3,4,5,6]$) approaches.}
\label{fig:verification_one_wall_1}
\end{figure}

\begin{figure}[!htb]
\centering
\iftoggle{tikzExternal}{
\input{./tikz/00__verification/one_wall_2.tikz}}{
\includegraphics{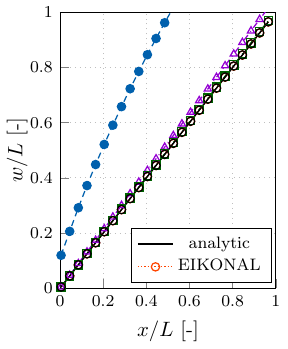}
\includegraphics{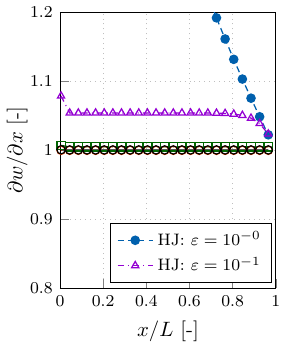}
\includegraphics{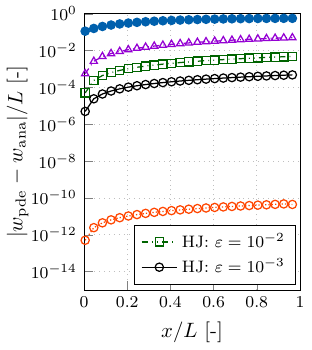}}
\caption{Generic 1D single wall case -- wall left ($x/L=0$), open end right ($x/L=1$): Verification and validation of the predicted, non-dimensional wall distance (left), its gradient (center), and its error against the analytic solution (right) for the Eikonal and four Hamilton-Jacobi ($\varepsilon = [1, 10^{-1}, 10^{-2}, 10^{-3}]$) approaches.}
\label{fig:verification_one_wall_2}
\end{figure}

\begin{figure}[!htb]
\centering
\iftoggle{tikzExternal}{
\input{./tikz/00__verification/one_wall_3.tikz}}{
\includegraphics{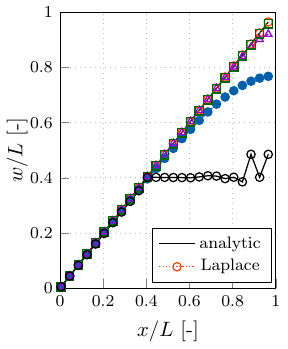}
\includegraphics{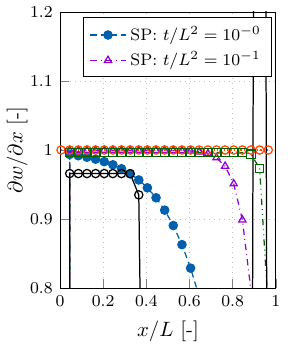}
\includegraphics{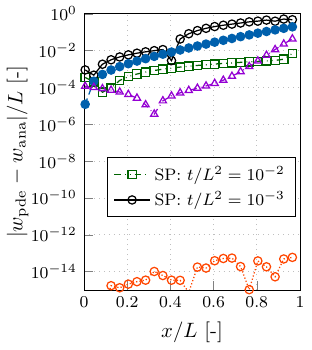}}
\caption{Generic 1D single wall case -- wall left ($x/L=0$), open end right ($x/L=1$): Verification and validation of the predicted, non-dimensional wall distance (left), its gradient (center), and its error against the analytic solution (right) for the Laplace and four Screened-Poisson ($t/L^2 = [1, 10^{-1}, 10^{-2}, 10^{-3}]$) approaches.}
\label{fig:verification_one_wall_3}
\end{figure}

In this subsection, the wall distance models from Sec. \ref{subsec:continuous_models} and their discrete approximation's implementation from Sec. \ref{subsec:discrete_approximation} are briefly validated and verified, respectively. Additionally, basic prediction quality aspects are considered, including a brief discussion on potential strengths and weaknesses. For this purpose, two one-dimensional situations are examined, which discretize the $x_1$-coordinate in the interval $x_1=[0, L]$ with one hundred finite volumes. In both cases, a wall boundary is placed at $x_1/L=0$, but only once at $x_1/L=1$. In the other case, an open boundary condition is specified, i.e., $\partial w/ \partial x_1 = 1$ at $x_1/L = 1$, cf. Eqn. \eqref{equ:eikonal_far}. Five p-Poisson ($p=[2,3,4,5,6]$), four Hamilton-Jacobi ($\varepsilon = [1, 10^{-1}, 10^{-2}, 10^{-3}]$), and four Screened-Poisson ($t/L^2 = [1, 10^{-1}, 10^{-2}, 10^{-3}]$) studies are examined. Additionally, the Eikonal and Laplace approaches are considered. Results are shown in Figs. \ref{fig:verification_one_wall_1} - \ref{fig:verification_one_wall_3} for the single-walled case, with the non-dimensionalized wall distance $w/L$ on the left, the corresponding gradient $\partial w/\partial x_1$ in the center, and the error with respect to the analytical solution $w=x_1$. i.e., $|w-x_1|/L$, on the right, all over the non-dimensionalized coordinate $x_1/L$. All wall distances predicted by the p-Poisson methods align with the analytical solution, and differences can hardly be distinguished visually in Fig. \ref{fig:verification_one_wall_1}. Discrepancies become apparent in the gradient field and the error curve, whereby the linear $p=2$ method outperforms all other non-linear tests. Nevertheless, the highest errors are still in the range of $\mathcal{O}(10^{-2})$ percent. The Hamilton-Jacobi approaches' results in Fig. \ref{fig:verification_one_wall_2} show partly significant deviations from the analytical solution, especially for considerable viscosities, i.e., $\varepsilon \geq 10^{-1}$. The error curves show that the Eikonal method best approximates the wall distance. Similar significant differences arise for certain wall distances based on the Screened-Poisson method in Fig. \ref{fig:verification_one_wall_3}, whereby the extreme cases with small and large model parameters lead to solid errors. The Laplace method provides an exceptionally accurate prediction that is close to numerical accuracy.

\begin{figure}[!htb]
\centering
\iftoggle{tikzExternal}{
\input{./tikz/00__verification/two_wall_1.tikz}}{
\includegraphics{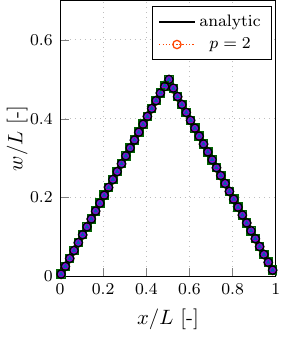}
\includegraphics{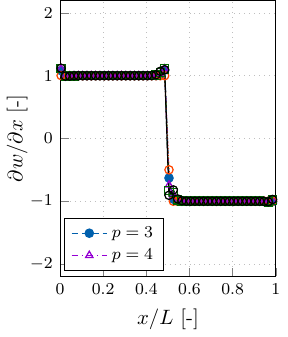}
\includegraphics{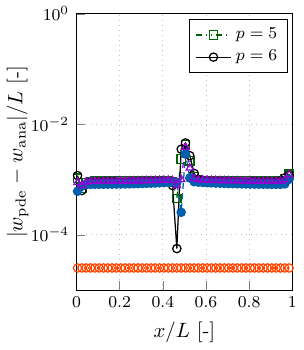}}
\caption{Generic 1D opposite positioned wall case -- wall left ($x/L=0$), wall right ($x/L=1$): Verification and validation of the predicted, non-dimensional wall distance (left), its gradient (center), and its error against the analytic solution (right) for five considered p-Poisson ($p=[2,3,4,5,6]$) approaches.}
\label{fig:verification_two_wall_1}
\end{figure}

\begin{figure}[!htb]
\centering
\iftoggle{tikzExternal}{
\input{./tikz/00__verification/two_wall_2.tikz}}{
\includegraphics{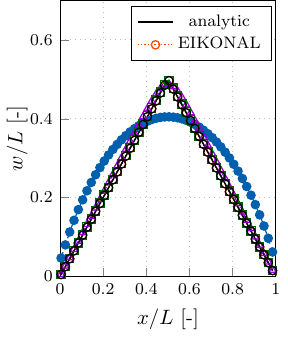}
\includegraphics{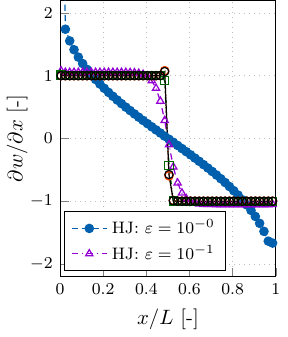}
\includegraphics{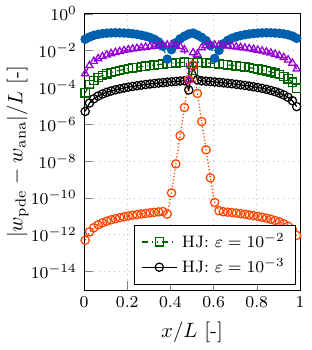}}
\caption{Generic 1D opposite positioned wall case -- wall left ($x/L=0$), wall right ($x/L=1$): Verification and validation of the predicted, non-dimensional wall distance (left), its gradient (center), and its error against the analytic solution (right) for the Eikonal and four Hamilton-Jacobi ($\varepsilon = [1, 10^{-1}, 10^{-2}, 10^{-3}]$) approaches.}
\label{fig:verification_two_wall_2}
\end{figure}

\begin{figure}[!htb]
\centering
\iftoggle{tikzExternal}{
\input{./tikz/00__verification/two_wall_3.tikz}}{
\includegraphics{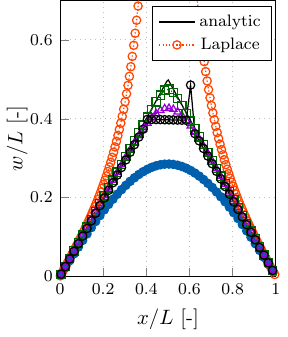}
\includegraphics{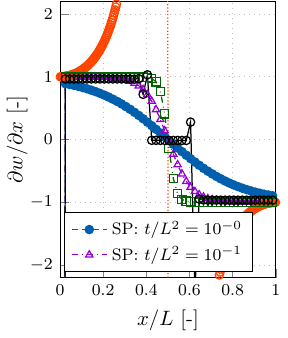}
\includegraphics{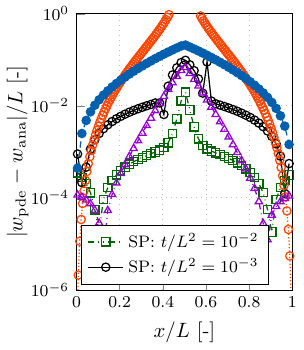}}
\caption{Generic 1D opposite positioned wall case -- wall left ($x/L=0$), wall right ($x/L=1$): Verification and validation of the predicted, non-dimensional wall distance (left), its gradient (center), and its error against the analytic solution (right) for the Laplace and four Screened-Poisson ($t/L^2 = [1, 10^{-1}, 10^{-2}, 10^{-3}]$) approaches.}
\label{fig:verification_two_wall_3}
\end{figure}

Perceptions of the results of the double-walled scenario in Figs. \ref{fig:verification_two_wall_1}-\ref{fig:verification_two_wall_3} are generally similar to those of the previous investigations. However, note that the gradient field changes abruptly at $x_1/L=1/2$, which can lead to numerical inconveniences as described in Sec. \ref{subsec:stabilization}. Again, the p-Poisson methods provide solid predictions, with the $p=2$ method performing better than its non-linear alternatives. High viscosities within the Hamilton-Jacobi approaches lead to highly distorted wall distance fields, with the Eikonal method providing the overall best predictions. In contrast to the single-walled case, the zero crossing in the gradient field now triggers singularities within the Laplace method, cf. Sec. \ref{subsec:stabilization}, which is why this method performs the weakest of all considered approaches and generates errors over 1 percent. 

\subsection{3D Engineering Case}
\label{subsec:error_over_rotation}
\begin{figure}[!htb]
\centering
\includegraphics[width=0.9\textwidth]{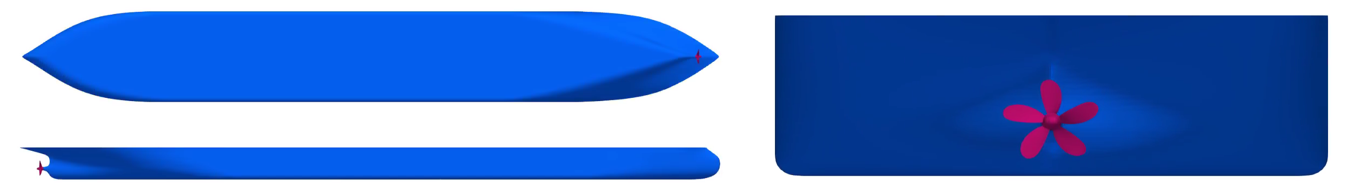}
\caption{Japan Bulk Carrier: Impressions of the geometry below the calm, free water surface, with the installed propeller highlighted in purple.}
\label{fig:propulsion__arrangement}
\end{figure}

\begin{figure}[!htb]
\centering
\includegraphics[width=0.9\textwidth]{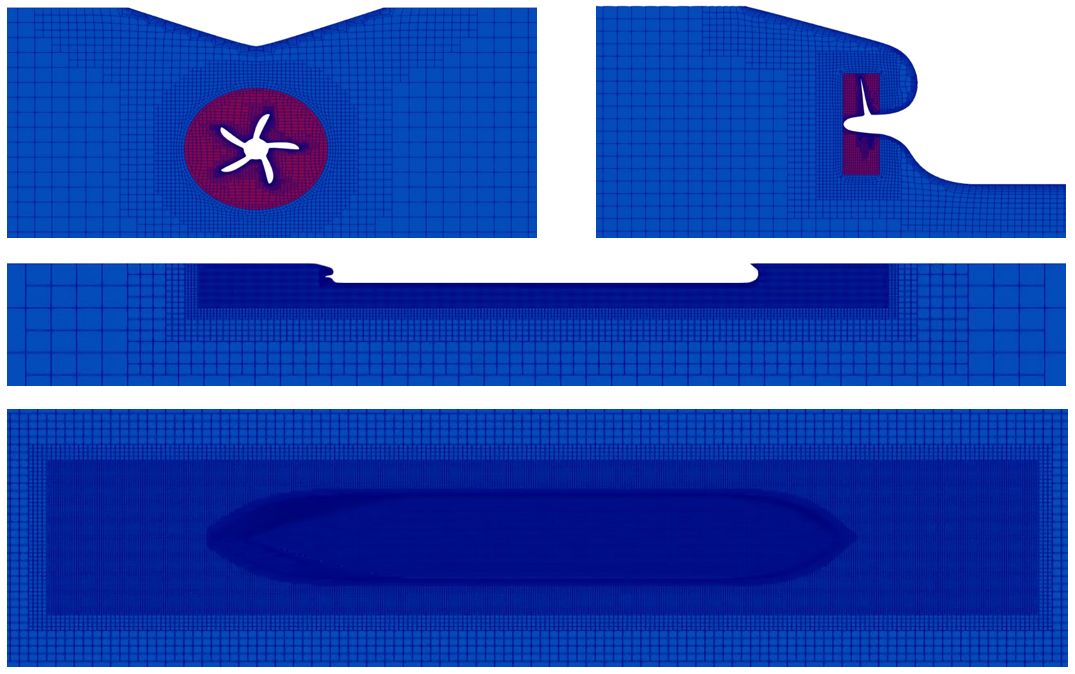}
\caption{Japan Bulk Carrier: Illustration of the numerical grids employed, with the different domains highlighted in color.}
\label{fig:propulsion__grid}
\end{figure}

The Japan Bulk Carrier (JBC) of the 2015 Tokyo CFD workshop (\cite{hino2020numerical}) serves as a real-world geometrical configuration that features strongly curved as well as partially pointy and the same time relative to each other moving walls like hull and propeller. A hull model of scale 1:40 is considered, and the analyses employ the workshop's 5-blade Mau-type marine propeller, cf. \cite{jbc} for further geometrical details. Impressions of the geometry below the calm, free water surface are shown in Fig. \ref{fig:propulsion__arrangement}, with the installed propeller highlighted in purple.
The wetted region around the geometry of Fig. \ref{fig:propulsion__arrangement} is discretized with around \SI{4E+06}{} unstructured hexahedral grids, which are refined towards the hull and propeller. The fluid domain extends over $6L_\mathrm{pp}$, $4L_\mathrm{pp}$, and $3L_\mathrm{pp}$ in longitudinal, lateral, and vertical directions. The free water surface is not resolved. Instead, a symmetry boundary condition along the undeflected calm water surface is used in line with the double-body procedure. No-slip walls along the hull [propeller] are discretized homogeneously with discrete surface elements of size $\Delta x / L_\mathrm{pp} = 2/700$ [$\Delta x / D = 4/100$], where $D$ refers to the propeller diameter. Impressions of the unstructured numerical grid are given in Fig. \ref{fig:propulsion__grid}. Rotation of the propeller is resolved via a sliding grid interface, whereby the time step is chosen such that one propeller's revolution is sampled with 360 time steps. The grid interface regions can be deduced from the differently colored grids in Fig. \ref{fig:propulsion__grid}.

\begin{figure}[!htb]
\centering
\iftoggle{tikzExternal}{
\input{./tikz/03__jbc_error_over_rotation/error_over_rotation.tikz}}{
\includegraphics{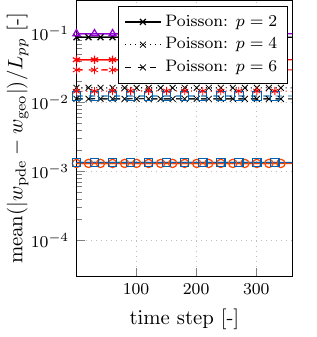}
\includegraphics{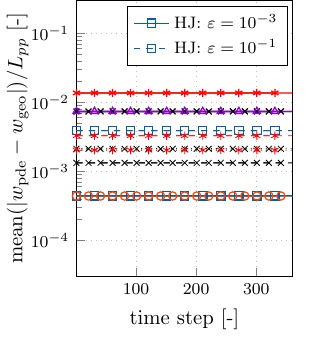}
\includegraphics{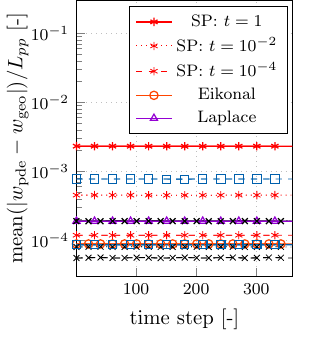}
}
\caption{Japan Bulk Carrier: Deviation between different PDE-based wall distances $w_\mathrm{pde}$ and the exact geometric distance $w_\mathrm{geo}$ over one rotation discretized with 360 time-steps for (left) the entire domain, (center) the region of $w_\mathrm{geo}/L_\mathrm{pp} < 1/10$, and (right) $w_\mathrm{geo}/L_\mathrm{pp} < 1/100$.}
\label{fig:jbc_error_over_rotation}
\end{figure}

The error of all introduced distance approximation strategies is compared against the exact geometric benchmark for one propeller rotation, i.e., for 360 time steps. The approximation of the fluid balance equations (cf. App. \ref{app:governing_equations}) is deactivated, and only the wall distance is calculated using (a) the model equations from Sec. \ref{subsec:continuous_models} and (b) geometrically utilizing the accelerated KD tree approach introduced in Sec. \ref{subsec:goemetric_reference}. The p-Poisson method for $p=[2,4,6]$, the Eikonal and Laplace methods, as well as the Hamilton-Jacobi for $\varepsilon=[10^{-3}, 10^{-1}]$ and the Screened-Poisson method for $t/L_\mathrm{pp}^2=[1, 10^{-2}, 10^{-4}]$ are considered. The normalized mean deviation between PDE-based wall distance $w_\mathrm{pde}$ and its exact geometric companion $w_\mathrm{geo}$, i.e., $\mathrm{mean}(|w_\mathrm{pde} - w_\mathrm{geo})|L_\mathrm{pp}$ is recorded over time, and results are shown in Fig. \ref{fig:jbc_error_over_rotation} for the entire fluid domain (left) and two regions closer to the wall, i.e., $w_\mathrm{geo}/L_\mathrm{pp} \leq 10$ (center) and $w_\mathrm{geo}/L_\mathrm{pp} \leq 1$ (right).
Generally, all equation-based methods converge to the exact solution in regions close to the wall. The p-Poisson results (black) follow the theory that higher p-values provide a more accurate wall distance. However, the pure Eikonal approach (orange) is more accurate in almost all cases; only in the case of $w_\mathrm{geo}/L_\mathrm{pp} < 1/100$ the deviations are in the range of the p-Poisson method with $p=4$; $p=6$ even outperforms the Eikonal approach. Hamilton-Jacobi solutions (blue) for $\varepsilon = 10^{-3}$ align with those of the Eikonal method, so the additional viscosity is almost ineffective. As theoretically expected, higher viscosities within the Hamilton-Jacobi strategy provide significant deviations from the Eikonal equation and, thus, the actual wall distance. The deviation of the Laplace approach (purple) tends to be slightly above the overall mean value but is in no case the most incorrect method. This is the Screened-Poisson method (red) close to the wall for $t/L_\mathrm{pp}^2 = 1$. A reduction to $t/L_\mathrm{pp}^2 = 10^{-2}$ reduces the approximation error, but a further decrease to $t/L_\mathrm{pp}^2 = 10^{-4}$ again increases the approximation error.

\begin{figure}[!htb]
\centering
\subfigure[]{
\includegraphics[width=0.31\textwidth]{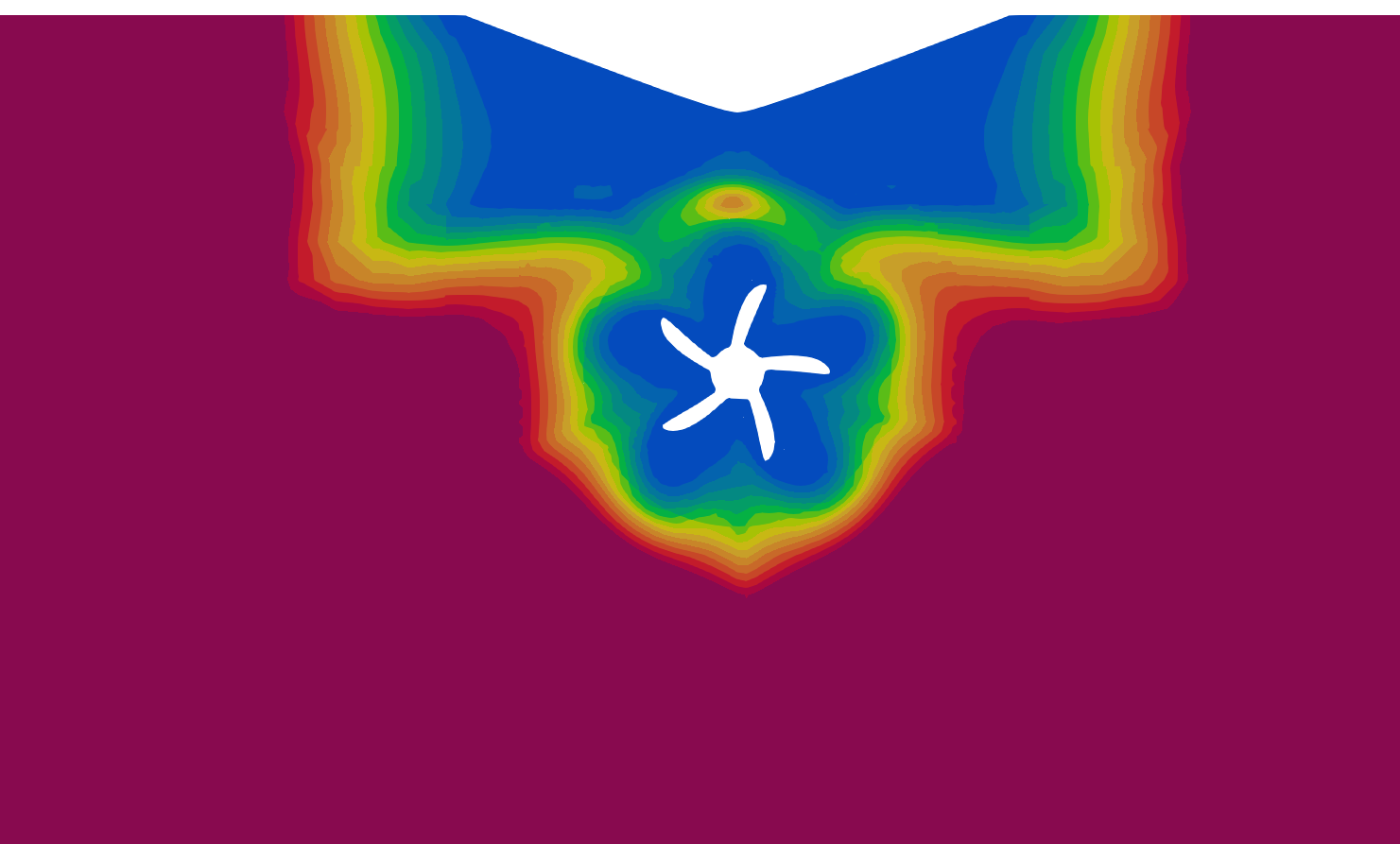}
}
\subfigure[]{
\includegraphics[width=0.31\textwidth]{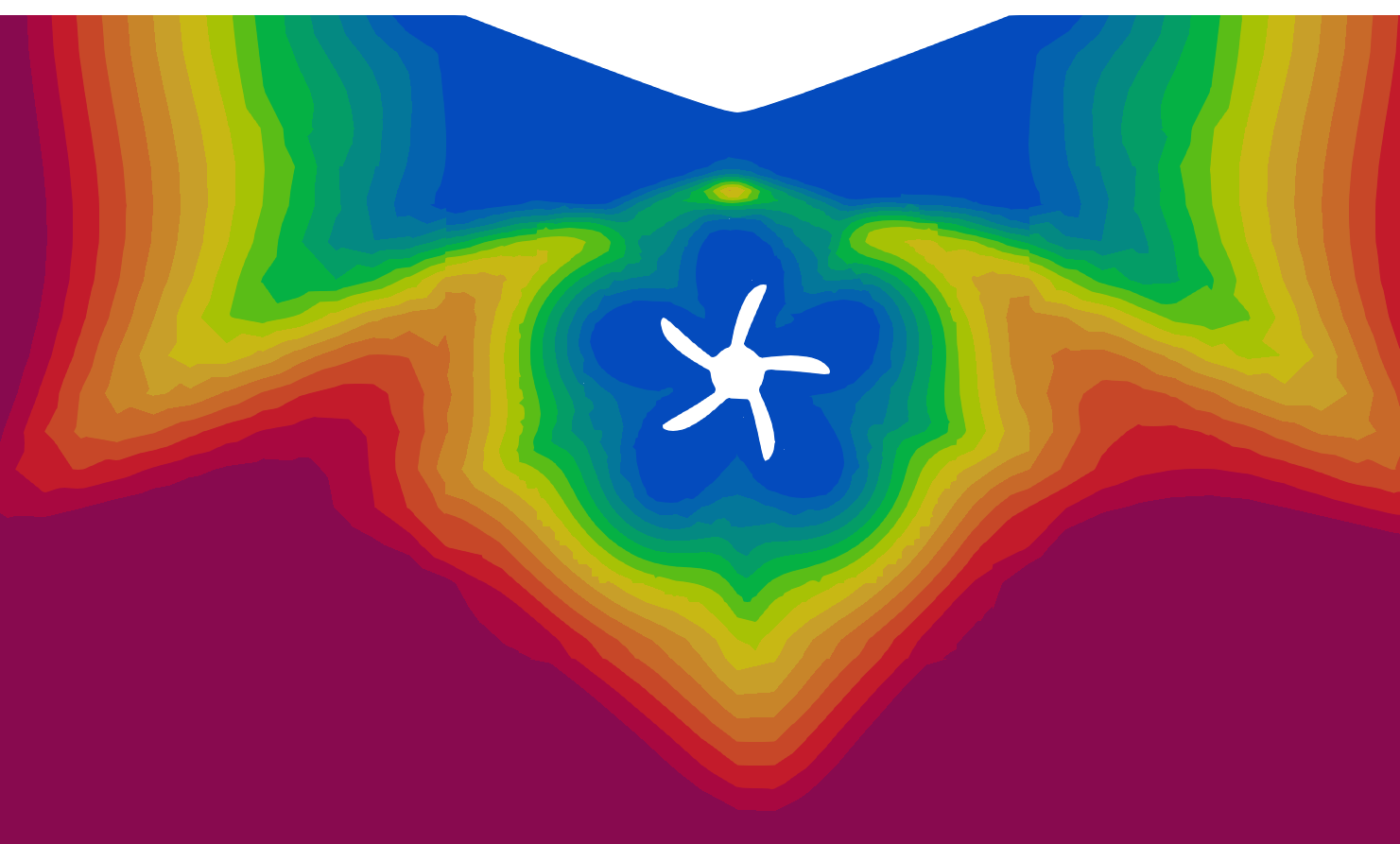}
}
\subfigure[]{
\includegraphics[width=0.31\textwidth]{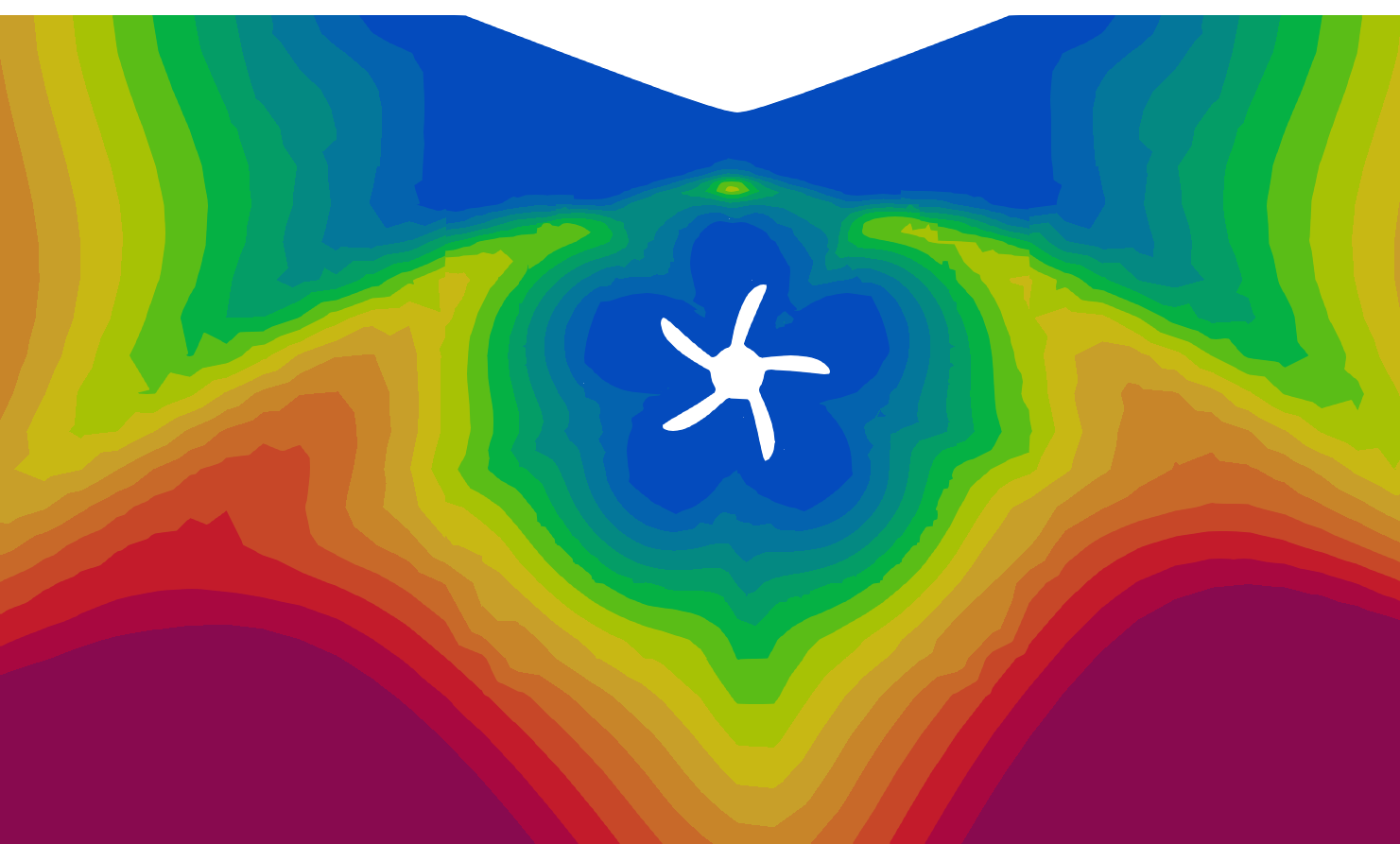}
}
\subfigure[]{
\includegraphics[width=0.31\textwidth]{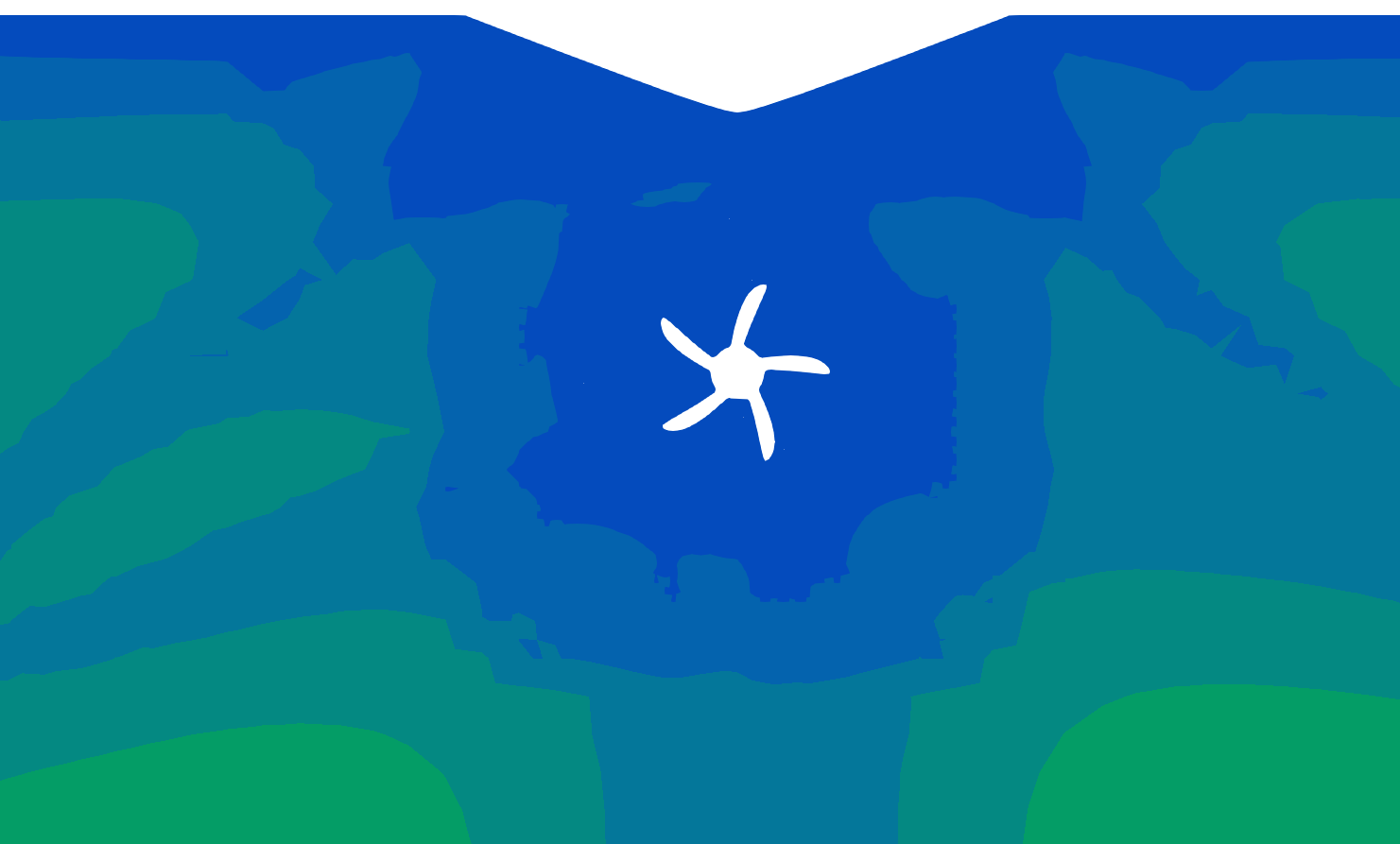}
}
\subfigure[]{
\includegraphics[width=0.31\textwidth]{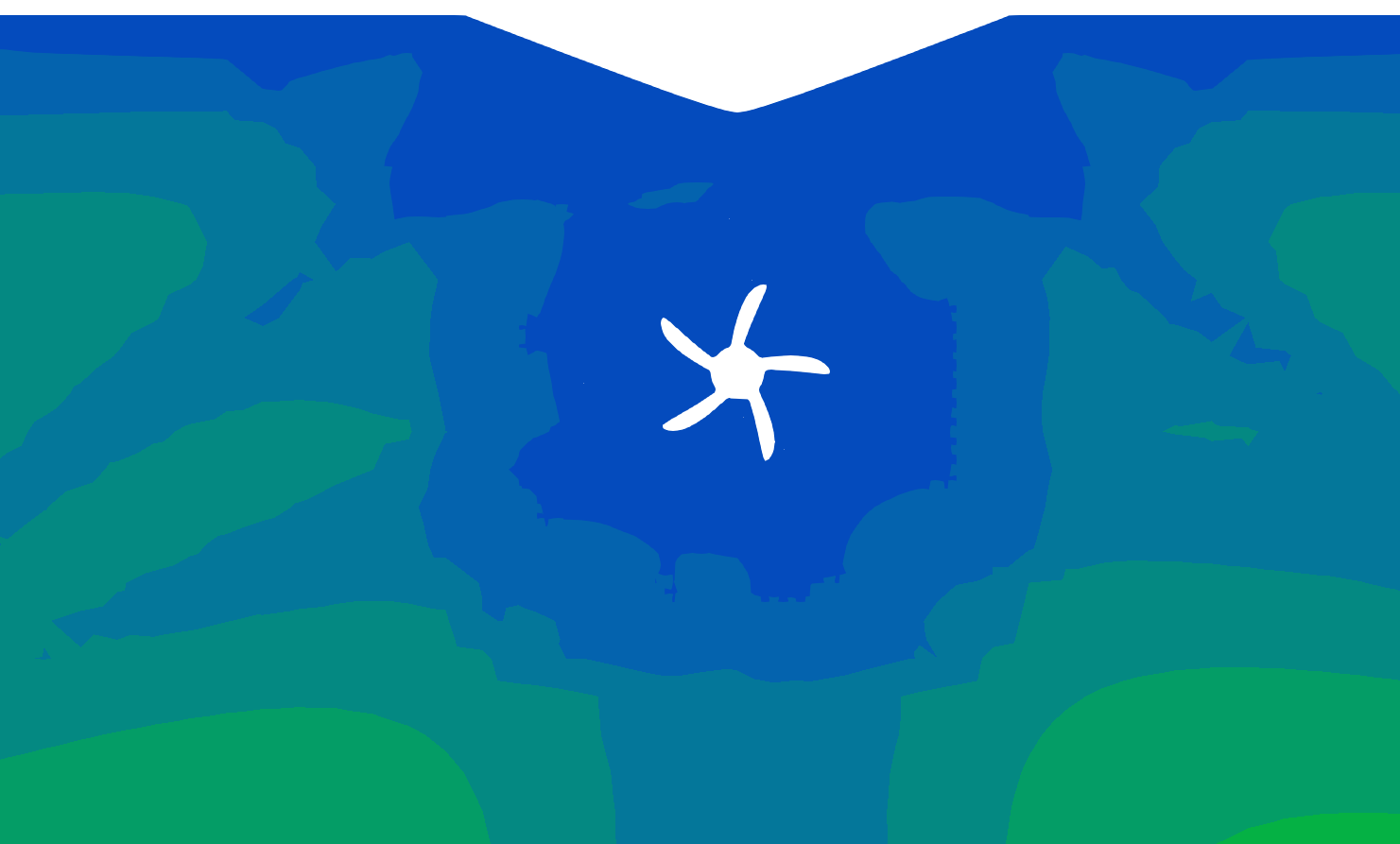}
}
\subfigure[]{
\includegraphics[width=0.31\textwidth]{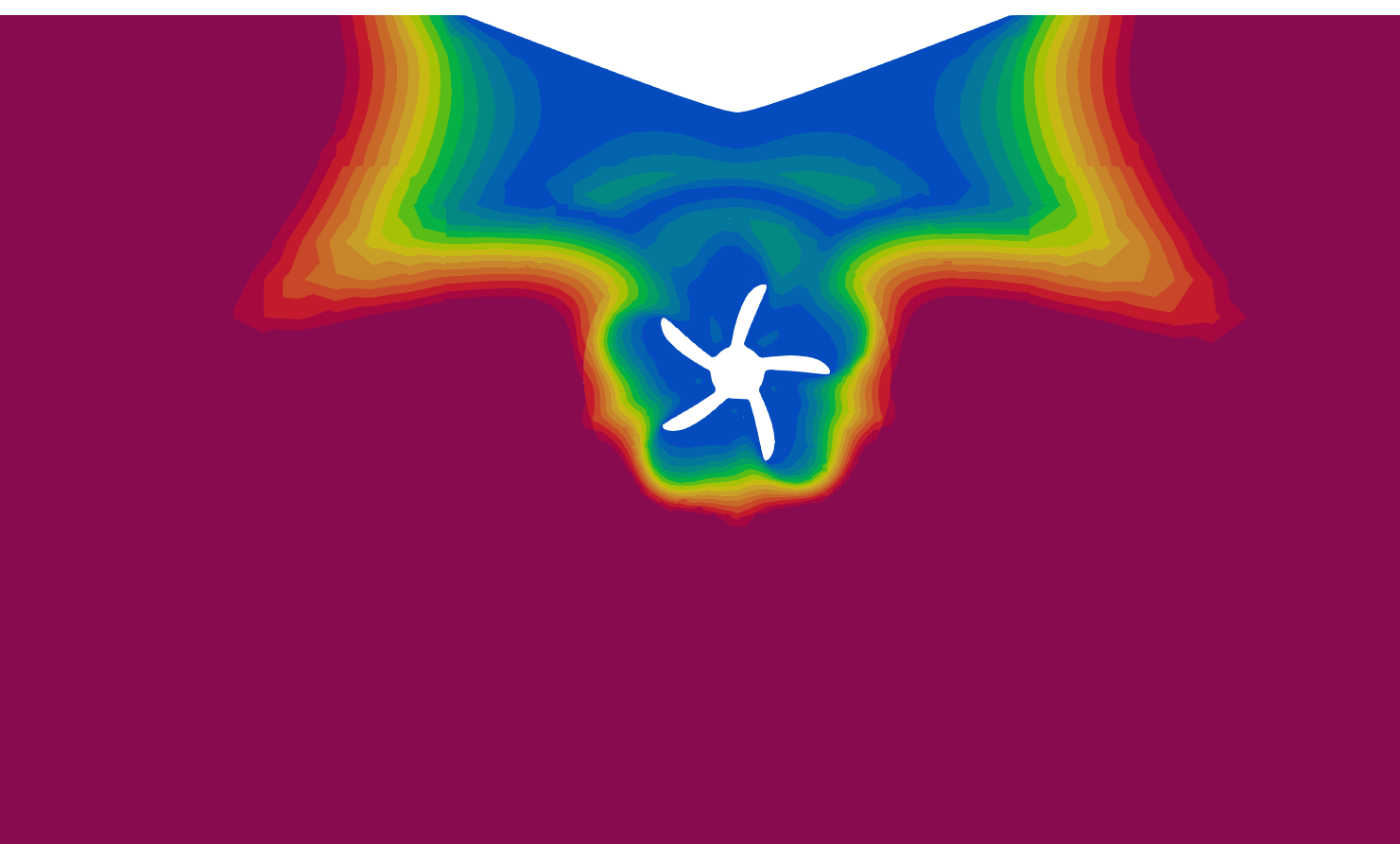}
}
\subfigure[]{
\includegraphics[width=0.31\textwidth]{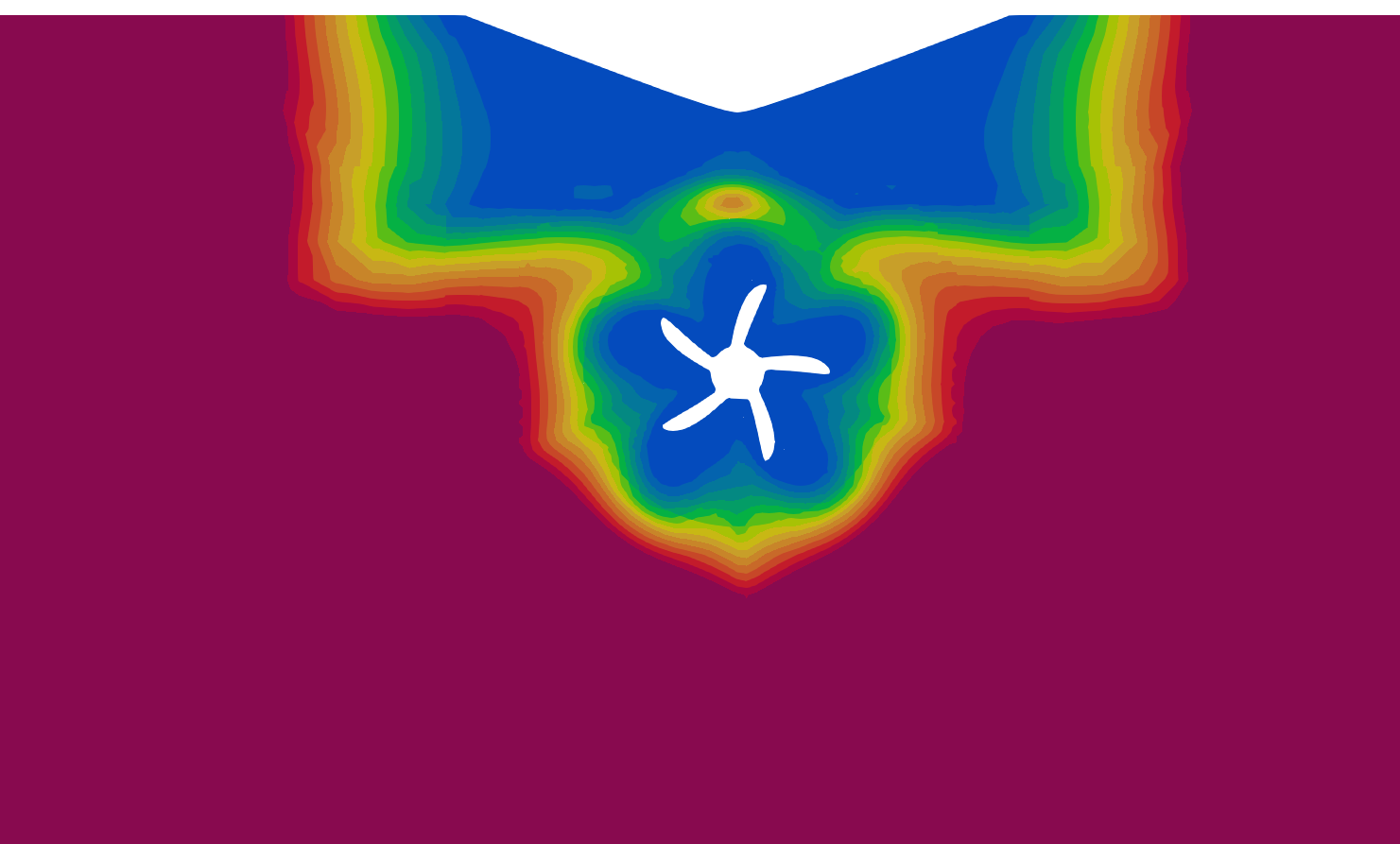}
}
\subfigure[]{
\includegraphics[width=0.31\textwidth]{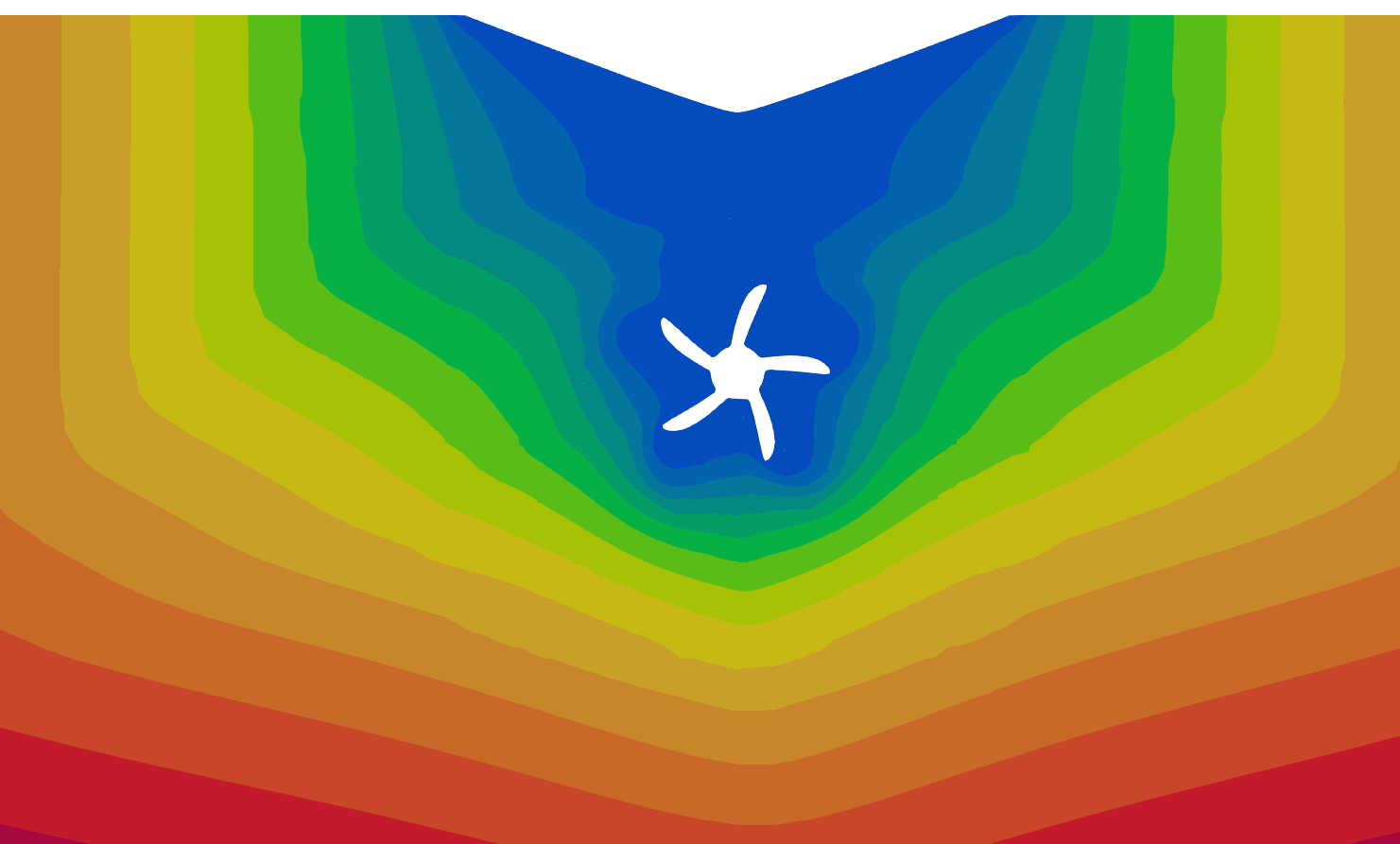}
}
\subfigure[]{
\includegraphics[width=0.31\textwidth]{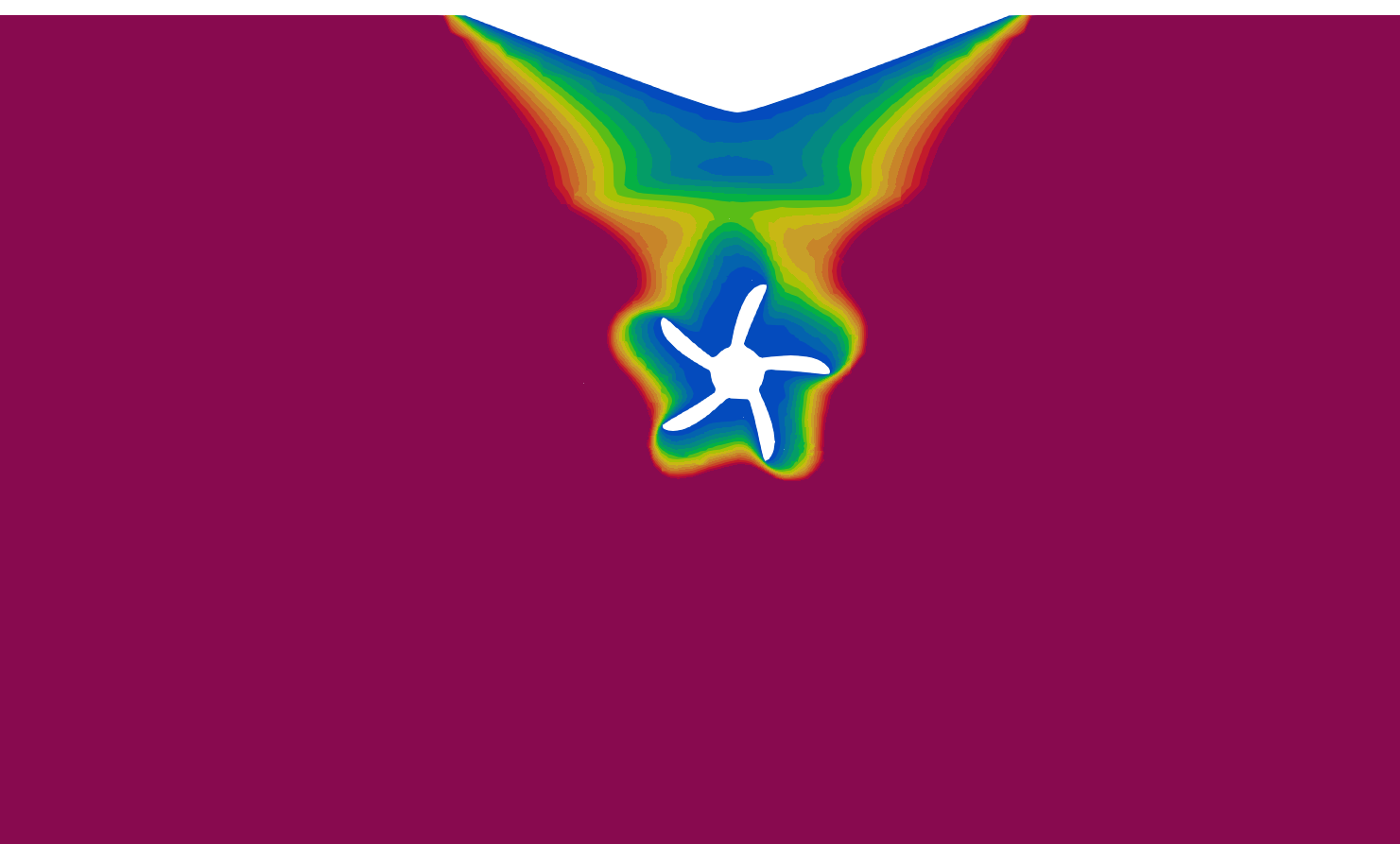}
}
\subfigure[]{
\includegraphics[width=0.31\textwidth]{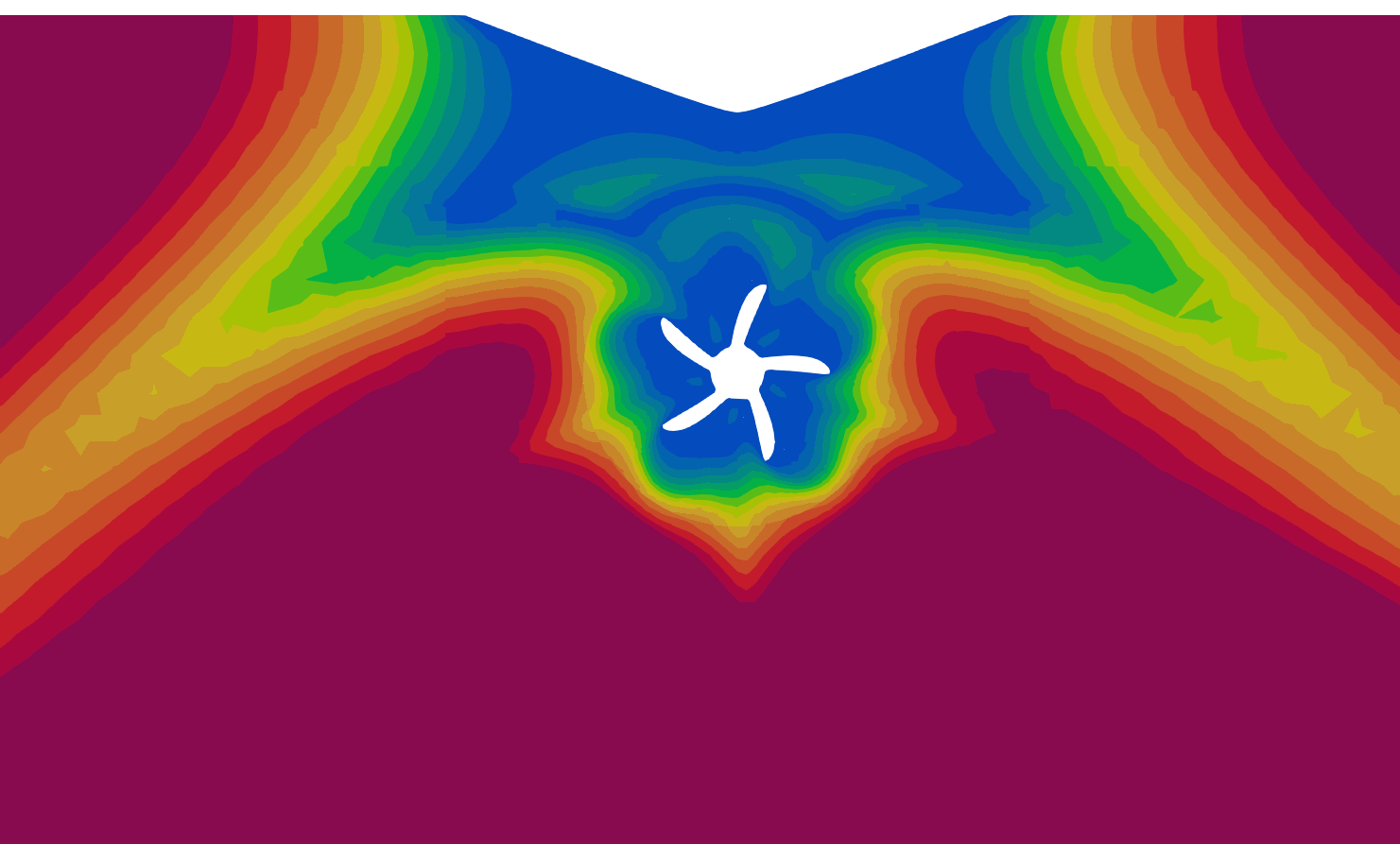}
}
\iftoggle{tikzExternal}{
\input{./tikz/03__jbc_error_over_rotation/legend_x_slices.tikz}}{
\includegraphics{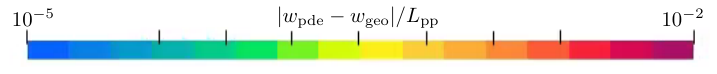}
}
\caption{Japan Bulk Carrier: Approximation difference $|w_\mathrm{pde} - w_\mathrm{geo}|L_\mathrm{pp}$ on a slice through the propeller plane, where (a)-(c) depicts the $p=[2,4,6]$ Poisson, (d) the pure Eikonal, (e) as well as (f) the ($\varepsilon=[10^{-3}, 10^{-1}]$) Hamilton Jacobi, (g) the Laplace, and (h)-(j) all three Screened-Poisson results for $t/L_\mathrm{pp}^2=[1,10^{-2},10^{-4}]$.}
\label{fig:jbc_error_over_rotation__x_slices}
\end{figure}

\begin{figure}[!htb]
\centering
\subfigure[]{
\includegraphics[width=0.31\textwidth]{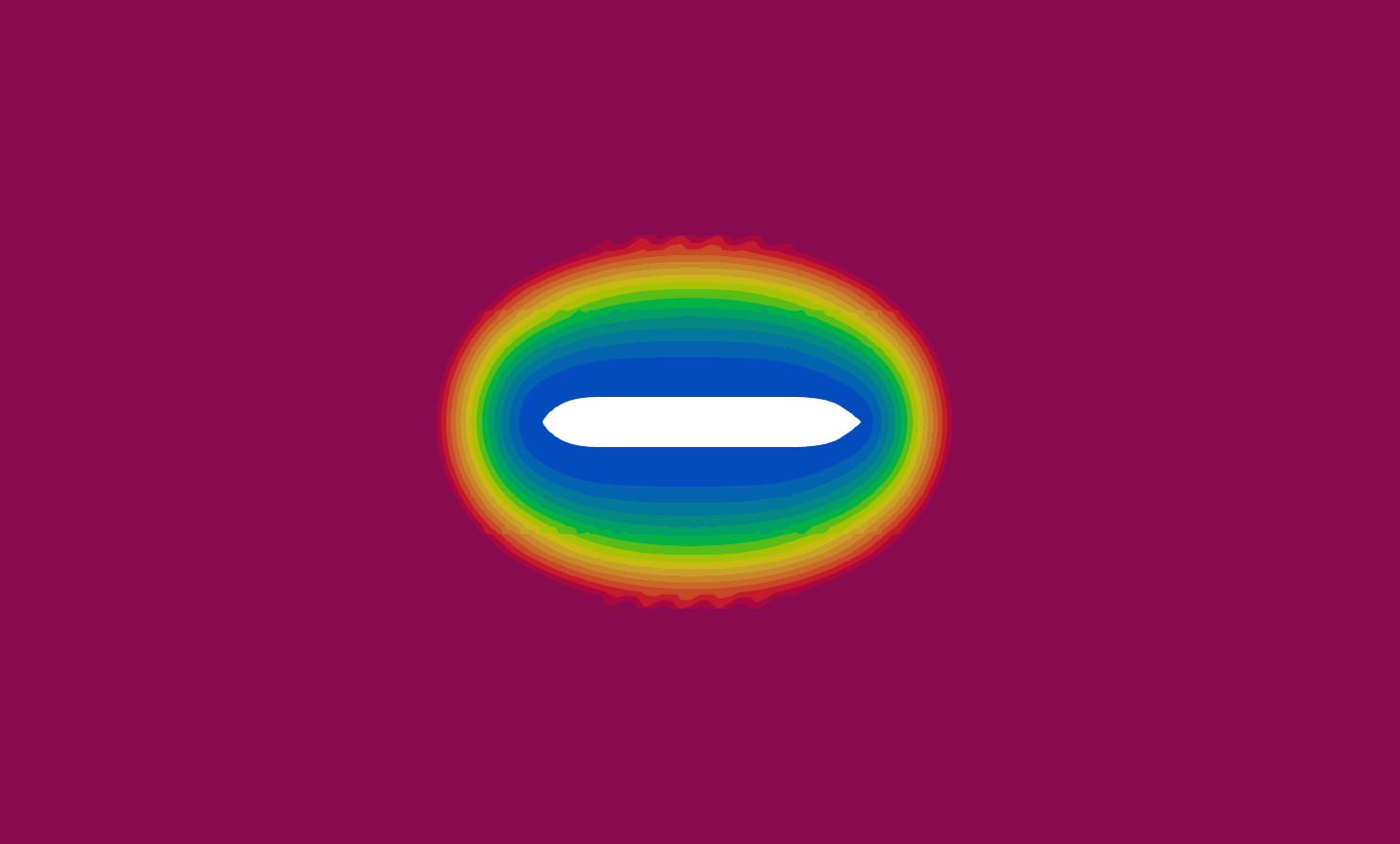}
}
\subfigure[]{
\includegraphics[width=0.31\textwidth]{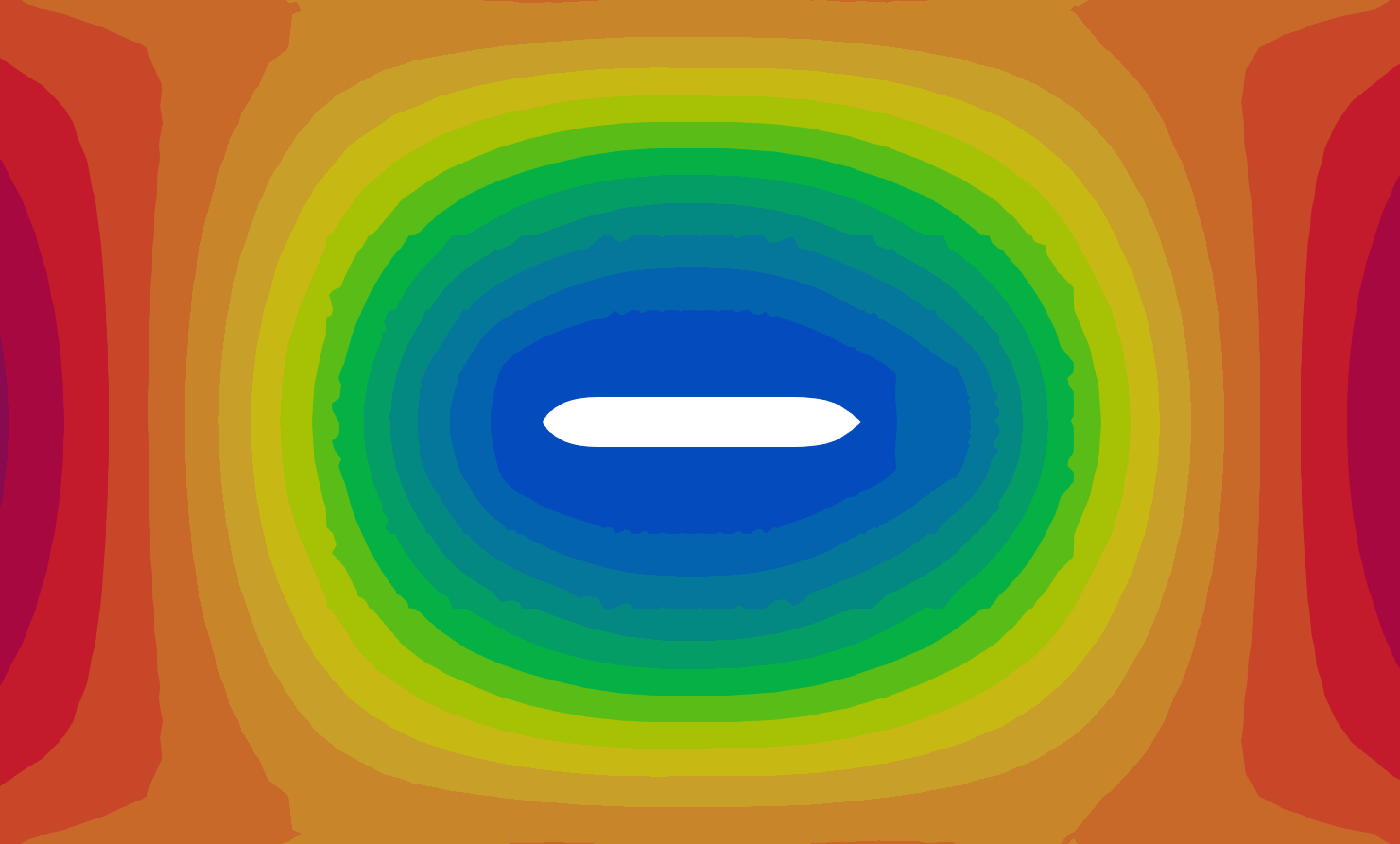}
}
\subfigure[]{
\includegraphics[width=0.31\textwidth]{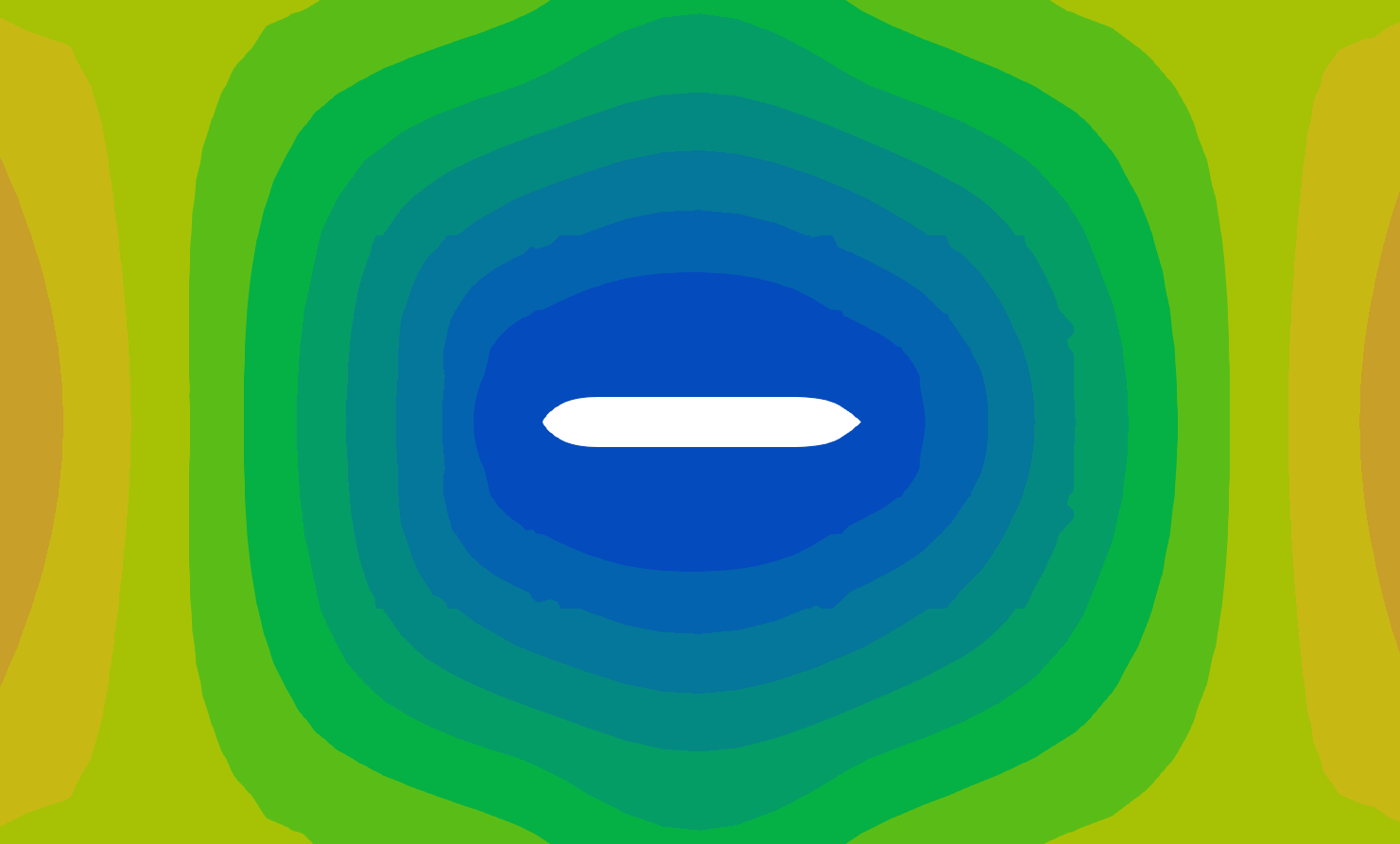}
}
\subfigure[]{
\includegraphics[width=0.31\textwidth]{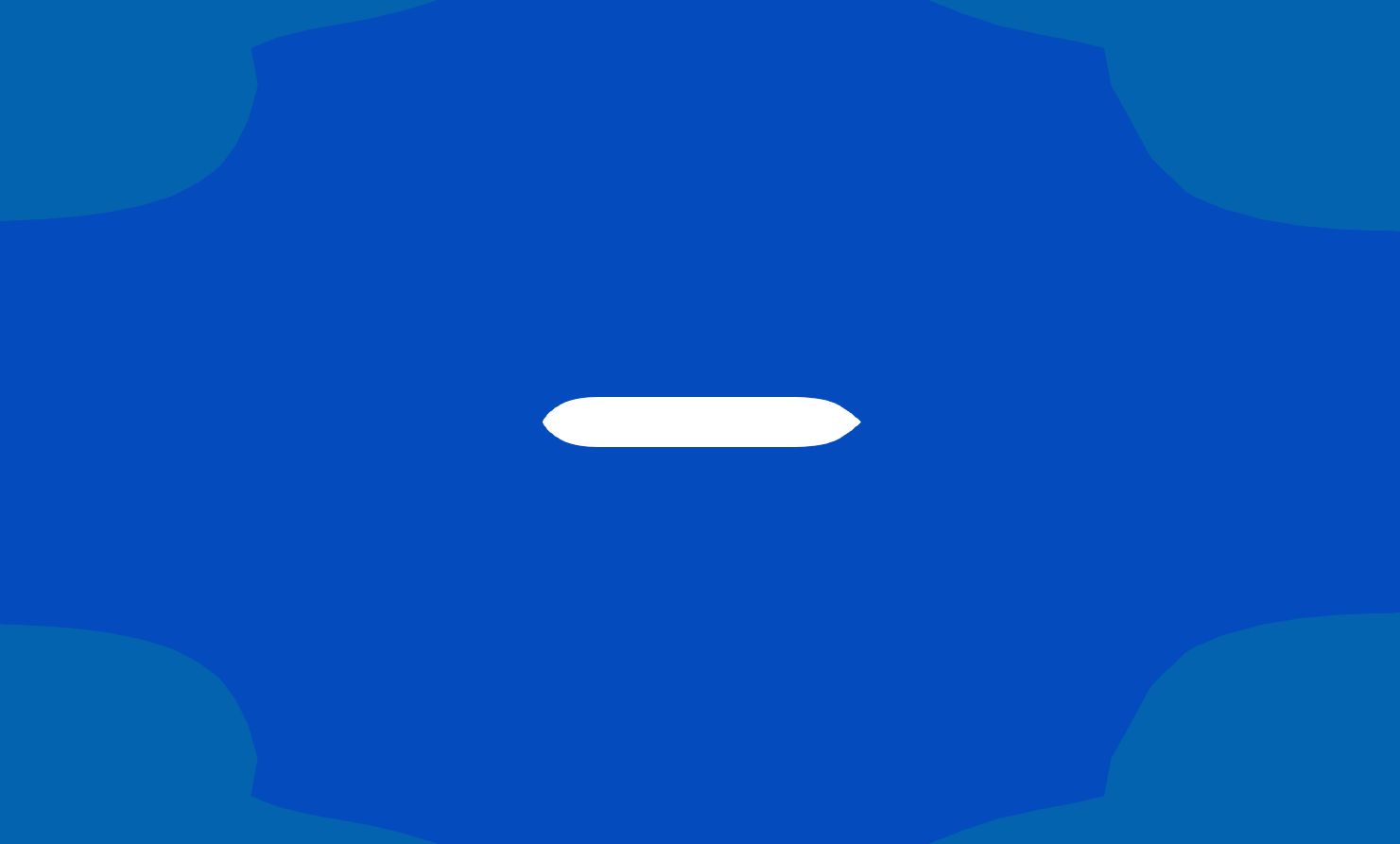}
}
\subfigure[]{
\includegraphics[width=0.31\textwidth]{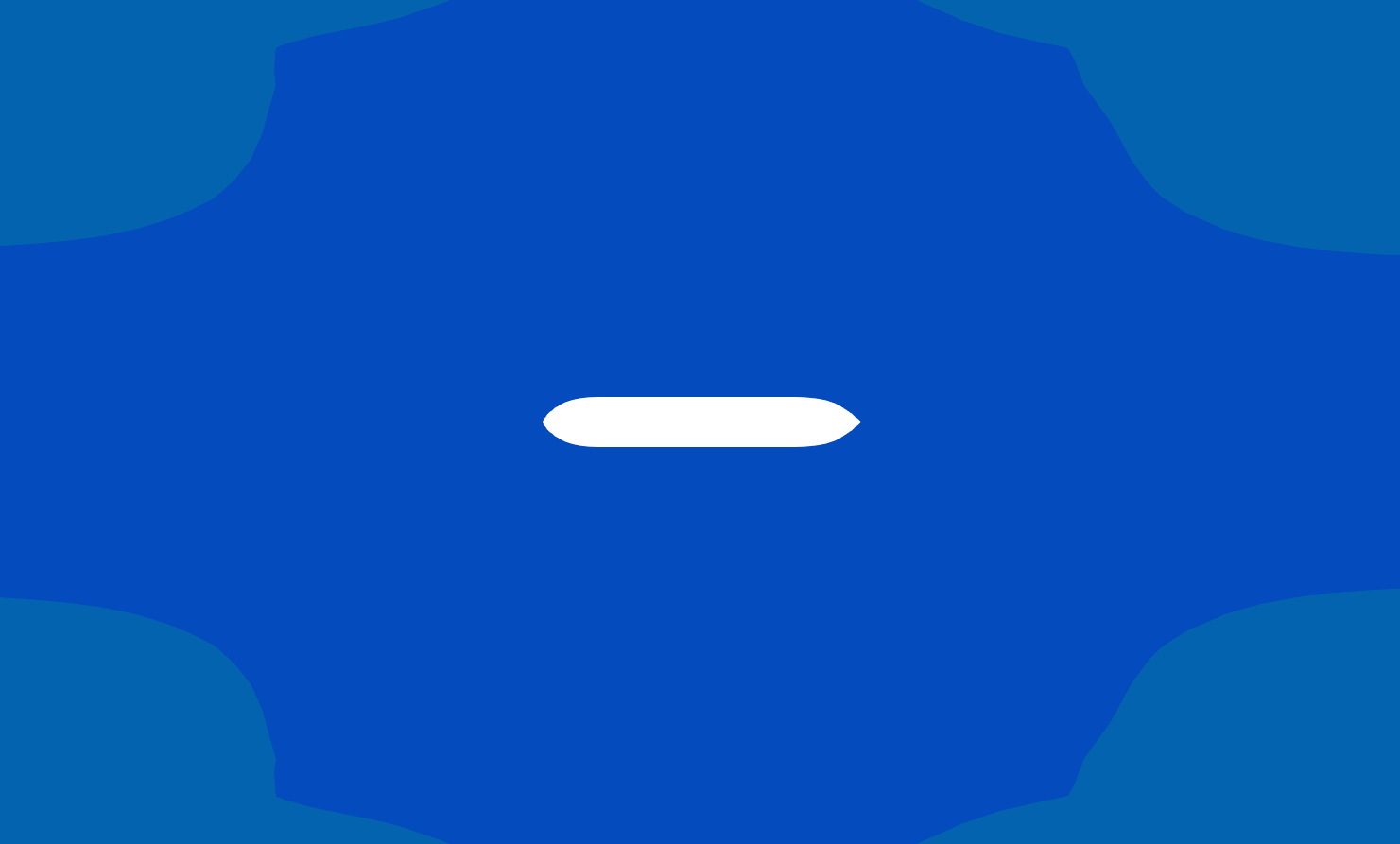}
}
\subfigure[]{
\includegraphics[width=0.31\textwidth]{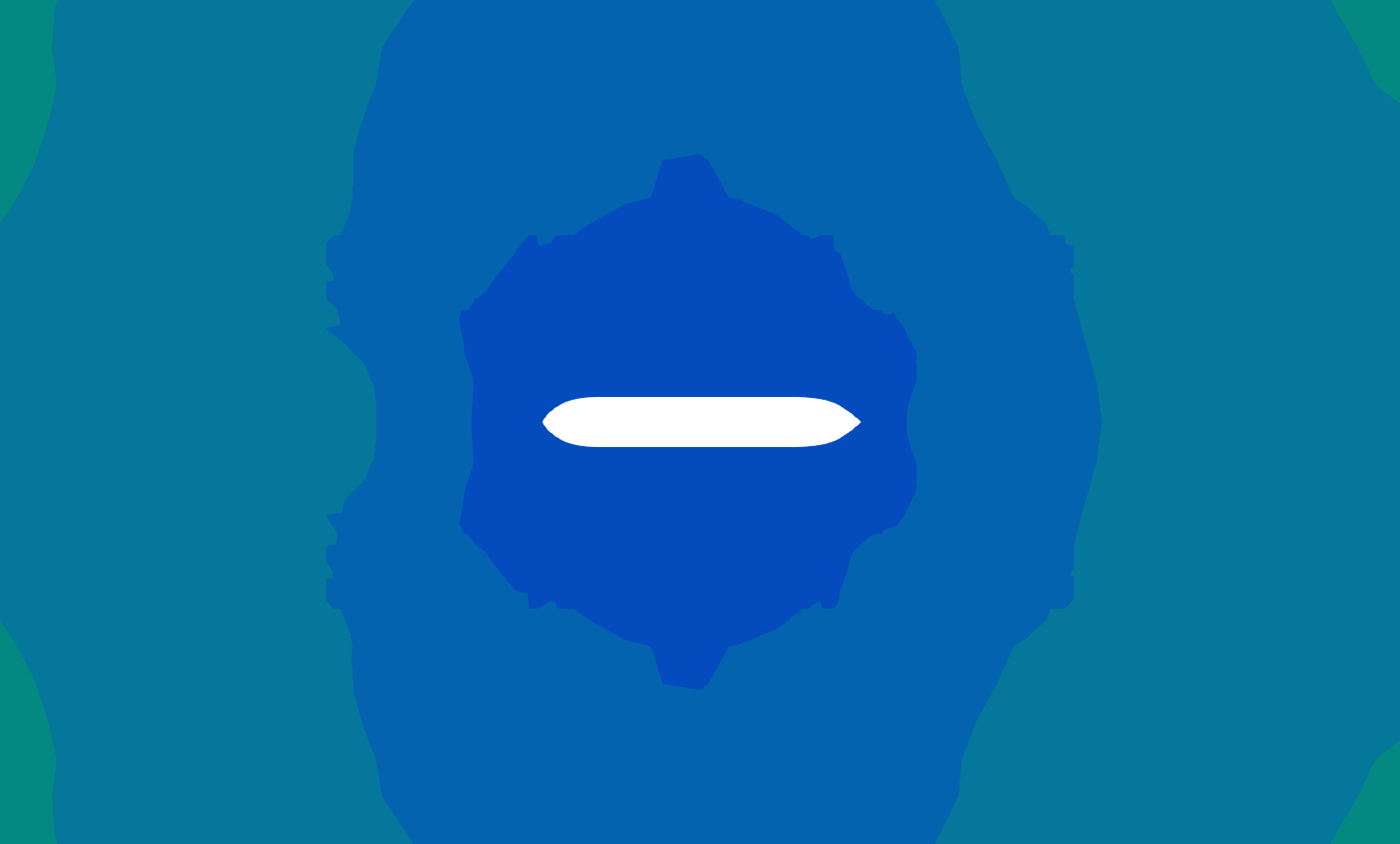}
}
\subfigure[]{
\includegraphics[width=0.31\textwidth]{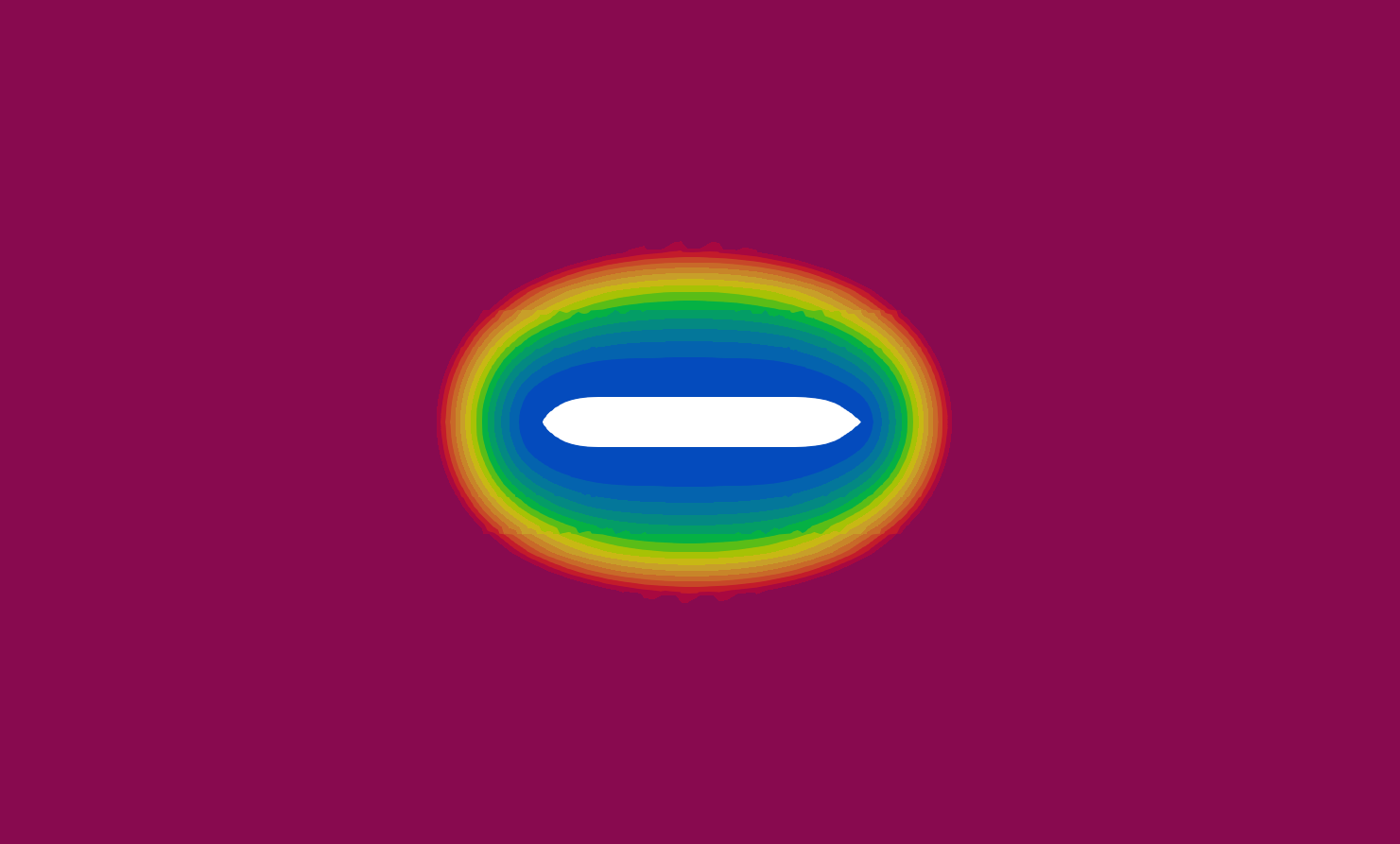}
}
\subfigure[]{
\includegraphics[width=0.31\textwidth]{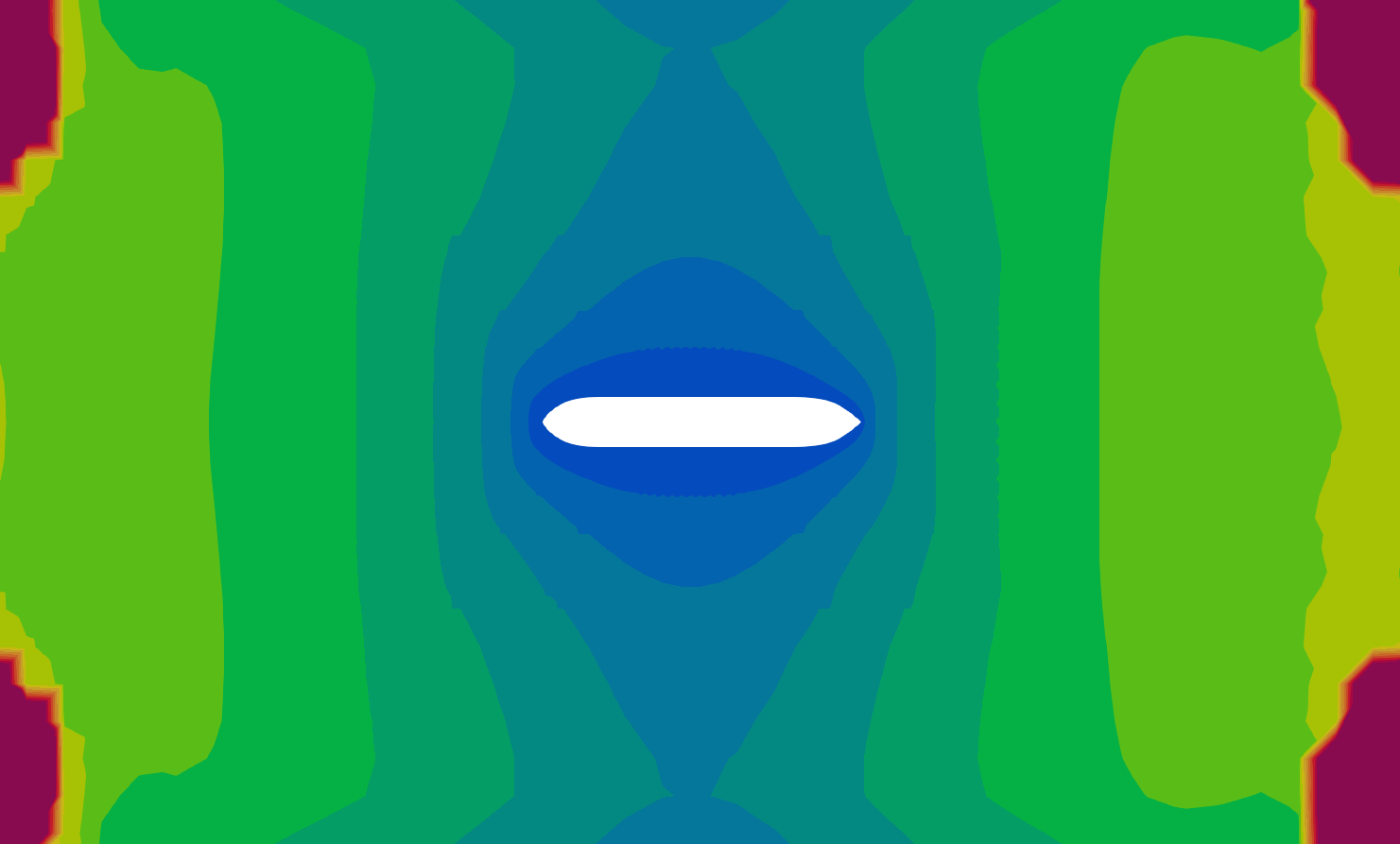}
}
\subfigure[]{
\includegraphics[width=0.31\textwidth]{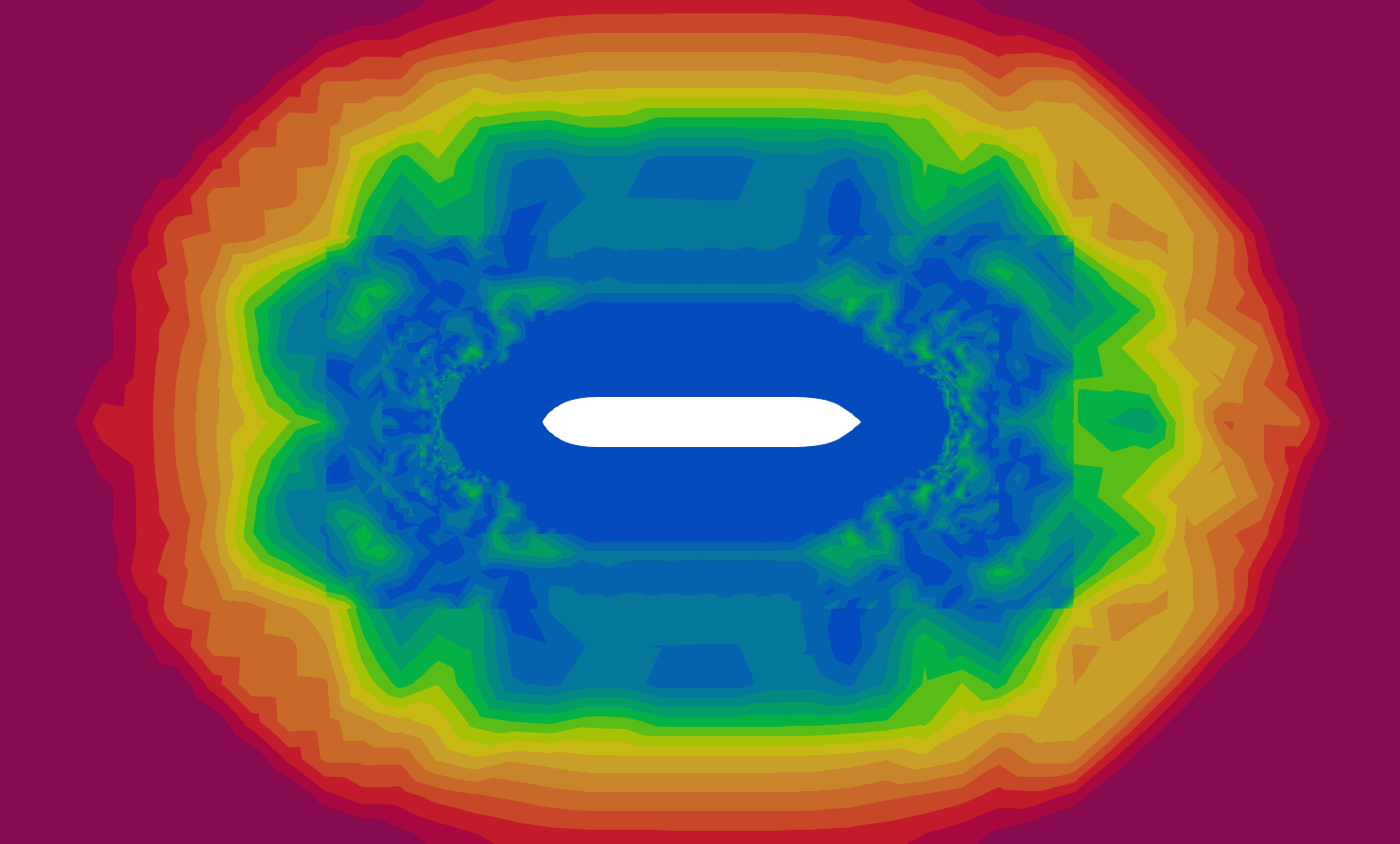}
}
\subfigure[]{
\includegraphics[width=0.31\textwidth]{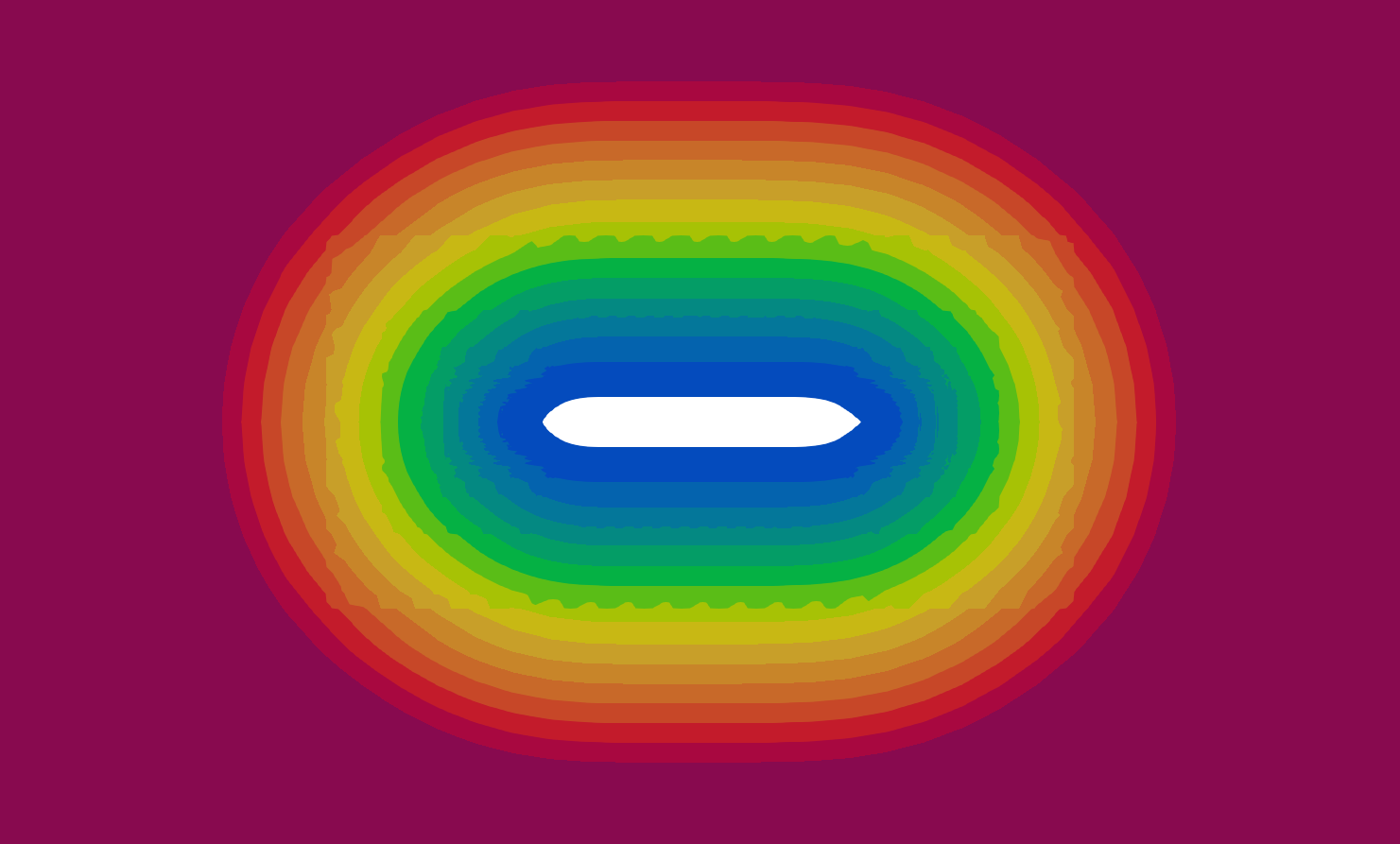}
}
\iftoggle{tikzExternal}{
\input{./tikz/03__jbc_error_over_rotation/legend_z_slices.tikz}}{
\includegraphics{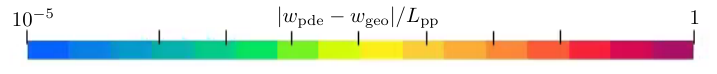}
}
\caption{Japan Bulk Carrier: Approximation difference $|w_\mathrm{pde} - w_\mathrm{geo}|L_\mathrm{pp}$ along the upper symmetry plane, where (a)-(c) depicts the $p=[2,4,6]$ Poisson, (d) the pure Eikonal, (e) as well as (f) the ($\varepsilon=[10^{-3}, 10^{-1}]$) Hamilton Jacobi, (g) the Laplace, and (h)-(j) all three Screened-Poisson results for $t/L_\mathrm{pp}^2=[1,10^{-2},10^{-4}]$.}
\label{fig:jbc_error_over_rotation__z_slices}
\end{figure}

In addition to the integral measures, Figs. \ref{fig:jbc_error_over_rotation__x_slices}-\ref{fig:jbc_error_over_rotation__z_slices} provide impressions of the local difference for longitudinal slices through the propeller plane as well as along the upper symmetry plane. Different scales were used for improved visualization.
In the figures, (a)-(c) represent the $p=[2,4,6]$ Poisson results, (d) depicts the pure Eikonal response, and (e) as well as (f) show the Hamilton Jacobi results for $\varepsilon=[10^{-3}, 10^{-1}]$ respectively. The Laplace result is shown in (g), followed by the three Screened-Poisson results in (h)-(j) for $t/L_\mathrm{pp}^2=[1,10^{-2},10^{-4}]$. The integral results from Fig. \ref{fig:jbc_error_over_rotation} are reflected in the fact that the p-Poisson results (a)-(c) for higher $p$'s provide a reduced approximation error. The smallest errors are obtained for the Eikonal (d) and the (e) Hamilton-Jacobi method with low viscosity. The remaining strategies adequately approximate the distance function close to the ship, but deviate significantly in the far field. The theoretic deficiency of the Laplace approach with oppositely placed walls can be recognized in the slice through the propeller plane in Fig. \ref{fig:jbc_error_over_rotation__x_slices} (g) by an increased relative error in the area between the propeller and hull.

\begin{figure}[!htb]
\centering
\iftoggle{tikzExternal}{
\input{./tikz/03__jbc_error_over_rotation/timing.tikz}}{
\includegraphics{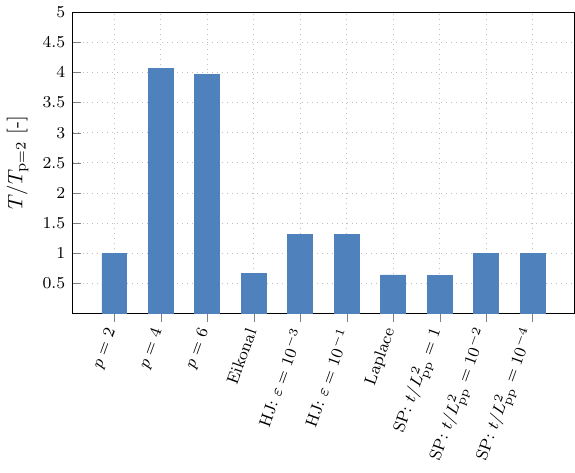}
}
\caption{Japan Bulk Carrier: The ratio of the wall clock time for several PDE-based wall distance approaches to compute the PDE-based wall distance field over the p=2 approach for one propeller rotation sampled with 360-time steps.}
\label{fig:jbc_error_over_rotation__timing}
\end{figure}

Figure \ref{fig:jbc_error_over_rotation__timing} below compares the numerical effort required in each case. For this purpose, the wall-clock time of all simulations is measured over the 360-time steps and related to the time of the linear p-Poisson $p=2$ case. The non-linear $p>2$ Poisson methods result in a significantly higher numerical effort, but all other approaches require a similar effort. All simulations were performed three times on the same numerical grid and computational hardware with identical parallelization and domain partitioning. Then, the mean value of the three simulations was used to evaluate the results.

\section{Application}
\label{sec:applicaiton} 

The paper's applications address technical flows in a maritime context and consider the influence of the introduced wall distance approximation strategies. Hydrodynamic studies utilize a Reynolds-Averaged Navier-Stokes (RANS, cf. App. \ref{app:rans}) approach on the Japan Bulk Carrier (JBC) case at model scale. Subsequently, the aerodynamics of a full-scale feeder vessel are analyzed using a scale-resolving, hybrid RANS Large-Eddy-Simulation (LES) in line with the Improved Delayed Detached-Eddy-Simulation (IDDES, cf. App. \ref{app:detached_eddy_simulation}) approach. The RANS and the DES methods employ shear stress transport (SST) strategies, which incorporate the distance to the nearest wall at several procedural points. In addition, grid morphing strategies access the (inverse) wall distance function for floatation position adaptations, cf. \cite{kuhl2021phd, lohner1996improved}.

The investigations do not seek to validate the flow solver against fluid experimental but apply a validated CFD solver, vary the wall distance approach, and check the influence against the geometrically determined reference wall distance results.

\subsection{Hydrodynamics of a Bulk Carrier}
The model scale bulk carrier from Sec. \ref{subsec:error_over_rotation} is analyzed in more detail in the following. The two-phase flow equations from App. \ref{app:governing_equations} are now considered in combination with the RANS framework, cf. App. \ref{app:rans}. Convective fluxes of the volume fraction field are approximated with a compressive High Resolution Interface Capturing (HRIC, cf. \cite{kuhl2021adjoint}) convection scheme. All other convective fluxes are approximated with the Quadratic Upstream Interpolation for Convective Kinematics (QUICK) method, and diffusive fluxes with CDS, cf. Sec. \ref{sec:approximation}. Temporal changes are represented by an implicit one or two time level method for pseudo-steady resistance and transient propulsion situations, respectively. The pressure-velocity coupling is realized with a modified Semi-implicit Method for Pressure Linked Equation (SIMPLE) method, cf. \cite{kuhl2022discrete}.

The Reynolds and Froude numbers are $\mathrm{Re} = V L_\mathrm{pp} / \nu = \SI{7.246E+06}{}$ and $\mathrm{Fn} = V / \sqrt{L_\mathrm{pp} \, G} = 0.142$, based on the towing speed $V = \SI{1.179}{m/s}$, the length between the perpendiculars $L_\mathrm{pp} = \SI{7}{m}$, the dynamic viscosity of water $\nu = \SI{1.139E-06}{m^2/s}$, and the gravitational constant of $G = \SI{9.81}{m/s^2}$. Drag studies feature no hull attachments, in particular, no energy-saving devices.
First, the hull resistance behavior is examined. Then the self-propulsion behavior is discussed.

\subsubsection{Resistance}
\begin{figure}[!h]
\centering
\subfigure[]{
\iftoggle{tikzExternal}{
\input{./tikz/04__jbc_resistance/scetch.tikz}}{
\includegraphics{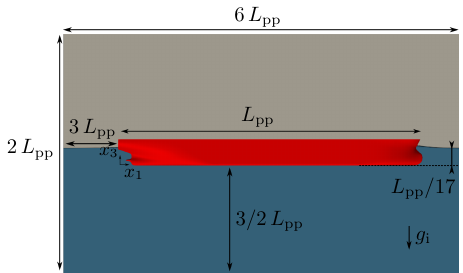}}
}
\subfigure[]{
\includegraphics[width=0.4\textwidth]{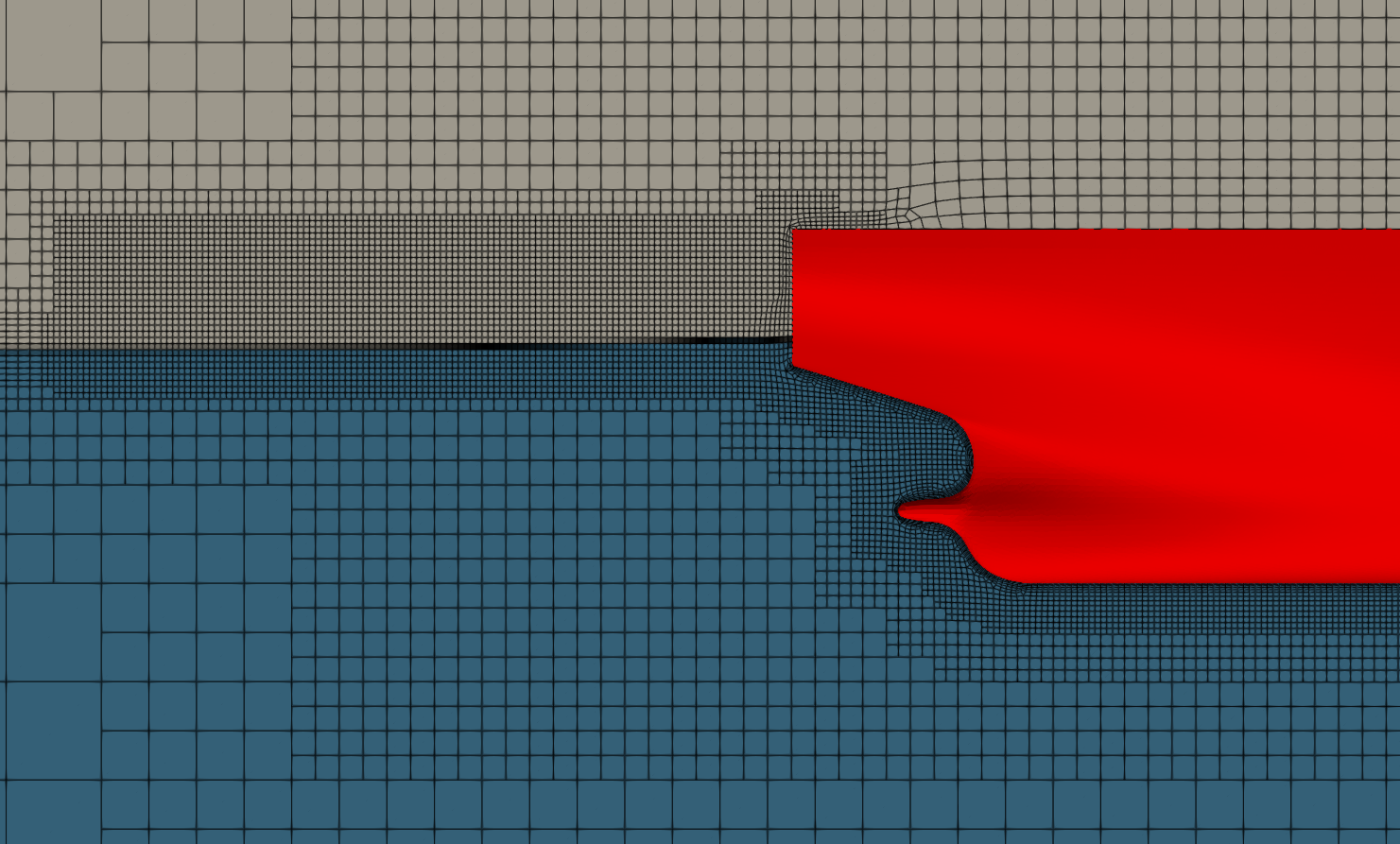}
}
\caption{Japan Bulk Carrier ($\mathrm{Re}_\mathrm{L} = \SI{7.246}{} \cdot 10^6$, $\mathrm{Fn}=0.142$): (a) Schematic drawing of the initial configuration and (b) unstructured numerical grid around the 
stern region.}
\label{fig:jbc_scetch}
\end{figure}

\begin{figure}[!h]
\centering
\includegraphics[width=0.9\textwidth]{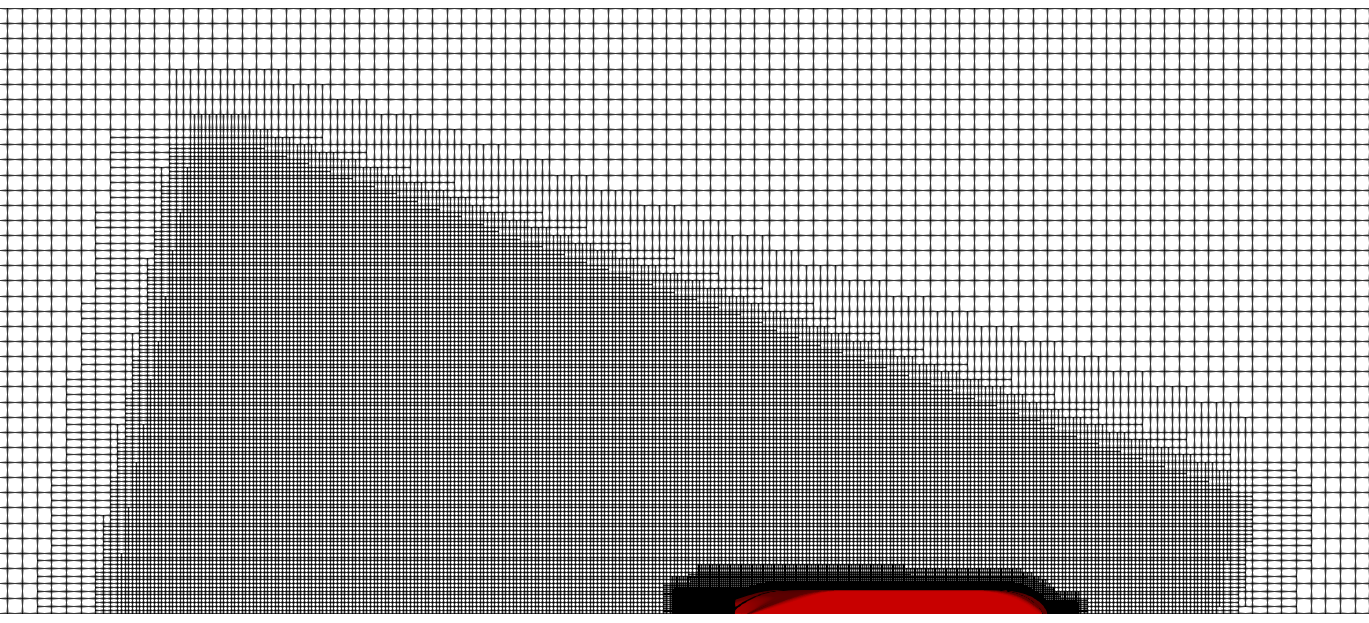}
\caption{Japan Bulk Carrier ($\mathrm{Re}_\mathrm{L} = \SI{7.246}{} \cdot 10^6$, $\mathrm{Fn}=0.142$): Numerical grid in the still water plane.}
\label{fig:jbc_free_surface}
\end{figure}

The analysis focuses on steady-state conditions in calm water. As shown in Fig. \ref{fig:jbc_scetch}, the origin of the Cartesian coordinate system is positioned beneath the transom at stern of the initial configuration. The free surface is initialized in the $x_\mathrm{1}-x_\mathrm{2}$ plane at a height of $x_\mathrm{3} / L_\mathrm{pp} = 1 / 17$.
The simulation domain extends over $6 \, L_\mathrm{pp}$ in length, $2 \, L_\mathrm{pp}$ in height, and $2 \, L_\mathrm{pp}$ in width. The outlet and bottom boundaries are located at distances of three and three-and-a-half hull lengths, respectively, from the origin. The expected dimensionless wavelength is given by $\lambda / L_\mathrm{pp} = 2 \, \pi \, \mathrm{Fn}^2 = 0.852$. The computational mesh consists of approximately $\SI{2.5}{} \times 10^6$ control volumes. A detailed view of the mesh near the transom is shown in Fig. \ref{fig:jbc_scetch} (b). Due to symmetry, only half of the geometry is modeled in the lateral ($x_\mathrm{2}$) direction. The simulations are fully turbulent, employing the wall-function-based $k-\omega$ SST model from \cite{menter2003ten}, with a non-dimensional wall-normal distance of $y^+ \approx \SI{30}{}$ for the first grid layer adjacent to the hull. To accurately capture the wave field generated by the vessel, the horizontal resolution in the free surface region is refined within the Kelvin wedge (see Fig. \ref{fig:jbc_free_surface}). The free surface resolution corresponds to approximately $\Delta x_\mathrm{1} / \lambda = \Delta x_\mathrm{2} / \lambda = 1/50$ cells in the horizontal directions and $\Delta x_\mathrm{3} / \lambda = 1/500$ cells in the vertical direction. The simulations progress towards a steady-state solution in pseudo-time, following an implicit Euler method with an average Courant number condition of $\mathrm{Co} \leq 0.9$. The simulations are performed for 5000 time steps and are truncated if the residuals of the flow solver no longer change between two time steps. A uniform horizontal bulk flow is applied along the outer boundaries for both phases, ensuring a calm water concentration distribution. A hydrostatic pressure condition is imposed at the upper boundary. To dampen the wave field and satisfy the outlet condition, the grid is stretched near the outlet. Additionally, a symmetry condition is enforced along the midship plane.

Floatation is considered using an equilibrium floating approach that mimics calm-water towing tank conditions. The method adjusts the hull's trim and sinkage by distorting the near-ship grid using a grid morphing approach via free-form deformation. Strictly speaking, the method generates a new simulation grid after each adjustment. However, the approach retains the grid's topology and relative geometry so that the simulation can be continued directly. Due to the pseudo-stationary simulation process, the floating is not adjusted each time step but is only triggered as soon as relevant forces on the hull are converged. This typically results in approx. $\mathcal{O}(20)$ floatation adjustments per simulation. Interested readers might consult \cite{lohner1996improved} or \cite{kuhl2021phd} for further information on the approach.

\begin{figure}[!htb]
\centering
\iftoggle{tikzExternal}{
\input{./tikz/04__jbc_resistance/resistance.tikz}}{
\includegraphics{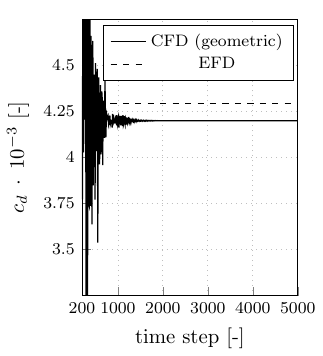}
\includegraphics{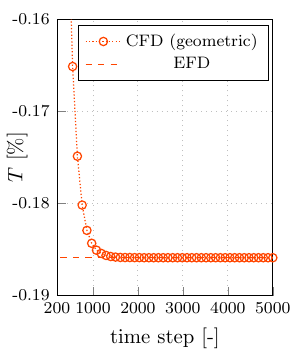}
\includegraphics{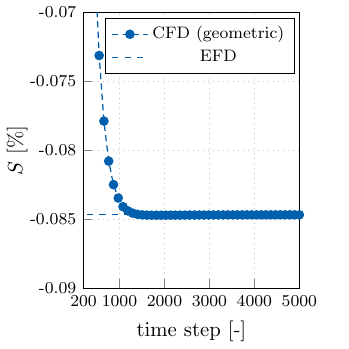}
}
\caption{Japan Bulk Carrier ($\mathrm{Re}_\mathrm{L} = \SI{7.246}{} \cdot 10^6$, $\mathrm{Fn}=0.142$): Development of the drag coefficient ($c_\mathrm{d} \cdot 10^{-3}$, left), the relative trim angle ($T$, center) and the relative sinkage ($S$, right) over the simulated time steps.}
\label{fig:jbc_resistance_results}
\end{figure}

First, the reference simulation based on a geometric wall distance determination is considered. Figure \ref{fig:jbc_resistance_results} shows the development of the drag coefficient ($2 R_T / (\rho V^2 A_\mathrm{w})$, left), the relative trim ($(\tau_\mathrm{a} - \tau_\mathrm{f}))/L_\mathrm{pp} \cdot 100$, center), as well as the relative sinkage ($-0.5(\tau_\mathrm{a} + \tau_\mathrm{f}))/L_\mathrm{pp} \cdot 100$, right) as solid lines and, in the case of the floating position parameters, provided with markers over the simulated time steps. Therein, $R_T$, $\tau_\mathrm{a}$, $\tau_\mathrm{f}$, and $A_\mathrm{w}$ refer to the hull's total resistance, its dip at the aft and front as well as wetted surface, respectively. The analysis's credibility is underlined by comparing experimental data (\cite{hino2020numerical}), represented by constant, dashed lines. The markings within the sinkage and trim development indicate the adjustment time steps of the floatation position, whereby the integral variables under consideration virtually no longer change after approx. 2000 simulation time steps. The difference between experimental as well as numerical sinkage and trim prediction is hardly visible. The drag coefficient, however, shows a deviation of approx. 1\% between the experiment and simulation. Due to several possible influences (e.g., flow turbulence modeling), the geometric wall distance approach is considered acceptable and is assumed to be the reference solution in the following.

\begin{figure}[!htb]
\centering
\iftoggle{tikzExternal}{
\input{./tikz/04__jbc_resistance/resistance_errors.tikz}}{
\includegraphics{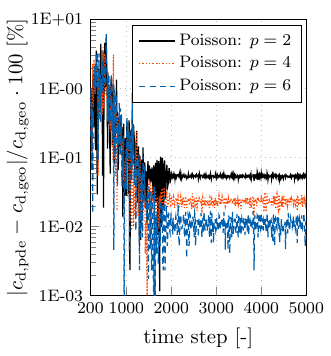}
\includegraphics{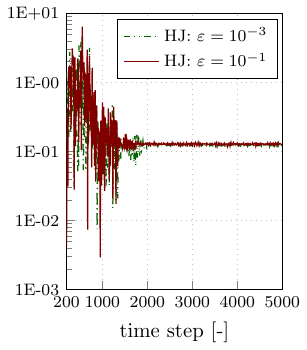}
\includegraphics{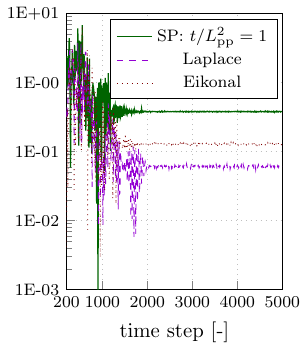}
}
\caption{Japan Bulk Carrier: Development of the drag coefficient's relative error ($|c_\mathrm{d,pde} -  c_\mathrm{d,geo}|/c_\mathrm{d,geo} \cdot 100$ [\%]) for a resistance prediction based on a PDE-based wall distance against the geometrical reference solution for three p-Poisson ($p = [2,4,6]$, left) and two Hamilton-Jacobi approaches ($\varepsilon = [10^{-1}, 10^{-3}]$, center), as well as one Screened-Poisson with $t/L_\mathrm{pp}^2$, the Laplacian and the Eikonal approach (right) over the simulated time steps.}
\label{fig:jbc_resistance_errors}
\end{figure}

\begin{figure}[!htb]
\centering
\iftoggle{tikzExternal}{
\input{./tikz/04__jbc_resistance/trim_errors.tikz}}{
\includegraphics{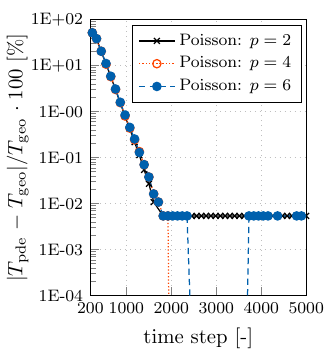}
\includegraphics{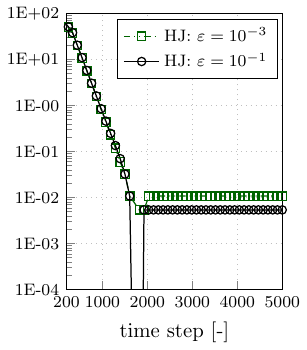}
\includegraphics{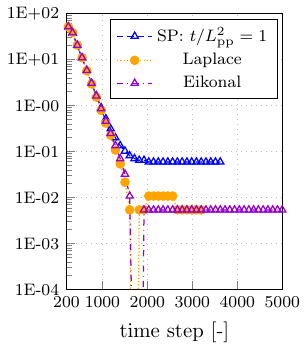}
}
\caption{Japan Bulk Carrier ($\mathrm{Re}_\mathrm{L} = \SI{7.246}{} \cdot 10^6$, $\mathrm{Fn}=0.142$): Development of the relative trim error ($|T_\mathrm{pde} -  T_\mathrm{geo}|/T_\mathrm{geo} \cdot 100$ [\%]) for a resistance prediction based on a PDE-based wall distance against the geometrical reference solution for three p-Poisson ($p = [2,4,6]$, left) and two Hamilton-Jacobi approaches ($\varepsilon = [10^{-1}, 10^{-3}]$, center), as well as one Screened-Poisson with $t/L_\mathrm{pp}^2$, the Laplacian and the Eikonal approach (right) over the simulated time steps.}
\label{fig:jbc_trim_errors}
\end{figure}

\begin{figure}[!htb]
\centering
\iftoggle{tikzExternal}{
\input{./tikz/04__jbc_resistance/sinkage_errors.tikz}}{
\includegraphics{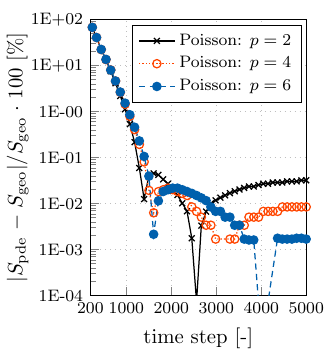}
\includegraphics{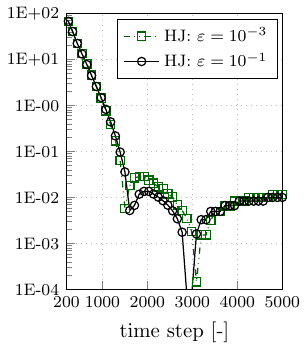}
\includegraphics{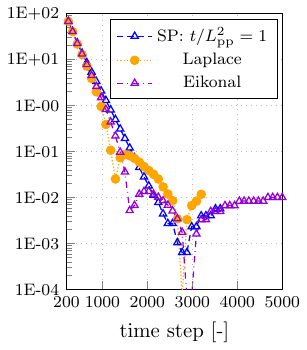}
}
\caption{Japan Bulk Carrier ($\mathrm{Re}_\mathrm{L} = \SI{7.246}{} \cdot 10^6$, $\mathrm{Fn}=0.142$): Development of the relative sinkage error ($|S_\mathrm{pde} -  S_\mathrm{geo}|/S_\mathrm{geo} \cdot 100$ [\%]) for a resistance prediction based on a PDE-based wall distance against the geometrical reference solution for three p-Poisson ($p = [2,4,6]$, left) and two Hamilton-Jacobi approaches ($\varepsilon = [10^{-1}, 10^{-3}]$, center), as well as one Screened-Poisson with $t/L_\mathrm{pp}^2$, the Laplacian and the Eikonal approach (right) over the simulated time steps.}
\label{fig:jbc_sinkage_errors}
\end{figure}

The resistance tests are repeated, and the determination of the wall distance function is consistently varied. All approaches from Sec. \ref{subsec:error_over_rotation} are considered; only the two smallest Screened-Poisson approaches with $t/L_\mathrm{pp}^2 = [10^{-2}, 10^{-4}]$ lead to a divergence of the flow solver and are, therefore, not considered in the following. All other approaches converge similarly to the reference investigation based on the geometric approach. Figures \ref{fig:jbc_resistance_errors}-\ref{fig:jbc_sinkage_errors} show the relative change in the total resistance coefficient  ($|c_\mathrm{d,pde} -  c_\mathrm{d,geo}|/c_\mathrm{d,geo} \cdot 100$ [\%]), the trim angle ($|T_\mathrm{pde} -  T_\mathrm{geo}|/T_\mathrm{geo} \cdot 100$ [\%]), and the sinkage ($|S_\mathrm{pde} -  S_\mathrm{geo}|/S_\mathrm{geo} \cdot 100$ [\%]) between different resistance analyses based on PDE-based wall distance fields compared to the reference resistance analysis with a purely geometric wall distance. The three p-Poisson results for $p = [2,4,6]$ are shown on the left, the two Hamilton-Jacobi results for $\varepsilon = [10^{-1}, 10^{-3}]$ in the center, and the Screened-Poisson with $t/L_\mathrm{pp}^2$, the Laplace, as well as the Eikonal results on the right.
Based on the errors in the resistance coefficient, the errors are well below one percent, instead in the order of $\mathcal{O}(10^{-1}\%)$. As expected, the relative error becomes smaller for higher p-values within the p-Poisson method, with $p=6$ outperforming all other PDE-based approaches and generating the smallest error of $10^{-2}$ percent. The Laplace method predicts smaller errors than the Eikonal and the Hamilton-Jacobi approaches and is about as good as the linear $p=2$ Poisson approach. The errors based on the Eikonal approach are identical to both Hamilton-Jacobi tests, and the highest error results stem from the Screened-Poisson approach.
The floatation parameters allow similar conclusions to be drawn, particularly the behavior of the relative errors in the sinkage, where all methods achieve errors of around $\mathcal{O}(10^{-2}\%)$ percent. The differences in the relative error in the trim angle vary significantly less pronounced between the different methods and are almost the same at around $\mathcal{O}(10^{-2}\%)$ percent; only the SP method performs less well with $\mathcal{O}(10^{-1}\%)$ percent.

\subsubsection{Propulsion}
\label{subsec:propulsion}

The following examines the influence of wall distance modeling on the predictive behavior of the JBC's propulsion. The operating point and related fluid dynamic characteristics are consistent with the previous resistance analysis. For this purpose, the grid already presented in Sec. \ref{subsec:error_over_rotation} is reused. The influence of the free water surface and effects due to trim and sinkage are therefore not considered due to the double-body assumption.
The flow solver is configured analogously to the previous resistance calculations, but the approximation of the volume fraction (see Appendix \ref{app:governing_equations}) is neglected. One propeller revolution is discretized with 360-time steps that decreases the average Courant number by roughly one order of magnitude, so a relatively large number of smaller time steps are necessary for the convergence of the propulsion computation compared to the resistance simulations. All propulsion computations have shown convergence after approximately 25000-time steps and were carried out for 27500-time steps.

The propulsion point --i.e., the rotational speed that ensures a force equilibrium between hull resistance and propeller thrust-- is identified based on two simulations for each wall distance model with a fixed, specified propeller rotation rate slightly below ($n_1$) and above ($n_2$) an actual assumed correct ($\tilde{n}$) rotational speed. This allows the subsequent interpolation of the force equilibrium, viz.
\begin{align}
    \tilde{n} = n_1 + \frac{n_2 - n_1}{\bar{f}_1(n_1) - \bar{f}_2(n_2)} \bar{f}_1(n_1) \label{eqn:propulsion_point}
\end{align}
where $n_1 = \SI{9}{Hz}$ and $n_2 = \SI{11}{Hz}$ refer to the two prescribed, fixed rotational speeds and $\bar{f}_1(n_1) = \bar{f}_{\mathrm{hull}}(n_1) + \bar{f}_{\mathrm{prop,1}}(n_1)$ as well as $\bar{f}_2(n_2) = \bar{f}_{\mathrm{hull},2}(n_2) + \bar{f}_{\mathrm{prop},2}(n_2)$ are the respective total forces acting on hull and propeller, averaged over the simulation's last three rotations, i.e., 1080-time steps. The rotation rate values follow from a preceding, simplified propulsion analysis based on a volume force model using an actuator disk instead of a geometrically resolved propeller, that indicated on a rotational speed of $\tilde{n} \approx \SI{10}{Hz}$. Once the propulsion point is available, further relevant operative data is interpolated from the available propulsion point, i.e., forces acting on propeller or hull follow
\begin{align}
    \tilde{f} = f_1 + \frac{f_3 - f_1}{(n_3 - n_1)} (\tilde{n} - n_1) \, .
\end{align}

\paragraph{Temporally Fixed Wall Distance Fields} are considered in the following. Initially, the wall distance is only determined at the start of the simulation and is assumed to be constant, i.e., $w \not= w(t)$ during the propulsion analysis. This leads to an error in the wall distance field due to the relative movement between the propeller and the hull but results in an overall reduction in computational effort. Subsequently, at the end of this section, the propulsion behavior with a time variable, i.e., $w = w(t)$ wall distance function, is investigated.

\begin{figure}[!htb]
\centering
\iftoggle{tikzExternal}{
\input{./tikz/05__jbc_self_propulsion/forces.tikz}}{
\includegraphics{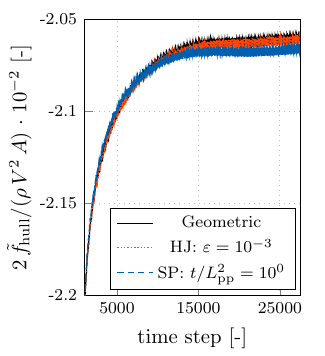}
\includegraphics{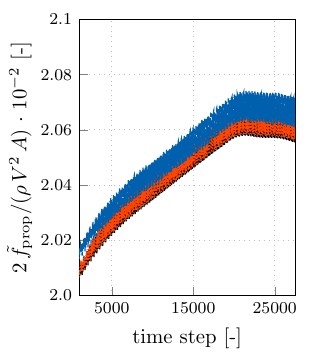}
\includegraphics{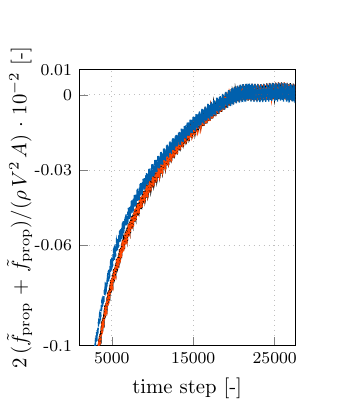}
}
\caption{Japan Bulk Carrier ($\mathrm{Re}_\mathrm{L} = \SI{7.246}{} \cdot 10^6$, $\mathrm{Fn}=0.142$): Non-dimensional forces acting on the hull (left) and the propeller (center), as well as their sum (right) for a geometric as well as a Hamilton-Jacobi ($\varepsilon = 10^{-3}$) and Screened-Poisson ($t/L_\mathrm{pp}^2 = 1$) based wall distance approach, interpolated into the respective propulsion point over the simulated time steps.}
\label{fig:jbc_self_propulsion_forces}
\end{figure}

Investigations are carried out for the p-Poisson approach with $p=[2,2.4]$, the Hamilton-Jacobi approach with $\varepsilon = [10^{-1}, 10^{-3}]$, the Screened-Poisson approach with $t/L_\mathrm{pp}^2 = [1, 10^{-2}, 10^{-4}]$, as well as the Eikonal, the Laplace, and the geometrical reference approach, which leads to a total of $2 \times 10 = 20$ simulations. The p-Poisson exponent is deliberately only slightly increased in order to keep the simulation times short and at the same time not to entirely neglect the non-linearity. The ten equilibrium rotational speeds interpolated according to Eqn. \eqref{eqn:propulsion_point} deviate only minimally from each other. For clarification, Fig. \ref{fig:jbc_self_propulsion_forces} shows the dimensionless forces interpolated into the propulsion point acting on the hull (left, $\tilde{f}_\mathrm{hull}$), the propeller (center, $\tilde{f}_\mathrm{prop}$), and their sum (right, $\tilde{f} = \tilde{f}_\mathrm{hull} + \tilde{f}_\mathrm{prop}$) for three exemplary wall distance scenarios (geometric as well as Hamilton-Jacobi and Screened-Poisson with $\varepsilon = 10^{-3}$ and $t/L_\mathrm{pp}^2 = 1$, respectively). Fluctuating forces are visible due to the resolved propeller rotation. The determination of the propulsion point is indicated by the convergence of the total force (right) towards the equilibrium. Convergence is approached from the negative side in the underlying coordinate system, i.e., the propeller needs a specific simulation time to balance the hull resistance. While all three converged total forces appear almost the same, the hull resistance and propeller thrust differences are recognizable. The Hamilton-Jacobi approach provides minimally more negative forces acting on the hull and, thus, an increased counterpart on the propeller. This trend is more pronounced for the Screened-Poisson approach.

\begin{figure}[!htb]
\centering
\iftoggle{tikzExternal}{
\input{./tikz/05__jbc_self_propulsion/resistance_errors.tikz}}{
\includegraphics{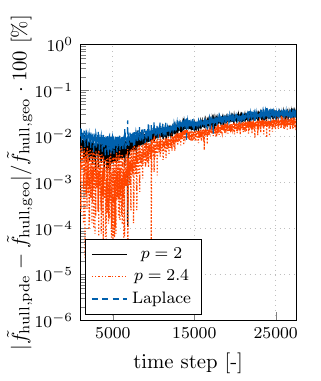}
\includegraphics{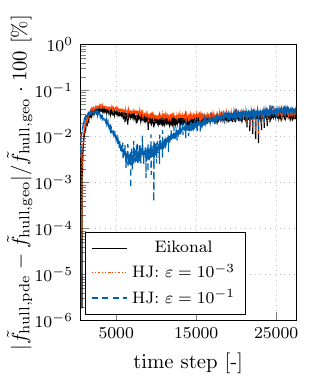}
\includegraphics{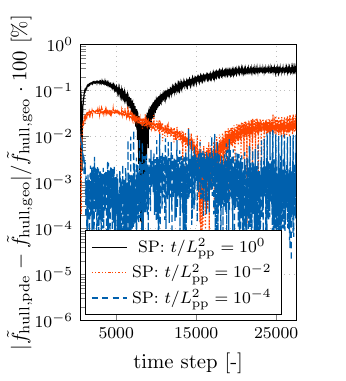}
}
\caption{Japan Bulk Carrier ($\mathrm{Re}_\mathrm{L} = \SI{7.246}{} \cdot 10^6$, $\mathrm{Fn}=0.142$): Relative error between hull force predictions for PDE-based against geometrically determined reference wall distances over the simulated time steps.}

\label{fig:jbc_self_propulsion_resistance_errors}
\end{figure}

\begin{figure}[!htb]
\centering
\iftoggle{tikzExternal}{
\input{./tikz/05__jbc_self_propulsion/thrust_errors.tikz}}{
\includegraphics{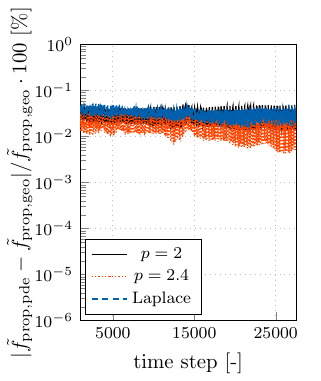}
\includegraphics{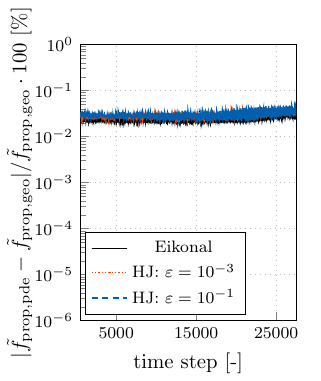}
\includegraphics{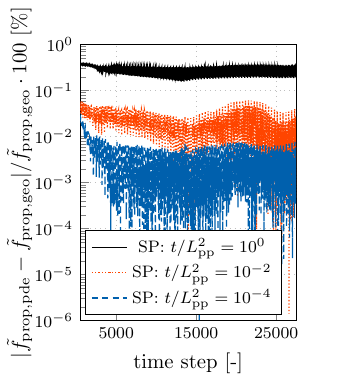}
}
\caption{Japan Bulk Carrier ($\mathrm{Re}_\mathrm{L} = \SI{7.246}{} \cdot 10^6$, $\mathrm{Fn}=0.142$): Relative error between propeller force predictions for PDE-based against geometrically determined reference wall distances over the simulated time steps.}
\label{fig:jbc_self_propulsion_thrust_errors}
\end{figure}

\begin{figure}[!htb]
\centering
\iftoggle{tikzExternal}{
\input{./tikz/05__jbc_self_propulsion/total_errors.tikz}}{
\includegraphics{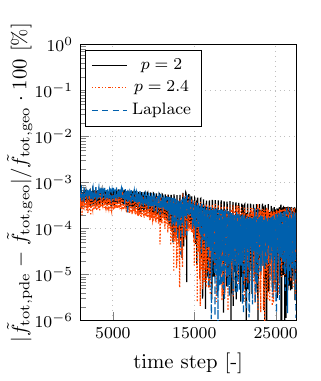}
\includegraphics{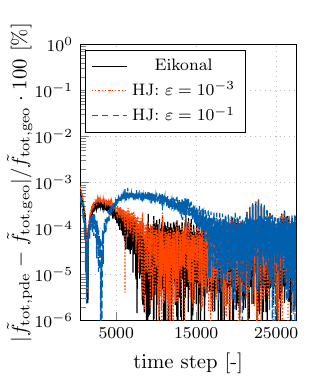}
\includegraphics{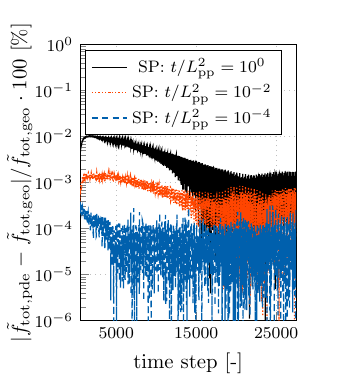}
}
\caption{Japan Bulk Carrier ($\mathrm{Re}_\mathrm{L} = \SI{7.246}{} \cdot 10^6$, $\mathrm{Fn}=0.142$): Relative error between hull and propeller force predictions for PDE-based against geometrically determined reference wall distances over the simulated time steps.}

\label{fig:jbc_self_propulsion_total_errors}
\end{figure}

To highlight the minor differences in predicted propulsion points due to various wall distance modeling strategies, the relative errors $|\tilde{f}_\mathrm{pde} - \tilde{f}_\mathrm{geo}|/\tilde{f}_\mathrm{geo} \cdot 100\%$ between the forces using PDE-based wall distances and those predicted by the geometric wall distance reference are presented. These errors, calculated for forces acting on the hull, the propeller and their sum, are shown in Figs. \ref{fig:jbc_self_propulsion_resistance_errors}, \ref{fig:jbc_self_propulsion_thrust_errors}, and \ref{fig:jbc_self_propulsion_total_errors}, respectively.
The results of the two p-Poisson and Laplace methods are shown on the left, those of the Eikonal and the two Hamilton-Jacobi methods in the middle, and those of the remaining three Screened-Poisson approaches on the right. The errors are almost without exception below $\mathcal{O}(10^{-1}\%)$ percent; only the Screened-Poisson approach for $t/L_\mathrm{pp}^2 = 1$ represents an outlier in the order of one percent, cf. Fig. \ref{fig:jbc_self_propulsion_forces}. The errors of the total force are generally approx. two orders of magnitude below the individual hull resistance or propeller thrust errors. The Screened-Poisson method appears to be particularly sensitive to its model parameter: While the case $t/L_\mathrm{pp}^2 = 1$ gives the highest error, $t/L_\mathrm{pp}^2 = 10^{-4}$ outperforms all other methods, especially for the individual hull and propeller forces, and the intermediate value $t/L_\mathrm{pp}^2 = 10^{-2}$ yields error values in between. The Eikonal and the two Hamilton-Jacobi approaches show the smallest differences. Finally, as expected, the p-Poisson errors for $p=2$ are higher than for $p=2.4$. 

\paragraph{Temporal Variable Wall Distance Fields} are considered in the following. The wall distance is recalculated after at the beginning of each time step, i.e., $w = w(t)$. The relative motion between the hull and the propeller results in slightly different wall distance fields between two time steps. This results in an additional numerical effort minimized by the solver's restart capabilities. Due to its disproportionately high computational effort, the geometric method is omitted. Only the linear wall distance models are considered, i.e., $p=2$, Eikonal, one Hamilton Jacobi for $\varepsilon = 10^{-3}$, Laplace, and one Screened-Poisson with $t/L_\mathrm{pp}^2 = 1$ approach are examined. Instead of re-interpolating the propulsion point from two simulations with a fixed rotational speed, only the simulations for one rotation rate $n_1$ are carried out. The resulting forces (i.e., $f(w = w(t))$) on the hull and propeller are compared directly against the respective results with constant wall distances (i.e., $f(w \not= w(t))$) from the previous investigation.

\begin{figure}[!htb]
\centering
\iftoggle{tikzExternal}{
\input{./tikz/05__jbc_self_propulsion/resistance_differences_var_walldistance.tikz}}{
\includegraphics{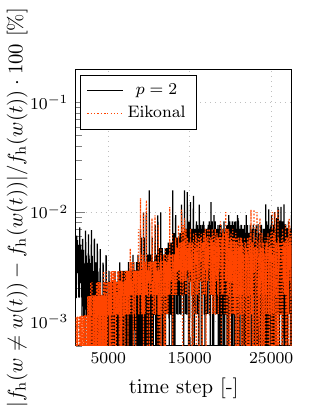}
\includegraphics{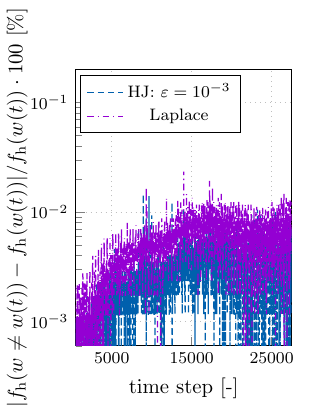}
\includegraphics{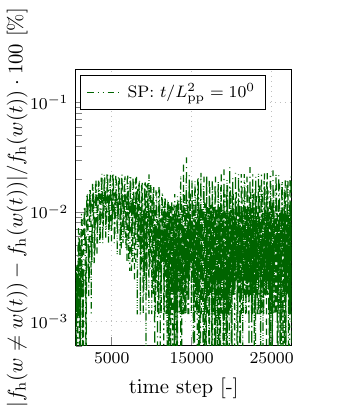}
}
\caption{Japan Bulk Carrier ($\mathrm{Re}_\mathrm{L} = \SI{7.246}{} \cdot 10^6$, $\mathrm{Fn}=0.142$): Relative error between forces acting on the hull from simulations with temporally fixed against variable wall distance fields for five different equation-based wall distance models over the simulated time steps.}
\label{fig:jbc_self_propulsion_resistance_differences_var_walldistance}
\end{figure}

\begin{figure}[!htb]
\centering
\iftoggle{tikzExternal}{
\input{./tikz/05__jbc_self_propulsion/thrust_differences_var_walldistance.tikz}}{
\includegraphics{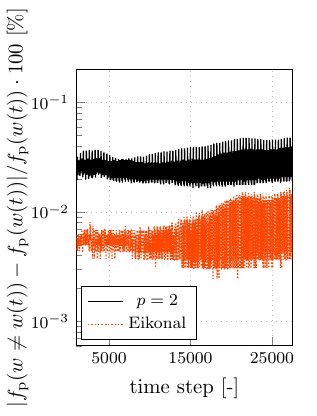}
\includegraphics{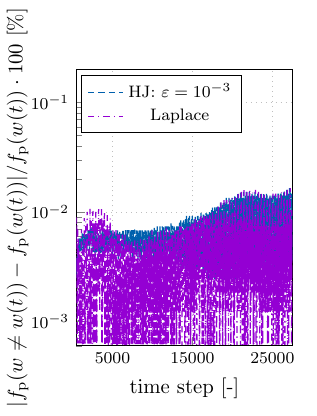}
\includegraphics{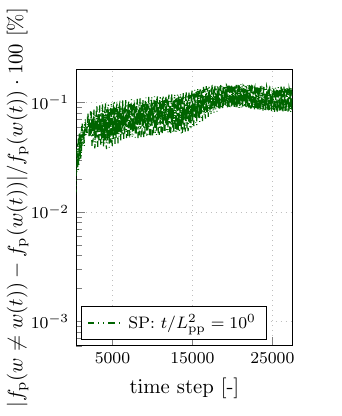}
}
\caption{Japan Bulk Carrier ($\mathrm{Re}_\mathrm{L} = \SI{7.246}{} \cdot 10^6$, $\mathrm{Fn}=0.142$): Relative error between forces acting on the prop from simulations with temporally fixed against variable wall distance fields for five different equation-based wall distance models over the simulated time steps.}
\label{fig:jbc_self_propulsion_thrust_differences_var_walldistance}
\end{figure}

The relative error $|f(w \not=w(t)) -  f(w(t))|/f(w(t)) \cdot 100$ [\%] between time-constant versus time-varying wall distances is considered. For the five cases investigated, the error of the hull force (i.e., $|f_\mathrm{h}(w \not=w(t)) -  f_\mathrm{h}(w(t))|/f_\mathrm{h}(w(t)) \cdot 100$ [\%]) is shown in Fig. \ref{fig:jbc_self_propulsion_resistance_differences_var_walldistance}, with results of the p-Poisson and Eikonal methods on the left, the Hamilton-Jacobi and Laplace methods in the center, and the Screened-Poisson method on the right. In all cases, the predicted forces appear comparatively insensitive to the time-resolved wall distance field, as the relative errors are in the order of $\mathcal{O}(10^{-2})$ percent, with the Screened-Poisson method having the comparatively largest error. An analogous illustration is given in Fig. \ref{fig:jbc_self_propulsion_thrust_differences_var_walldistance} for the relative errors in the predicted propeller force (i.e., $|f_\mathrm{p}(w \not=w(t)) -  f_\mathrm{p}(w(t))|/f_\mathrm{p}(w(t)) \cdot 100$ [\%]). Compared to the investigated hull force from Fig. \ref{fig:jbc_self_propulsion_resistance_differences_var_walldistance}, an increased sensitivity to a time-resolved wall distance is identified for the two Poisson-based methods, which show an increased error towards $\mathcal{O}(10^{-1})$ percent. The relative errors of the remaining three methods are still in the range of $\mathcal{O}(10^{-2})$ percent. The minor differences in the predicted forces result in almost identical propulsion points, which are not examined in detail to save space.

\begin{figure}[!htb]
\centering
\iftoggle{tikzExternal}{
\input{./tikz/05__jbc_self_propulsion/timing.tikz}}{
\includegraphics{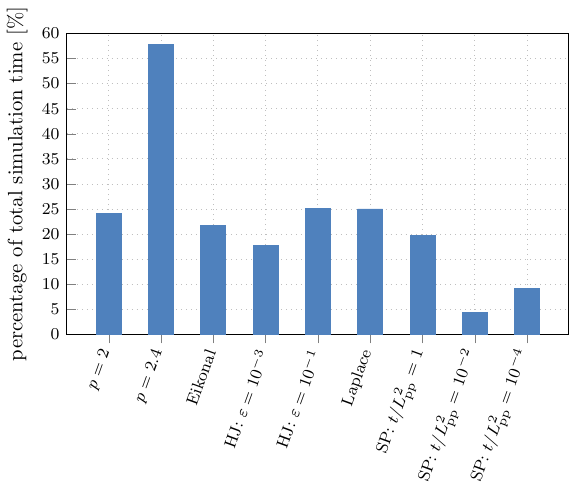}
}
\caption{Japan Bulk Carrier ($\mathrm{Re}_\mathrm{L} = \SI{7.246}{} \cdot 10^6$, $\mathrm{Fn}=0.142$): Percentage of the simulation time required to compute the wall distance PDE's for each considered wall distance model in relation to the flow solver's total wall clock time.}
\label{fig:jbc_self_propulsion_timing}
\end{figure}

Finally, the additional numerical effort to re-calculate the new wall distance each time step is examined. Figure \ref{fig:jbc_self_propulsion_timing} shows the relative percentage of the wall distance calculation in relation to the overall required flow solver's wall clock time for the (a) previously considered and (b) additional wall distance approaches. Except for some Poisson methods, the additional effort required to determine the wall distance is around $\mathcal{O}(20\%)$. While the non-linear p-Poisson method results in significant additional costs of approx. $\mathcal{O}(60\%)$, the Screened-Poisson costs decrease for smaller model parameters to $\mathcal{O}(10\%)$ percent, which follows from the reduced diffusion impact, cf. the balance equation from Eqn. \eqref{equ:screened_poisson}. 

\subsection{Aerodynamics of a Feeder Ship}
\begin{figure}[!ht]
\begin{minipage}[c][11cm][t]{0.4\textwidth}
	\vspace*{\fill}
	\centering
	\subfigure[]{
	\includegraphics[width=0.9\textwidth]{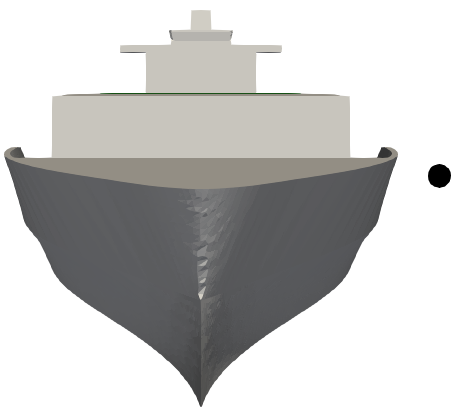}
	}
\end{minipage}
\begin{minipage}[c][11cm][t]{0.6\textwidth}
	\vspace*{\fill}
	\centering
	\subfigure[]{
	\includegraphics[width=1.0\textwidth]{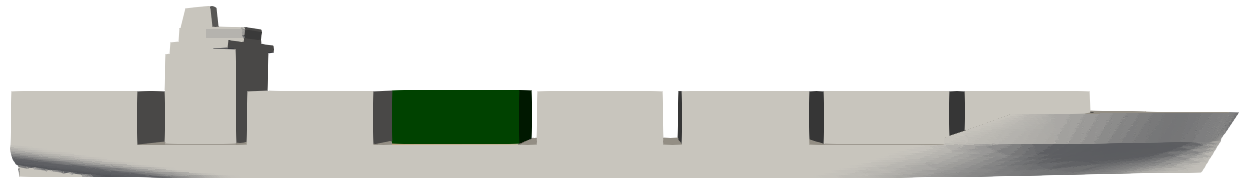}
	}
	\subfigure[]{
	\includegraphics[width=1.0\textwidth]{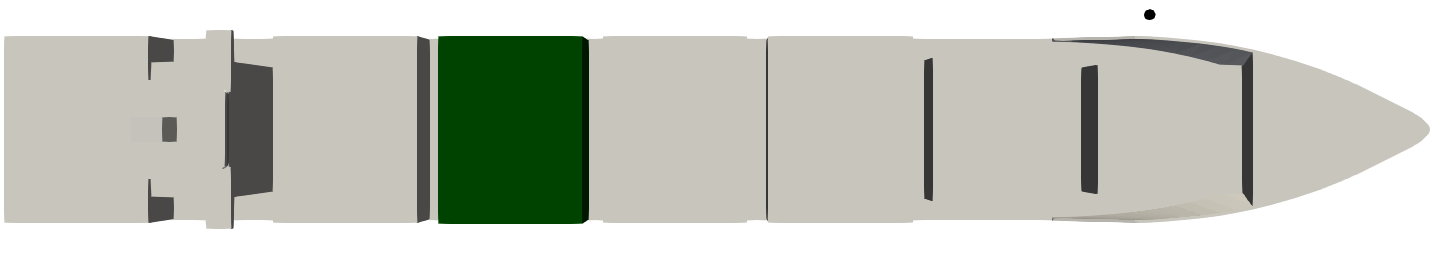}
	}
\end{minipage}
\caption{Feeder ship flow ($\mathrm{Re}_\mathrm{L} = 5.0 \cdot 10^8$): Representation of the geometry from (a) front, (b) starboard side, and (c) top, as well as the indication of the measuring point used for local evaluation. Additionally, one container to asses integral quantities is highlighted in green.}
\label{fig:kcs_propes}
\end{figure}

\begin{figure}[!ht]
\begin{minipage}[c][11cm][t]{0.3\textwidth}
	\vspace*{\fill}
	\centering
	\subfigure[]{
	\includegraphics[width=0.9\textwidth]{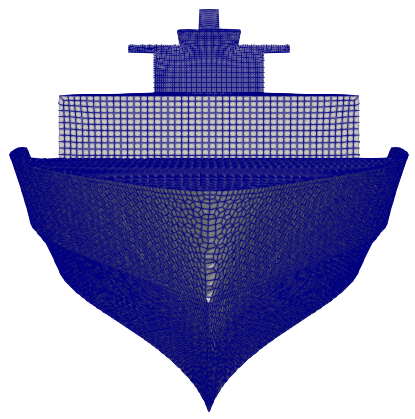}
	}
\end{minipage}
\begin{minipage}[c][11cm][t]{0.7\textwidth}
	\vspace*{\fill}
	\centering
	\subfigure[]{
	\includegraphics[width=1.0\textwidth]{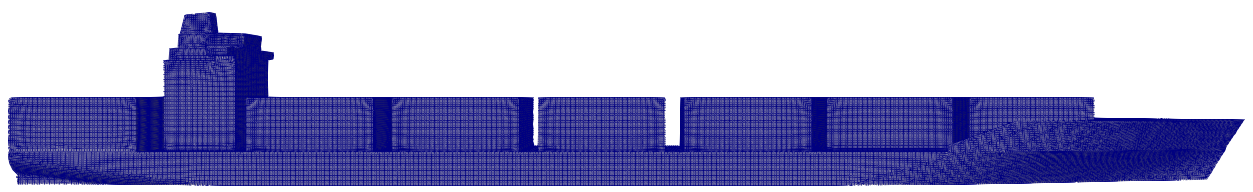}
	}
	\subfigure[]{
	\includegraphics[width=1.0\textwidth]{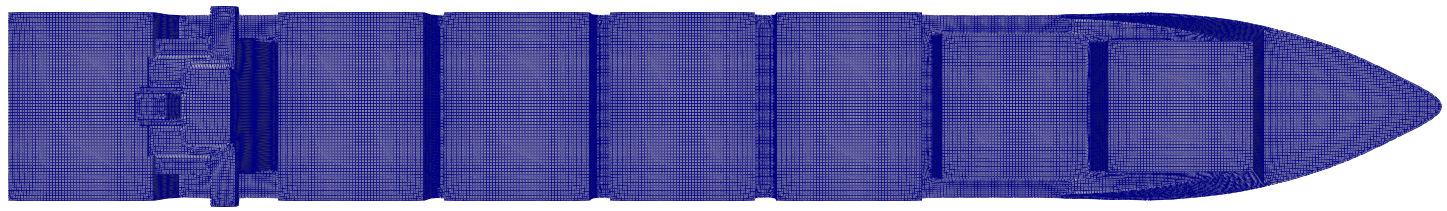}
	}
\end{minipage}
\caption{Feeder ship flow ($\mathrm{Re}_\mathrm{L} = 5.0 \cdot 10^8$): The utilized surface discretization presented from (a) front, (b) starboard side, and (c) top, cf. Fig. \ref{fig:kcs_propes}.}
\label{fig:kcs_grid}
\end{figure}

\begin{figure}[!ht]
\begin{minipage}[c][11cm][t]{1.0\textwidth}
	\vspace*{\fill}
	\centering
	\subfigure[]{
	\includegraphics[width=0.475\textwidth]{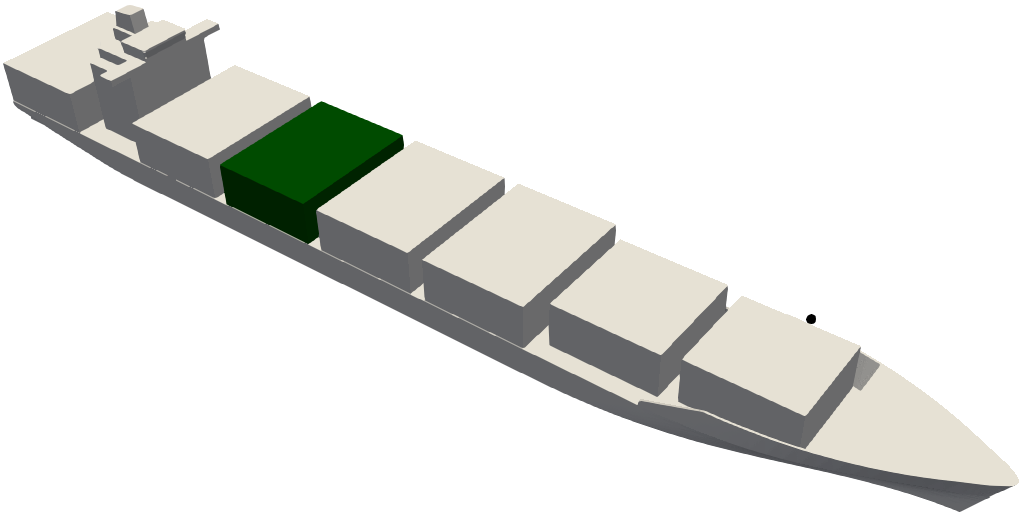}
	}
	\subfigure[]{
	\includegraphics[width=0.475\textwidth]{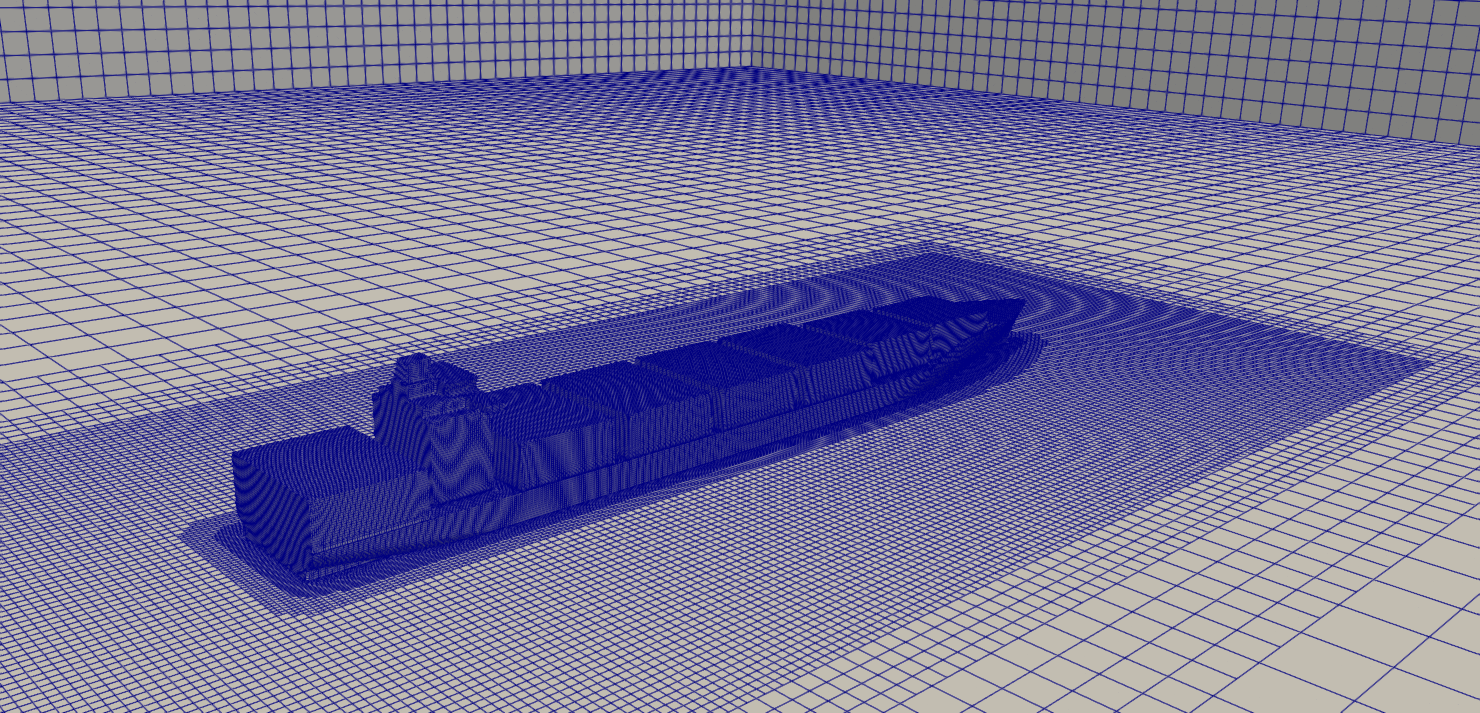}
	}
\end{minipage}
\caption{Feeder ship flow ($\mathrm{Re}_\mathrm{L} = 5.0 \cdot 10^8$): (a) Perspective view of the considered geometry, cf. Fig. \ref{fig:kcs_propes}, and (b) part of the numerical near and far-field grid, cf. Fig. \ref{fig:kcs_grid}.}
\label{fig:kcs_perspectives}
\end{figure}

Another industrial study examines the airflow around the full-scale superstructure of a feeder ship with a standardized container configuration, as depicted in Figs. \ref{fig:kcs_propes}-\ref{fig:kcs_perspectives}. Additional geometric details can be found in \cite{angerbauer2020hybrid, pache2022data, pache2023datenbasierte}. The investigation is performed at a Reynolds number of $\mathrm{Re}_\mathrm{L} = V L / \nu = 5.0 \cdot 10^8$, based on the ship length $L$ and the apparent wind velocity $V$ (air, $\nu$) approaching from a $20^\circ$ port-side direction.
To mitigate blockage effects, the computational domain extends ten, seven, and two ship lengths in the longitudinal ($x_1$), lateral ($x_2$), and vertical ($x_3$) directions, respectively. The superstructure is centrally positioned in the $x_1$-$x_2$ plane and aligned with the lowest $x_3$ coordinate.
An unstructured hexahedral grid with approximately four million control volumes is employed, optimized for an inflow angle of 20 degrees. The mesh includes a refinement zone in the anticipated wake region of the superstructure, with around 200 isotropic elements distributed along the ship’s length. To model near-wall turbulence effects, a universal wall function is applied, blending between the viscous sub-layer and a standard wall function (\cite{rung2001universal, gritskevich2017comprehensive}), maintaining a resolution of approximately $y^+ \approx 50$.
For spatial discretization, diffusive fluxes are approximated using a second-order Central Difference Scheme (CDS), while convective fluxes are computed using a hybrid scheme that blends CDS with a QUICK approach. A weighting of 80\% CDS and 20\% QUICK is chosen to balance numerical diffusion and stability. Additionally, a second-order implicit implicit three level time integration method is used, resulting in an an overall accurate representation of turbulence decay, as validated against experimental data in \cite{angerbauer2020hybrid, pache2023datenbasierte}. The simulation covers a time span of $(T , \Delta t) / (L/V) = 25$ characteristic flow periods, using a constant time step $\Delta t$ selected to maintain a local Courant number below unity, leading to a total of 30,000 time steps.

At the outer domain boundaries, an atmospheric wind profile—$v_i(x_3) = V_i (x_3/x_3^\mathrm{ref})^\alpha$, with $\alpha = 0.11$ and $x_3^\mathrm{ref} = L/8$—is imposed for velocity and turbulence variables. Pressure is specified at the domain's upper boundary, while the lower boundary employs a symmetry condition to represent a calm free-water surface. The initial conditions mirror the far-field boundary values, with a horizontal wind profile and a hydrostatic pressure distribution.
%
%
%
%

\begin{figure}[!ht]
\centering
\subfigure[]{
\iftoggle{tikzExternal}{
\input{./tikz/06__feeder/feeder_vortices.tikz}
}{
\includegraphics{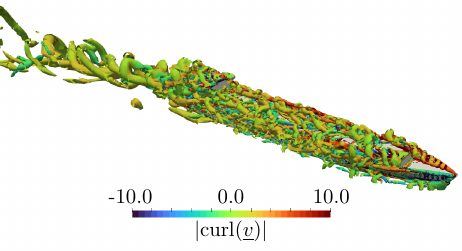}
}
}
\subfigure[]{
\iftoggle{tikzExternal}{
\input{./tikz/06__feeder/feeder_streamlines.tikz}
}{
\includegraphics{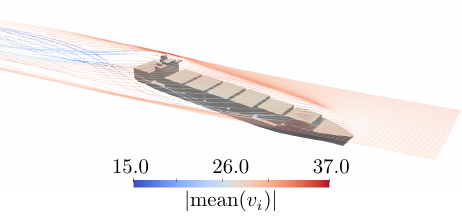}
}
}
\caption{Feeder ship flow ($\mathrm{Re}_\mathrm{L} = 5.0 \cdot 10^8$): Perspective view on (a) isolines of the $\lambda_2$-criterion for $\lambda_2 = -10$ highlighted by the vorticity magnitude and on (b) streamlines of the averaged velocity field seeded on the port-sided front ship colored by the averaged velocities magnitude.}
\label{fig:kcs_results}
\end{figure}

Figure \ref{fig:kcs_results} illustrates the flow characteristics by presenting (a) vortex structures identified using the $\lambda_2$ criterion ($\lambda_2 = -10$, colored by vorticity magnitude) and (b) streamlines seeded in the forward region of the ship, following the averaged velocity field and colored by its magnitude.

The following consideration analyzes the last 2/3 of the simulation period, i.e., the range $10000 < \leq T \leq 30000$, so that initial transient effects are not considered. This period is also used for the temporal averaging of the flow field. Twelve aerodynamic simulations are carried out, and the following wall distance strategies are employed at the start of each simulation: three p-Poisson ($p=[2,4,6]$) and Screened-Poisson ($t/L_\mathrm{pp}^2=[1,10^{-2},10^{-4}]$), three Hamilton-Jacobi methods ($\varepsilon=[1,10^{-1},10^{-3}]$), the Eikonal, the Laplace method, and the geometric method. The latter's results again serve as a reference solution.

\begin{figure}[!htb]
\centering
\iftoggle{tikzExternal}{
\input{./tikz/06__feeder/forces_total_f1.tikz}}{
\includegraphics{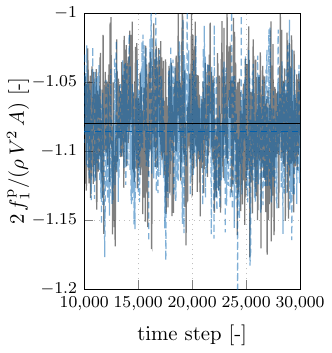}
\includegraphics{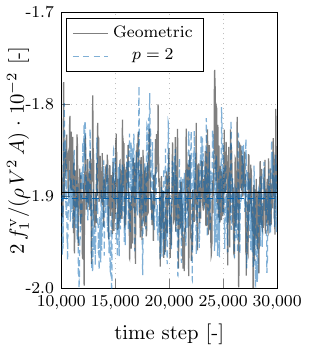}
\includegraphics{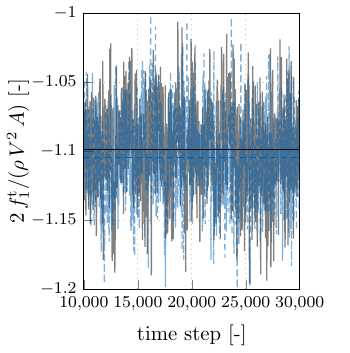}
}
\caption{Feeder ship flow ($\mathrm{Re}_\mathrm{L} = 5.0 \cdot 10^8$): Instantaneous and time-averaged frontal force prediction on the entire superstructure over the last 2/3 of the simulated time steps based on the geometric and a linear $p=2$ Poisson-based wall distance, with the pressure component of the total force shown on the left, the friction component in the center, and their sum on the right.)}
\label{fig:feeder_forces_total_f1}
\end{figure}

First, the influence of the wall distance model on global or integral variables is investigated. For this purpose, the frontal ($f_1$) and lateral ($f_2$) forces on the entire superstructure and the fifth container (see Figs. \ref{fig:kcs_propes} and \ref{fig:kcs_perspectives}, highlighted in green) are considered. Figure \ref{fig:feeder_forces_total_f1} shows the frontal force based on the geometric reference wall distance in black and the force signal based on the $p=2$ Poisson-based wall distance in blue dashed, whereby the pressure component ($f_1^\mathrm{p}$) is shown on the left, the friction component ($f_1^\mathrm{v}$) in the middle and the sum of both ($f_1^\mathrm{t} = f_1^\mathrm{p} + f_1^\mathrm{v}$) on the right, in each case non-dimensionalized with the air density, the apparent wind speed and the projected reference surface as seen from the wind. First, a solid fluctuation in the force signals can be observed due to the underlying scale resolving simulation. In addition to the instantaneous values, the temporal mean values are included as a horizontal line. As expected, the pressure component of the aerodynamic drag predominates. In some cases, the instantaneous values and the time-averaged values deviate visibly between the two wall distance scenarios under consideration.

\begin{figure}[!htb]
\centering
\iftoggle{tikzExternal}{
\input{./tikz/06__feeder/forces_container5_f1.tikz}}{
\includegraphics{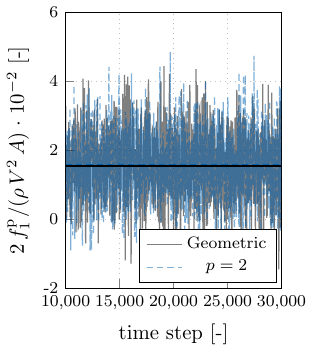}
\includegraphics{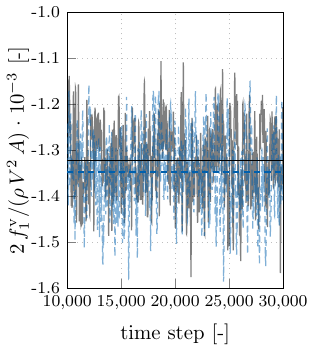}
\includegraphics{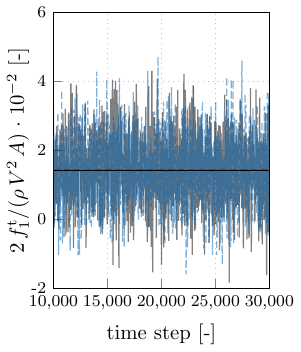}
}
\caption{Feeder ship flow ($\mathrm{Re}_\mathrm{L} = 5.0 \cdot 10^8$): Instantaneous and time-averaged frontal force prediction on the fifth container over the last 2/3 of the simulated time steps based on the geometric and a linear $p=2$ Poisson-based wall distance, with the pressure component of the total force shown on the left, the friction component in the center, and their sum on the right.}
\label{fig:feeder_forces_container5_f1}
\end{figure}

Figure \ref{fig:feeder_forces_container5_f1} gives an analogous impression of the resulting forces on the fifth front container, again distinguishing between the pressure component (left), the friction component (center), and the total force (right). In this case, the viscous component is only about one order of magnitude smaller than the pressure component, the pressure component is partially given a different sign, and the force predictions of the geometric wall distance and those based on the linear p-Poisson method again differ visibly from each other, which is also reflected in the time averages. However, differences in the latter are significantly more minor than in the overall superstructure from Fig. \ref{fig:feeder_forces_total_f1}, especially in the pressure component and due to its dominant role in the total resistance. However, the instantaneous results again indicate visible differences.

\begin{figure}[!htb]
\centering
\iftoggle{tikzExternal}{
\input{./tikz/06__feeder/error_forces_total_all_mean.tikz}}{
\includegraphics{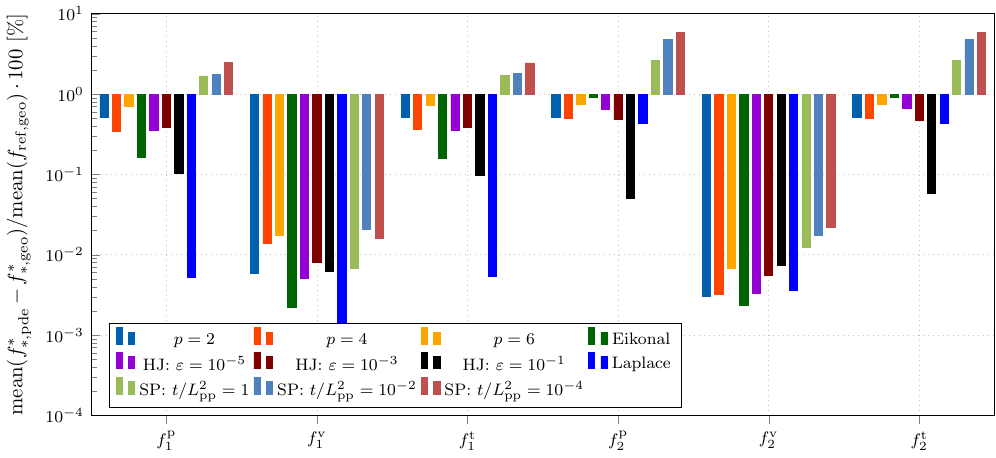}
}
\caption{Feeder ship flow ($\mathrm{Re}_\mathrm{L} = 5.0 \cdot 10^8$): Relative errors in the averaged force signals predicted with different wall distance models, divided into the pressure and friction components and the sum of both, in each case for the frontal and lateral force acting on the entire superstructure.}
\label{fig:feeder_error_forces_total_all_mean}
\end{figure}

\begin{figure}[!htb]
\centering
\iftoggle{tikzExternal}{
\input{./tikz/06__feeder/error_forces_container5_all_mean.tikz}}{
\includegraphics{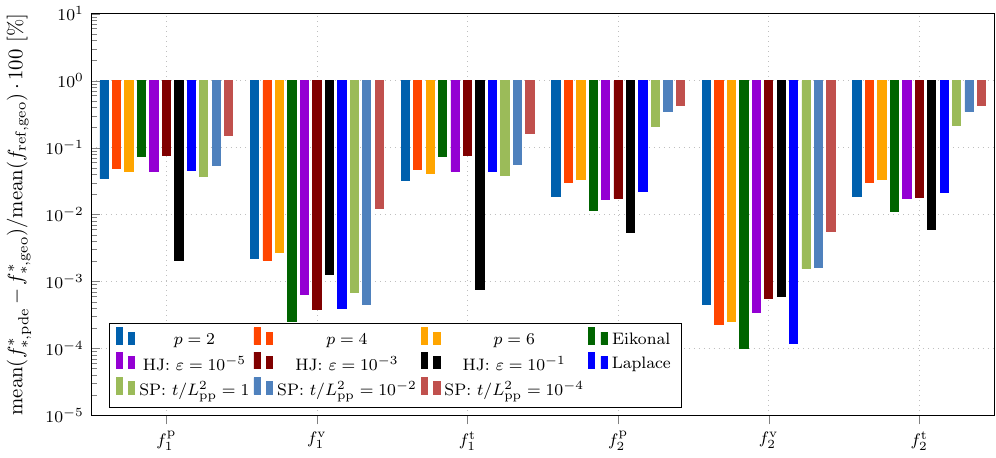}
}
\caption{Feeder ship flow ($\mathrm{Re}_\mathrm{L} = 5.0 \cdot 10^8$): Relative errors in the averaged force signals predicted with different wall distance models, divided into the pressure and friction components and the sum of both, in each case for the frontal and lateral force acting on the fifth container.}
\label{fig:feeder_error_forces_container5_all_mean}
\end{figure}

For a detailed analysis, only relative errors are considered below and, in line with the hydrodynamic investigations, the results on the feeder ship are now put into the following relative relation based on the temporally averaged forces: mean$(f_\mathrm{*,pde}^* - f_\mathrm{*,geo}^*)/\mathrm{mean}(f_\mathrm{ref,geo}) \cdot 100$ [\%]. Therein, $f_\mathrm{*,pde}^*$ and $f_\mathrm{*,geo}^*$ refer to the forces stemming from the PDE-based and geometrical determined wall distance values, respectively, and $\mathrm{mean}(f_\mathrm{ref,geo}) = \mathrm{mean}(f_\mathrm{1,geo})$ denotes a reference force. The latter is deliberately chosen to be the same for all methods and spatial directions, as otherwise, resulting reference forces close to zero tend to produce singularities in the error measures. The force's subscript and superscript characterize the direction and proportion. The results are compared in compressed form as a bar chart in Figures \ref{fig:feeder_error_forces_total_all_mean} and \ref{fig:feeder_error_forces_container5_all_mean} for the entire superstructure and the fifth container, respectively, cf. Figs. \ref{fig:feeder_forces_total_f1} - \ref{fig:feeder_forces_container5_f1}, again distinguishing between the pressure component, the friction component, and their sum. The different wall spacing methods for each bar group are sorted from left to right as follows: First, the three p-Poisson results with $p=2$ (dark blue), $p=4$ (orange), and $p=6$ (yellow), followed by the Eikonal (dark green) and the three Hamilton-Jacobi tests with $\varepsilon = 10^{-5}$ (purple), $\varepsilon = 10^{-3}$ (brown), and $\varepsilon = 10^{-1}$ (black), as well as the Laplace (light blue) and the three Screened-Poisson results for $t/L_\mathrm{pp}^2 = 1$ (light green), $t/L_\mathrm{pp}^2 = 10^{-2}$ (light blue), and $t/L_\mathrm{pp}^2 = 10^{-4}$ (light brown).

The averaged force errors of the entire ship from Fig. \ref{fig:feeder_error_forces_total_all_mean} are discussed first. Most wall spacing strategies result in less than one percent relative time-averaged errors. Only the three Screened-Poisson methods lead to errors above one percent for four out of six investigated forces. Thus, they are under performing. For the viscous forces, significantly lower error orders in the range of $10^{-3}\% \leq \mathrm{mean}(f_\mathrm{*,pde}^\mathrm{v} - f_\mathrm{*,geo}^\mathrm{v})/\mathrm{mean}(f_\mathrm{ref,geo}) \cdot 100  \leq 10^{-2}\%$ can be observed. However, the total force is pressure-dominated, with error values of around $\mathcal{O}(0.5\%)$ for the p-Poisson methods. Interestingly, the errors decrease for $p=2$ to $p=4$ but increase again for $p=6$. Errors of the Hamilton-Jacobi and Eikonal methods are slightly lower [higher] for the frontal [lateral] force than the errors based on the p-Poisson method, with the Hamilton-Jacobi method performing best for $\varepsilon = 10^{-1}$. The frontal force agrees particularly well with the geometric reference values for a wall distance based on the Laplace method.

The error behavior changes for the averaged forces on the fifth container in Figure \ref{fig:feeder_error_forces_container5_all_mean}. All errors are well below one percent, and the overall picture is more homogeneous, with the errors of the viscous force still being the smaller compared to those of the pressure companion. The Screened-Poisson methods still underperform with errors in the range of $\mathcal{O}(0.1\%)$ for the frontal and $\mathcal{O}(0.5\%)$ for the lateral pressure and total forces. Once again, the Hamilton-Jacobi method for $\varepsilon = 10^{-1}$ performs significantly better than the other Hamilton-Jacobi, the Eikonal, Laplace, and three p-Poisson methods with errors below $\mathcal{O}(10^{-2}\%)$ in the pressure and total forces. There is only a minimal difference between the individual errors based on different p-Poisson methods, and, again, there is no trend for improved results with increased non-linearity, i.e., higher $p$ values.

\begin{figure}[!htb]
\centering
\iftoggle{tikzExternal}{
\input{./tikz/06__feeder/probe_6_v1v2v3.tikz}}{
\includegraphics{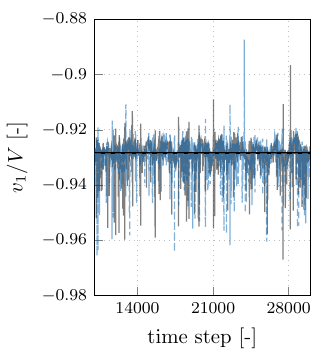}
\includegraphics{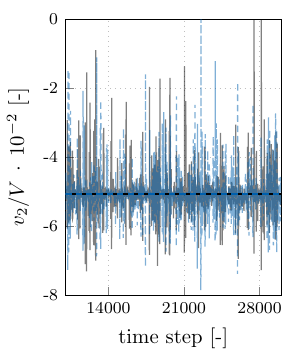}
\includegraphics{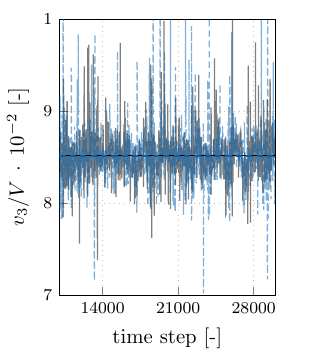}
}
\caption{Feeder ship flow ($\mathrm{Re}_\mathrm{L} = 5.0 \cdot 10^8$): Instantaneous and averaged fluid velocity components $v_1$ (left), $v_2$ (center), and $v_3$ (right) of the port-sided probe normalized by the apparent wind speed for two wall distance models over the last 2/3 of the simulated time steps.}
\label{fig:feeder_probe_6_v1v2v3}
\end{figure}

\begin{figure}[!htb]
\centering
\iftoggle{tikzExternal}{
\input{./tikz/06__feeder/probe_6_pwk.tikz}}{
\includegraphics{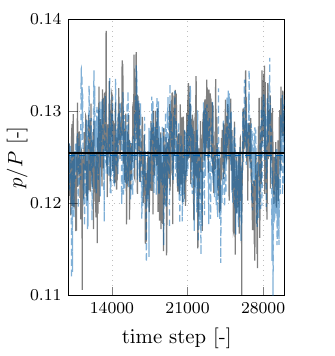}
\includegraphics{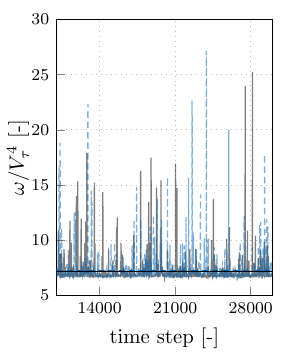}
\includegraphics{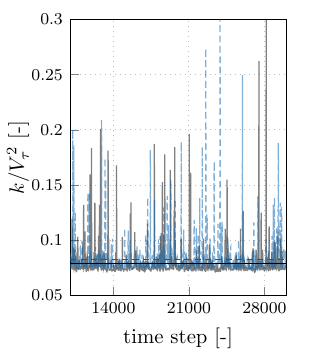}
}
\caption{Feeder ship flow ($\mathrm{Re}_\mathrm{L} = 5.0 \cdot 10^8$): Instantaneous and averaged, normalized fluid pressure $p$ (left), SDR $w$ (center), and TKE $k$ (right) of the port-sided probe for two wall distance models over the last 2/3 of the simulated time steps.}
\label{fig:feeder_probe_6_pwk}
\end{figure}

Next, the influence of the wall distance modeling on local flow variables at a port-side sample in the forward part of the ship at $x_i/L_\mathrm{pp} = [0.8, 0.08, 0.04]^T$ (see Figs. \ref{fig:kcs_propes} and \ref{fig:kcs_perspectives}) is investigated. All six flow variables following the balance equations from App. \ref{app:governing_equations} are recorded for the last 2/3 of the simulation period. These are the three components of the velocity vector $v_i$, the pressure $p$, the Turbulent Kinetic Energy (TKE) $k$, and its Specific Dissipation Rate (SDR) $\omega$. The development of all six variables are shown in Figs. \ref{fig:feeder_probe_6_v1v2v3} and \ref{fig:feeder_probe_6_pwk} normalized with respective reference variables over the number of simulated time steps for the $p=2$ Poisson-based (blue dashed) as well as the geometrically (black lines) determined wall distance field. Analogous to the integral force investigations, the temporal mean is also added as a horizontal line. In addition to the already introduced reference velocity $V$ based on the apparent wind speed, a dynamic reference pressure $P = \rho V^2 /2$ and a shear stress velocity $V_\tau$ are used for normalization. The latter follows empirical relationships on a flat plate, i.e., $V_\tau = \tau_\mathrm{w} / \rho$ with $\tau_w = c_f \rho V^2 / 2$ and $c_f = 0.026/\mathrm{Re}_\mathrm{L}^{1/7}$, cf. \cite{kuhl2025incremental}. In some cases, the instantaneous results deviate visibly from each other. In contrast, the time averages are very close, especially compared to the results of the integral force studies. Fundamental aspects are observed, e.g., positive TKE and SDR.

\begin{figure}[!htb]
\centering
\iftoggle{tikzExternal}{
\input{./tikz/06__feeder/error_probe_6_mean.tikz}}{
\includegraphics{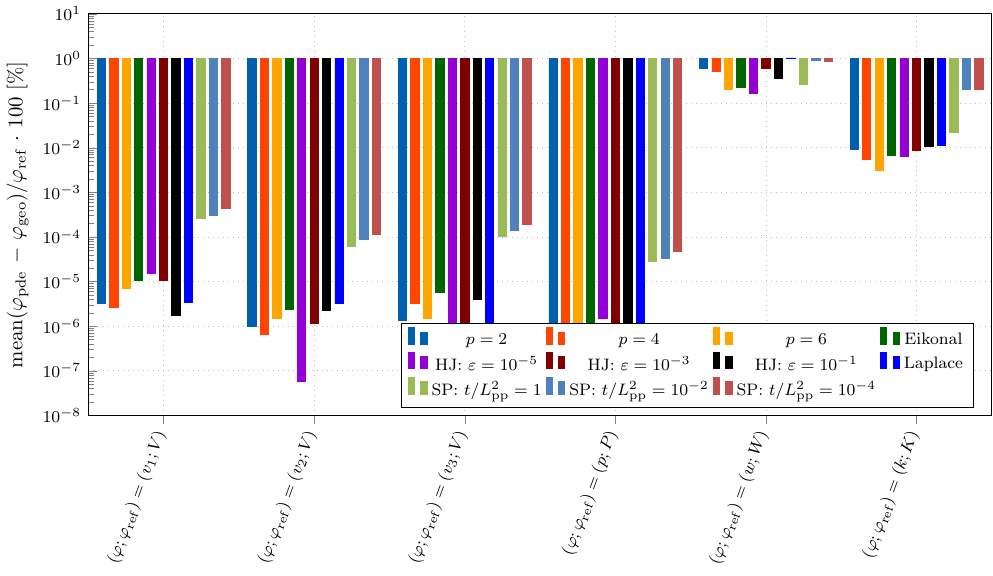}
}
\caption{Feeder ship flow ($\mathrm{Re}_\mathrm{L} = 5.0 \cdot 10^8$): Relative errors in the averaged flow variables predicted with different wall distance models and compared against the geometrical reference approach.}
\label{fig:feeder_probe_6_mean}
\end{figure}

In line with the integral force investigations, relative errors of the time-averaged, local flow variables are now considered via mean$(\varphi_\mathrm{pde} - \varphi_\mathrm{geo})/\varphi_\mathrm{ref} \cdot 100\%$, where field data based on the geometrical wall distance is considered as reference solution again. Combinations of all PDE-based wall distance approaches and the six flow variables are shown as a bar chart in Fig. \ref{fig:feeder_probe_6_mean}, using the same sorting and bar coloring as in Figs. \ref{fig:feeder_error_forces_total_all_mean} and \ref{fig:feeder_error_forces_container5_all_mean}. As can already be anticipated from the very similar time averages between the results of the $p=2$ Poisson method and the results based on the purely geometric wall distance from the previous Figs. \ref{fig:feeder_probe_6_v1v2v3} and \ref{fig:feeder_probe_6_pwk}, the relative errors are all comparatively small, especially in comparison to the relative errors of the integral investigations, cf. Figures \ref{fig:feeder_error_forces_total_all_mean} - \ref{fig:feeder_error_forces_container5_all_mean}. The errors range from $\mathcal{O}(10^{-6}\%)$ to $\mathcal{O}(10^{-3}\%)$ for the three velocity components and the pressure. However, the values increase significantly for the TKE and its SDR to $10^-{3}\% \leq  \mathrm{mean}(k_\mathrm{pde} - k_\mathrm{geo})/k_\mathrm{ref} \cdot 100 \leq 10^{-2}\%$ and $10^-{1}\% \leq  \mathrm{mean}(k_\mathrm{pde} - k_\mathrm{geo})/k_\mathrm{ref} \cdot 100 \leq 1\%$, respectively. This may be a direct consequence of the definition of the balance equations for determining the respective variables: The wall distance is only explicitly included in the turbulence equations; see, e.q., Eqn. \eqref{equ:w_in_eqn_0} in App. \ref{app:rans} or Eqns. \eqref{equ:w_in_eqn_1}, \eqref{equ:w_in_eqn_2}, \eqref{equ:w_in_eqn_3}, and \eqref{equ:w_in_eqn_4} in App. \ref{app:detached_eddy_simulation}. The velocity and pressure equations are implicitly affected by a potentially modified effective viscosity or effective pressure gradient. The Screened-Poisson approach again underperforms compared to all other methods, whereby errors in the range of $\mathcal{O}(10^{-3}\%)$ percent are still achieved for the three velocity components and the pressure. In the case of the latter quantities, there are only minor differences between the p-Poisson and Hamilton-Jacobi or Laplace and Eikonal methods. In contrast, errors in the TKE and SDR decrease for increased p-values, and $p=6$ features the most minor error in the TKE sample.

\begin{figure}[!htb]
\centering
\iftoggle{tikzExternal}{
\input{./tikz/06__feeder/feeder_les_region.tikz}}{
\includegraphics{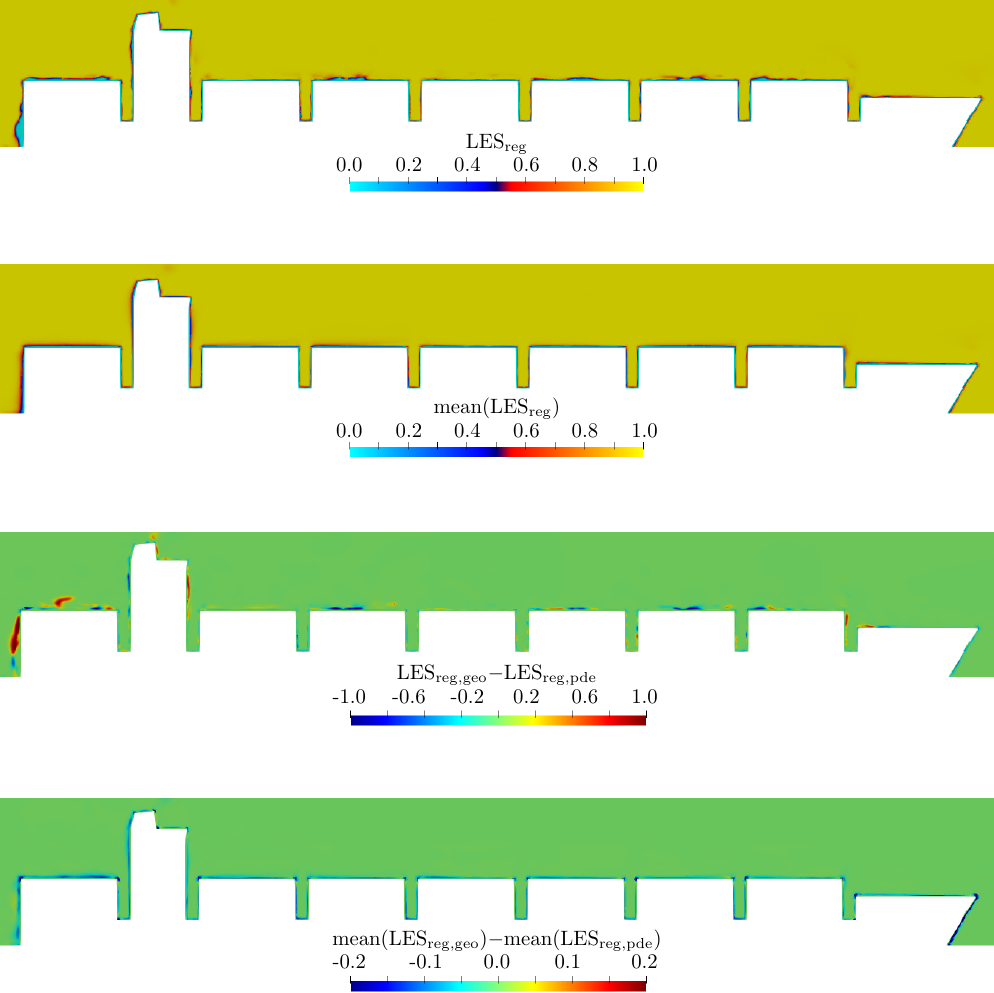}
}
\caption{Feeder ship flow ($\mathrm{Re}_\mathrm{L} = 5.0 \cdot 10^8$): From top to bottom: Instantaneous and time-averaged regional turbulence model indicator function from the reference solution with geometrically determined wall distance and its differences with corresponding Eikonal-based wall distance field results on a lateral slice at the center line.}
\label{fig:feeder_les_region}
\end{figure}

Finally, the influence of the wall distance modeling on the local or non-zonal mode of the hybrid RANS / LES turbulence model is examined. This can be roughly expressed by $\mathrm{LES}_\mathrm{reg} = \mathrm{min}(\mathrm{max}((1 - (l_\mathrm{DES} - l_\mathrm{LES})/(l_\mathrm{RANS} - l_\mathrm{LES}),0),1)$ based on Eqn. \eqref{equ:length_scales_des} in App. \ref{app:detached_eddy_simulation} and is evaluated in Fig. \ref{fig:feeder_les_region} on a lateral section through the ship's centerline for different scenarios. From top to bottom, an instantaneous indicator field originating from the last time step and an indicator field averaged over the last 2/3 of the simulation time based on a geometric wall distance determination are shown. The difference between the two upper indication functions and the corresponding results based on a wall distance determined using the Eikonal approach, both instantaneous and averaged over time, follow. The method primarily operates in scale resolving mode and only resorts to statistical turbulence modeling in certain areas close to the wall. However, there are some visible, especially instantaneous, differences between the operating modes of the turbulence model, which may be one of the reasons for the increased error values of the turbulence variables from Fig. \ref{fig:feeder_probe_6_mean}.


\section{Conclusion \& Outlook}
\label{sec:conclusion_outlook}

This paper dealt with modeling and simulating distance functions to describe wall distances based on partial differential equations (PDE). The distance to the nearest wall is required for many industrial problems in computational fluid dynamics (CFD). Standardized CFD strategies were utilized to approximate different wall distance models PDEs efficiently, and results were compared against the expensive but precisely determinable pure geometrical wall distance benchmark.

The first part of the manuscript addressed the investigation of modeling and simulation strategies. In addition to proven strategies based on linear elliptic PDEs, novel methods rarely used in industrial CFD were also considered. The following strategies have been studied: nonlinear and linear p-Poisson and Screened-Poisson methods, Eikonal as well as regularized Eikonal or Hamilton-Jacobi methods, and alternatives using Laplace equations. In addition to a clear definition of the field equations, boundary and initial conditions were also presented. The methods were implemented in a state-of-the-art unstructured Finite-Volume-Method, and the numerical method based on a Picard linearization has been presented in detail. Relevant measures to increase numerical robustness are described, such as using convective instead of diffusive approximation strategies in the case of the Eikonal equation combined with deferred correction approaches. All methods were first tested on a maritime application example with wall geometries moving relative to each other. Using the example of a ship with a rotating propeller, the implemented methods were verified and validated against the exact but expensive geometric method accelerated using a KD-tree. Theoretical assumptions were confirmed, e.g., that the p-Poisson method provides an improved approximation to the actual wall distance for increasing p-values or that the Laplace method is sensitive to singularities in the case of oppositely placed walls and thus locally vanishing wall distance gradients. For the considered ship case, the Eikonal approach and variants of the Hamilton-Jacobi approach with minor diffusivity produced the most satisfactory results while requiring a fair numerical effort. In contrast, the nonlinear p-Poisson investigations displayed a significantly increased numerical effort, among others, due to the linearized nature of the employed numerical process.

In the second part, the methods presented were applied to hydrodynamic and aerodynamic flow applications from maritime engineering, each relying on Shear Stress Transport (SST) strategies for turbulence modeling that require the distance to the nearest wall at different procedural points. First, the hydrodynamic behavior of a bulk carrier at model scale at $\mathrm{Re}_\mathrm{L} = \SI{7.246}{} \cdot 10^6$ and $\mathrm{Fn}=0.142$ was investigated on the influence of the wall distance formulation for predicting resistance and propulsion behavior in conjunction with statistical turbulence modeling based on a classical Reynolds-Averaged Navier-Stokes (RANS) method. It was shown that the different wall distance modeling barely influences relevant integral hydrodynamic quantities such as drag, trim, and sinkage. Related errors are in the range of $\mathcal{O}(10^{-1}\%)$ and, therefore, significantly below typical modeling, discretization, and approximation errors. Some methods, especially the Poisson approaches, were found to be very sensitive to their model parameters. The observations from the resistance investigations were confirmed for the subsequent consideration of a self-propulsion analysis with a geometrically resolved propeller. It was shown that the assumption of a temporally constant wall distance by calculating the wall distance once at the beginning of the simulation is sufficient for an adequately accurate propulsion calculation, even though the wall distance is theoretically a function of time due to the relative movement of the ship's propeller and hull and must therefore be re-determined in each time step. The latter was performed for all methods, whereby the additional numerical effort for continuously updating the wall distance field already amounted to approximately $\mathcal{O}(20\%)$ of the simulation time for the linear methods and more than $\mathcal{O}(50\%)$ percent for nonlinear wall distance strategies. Finally, the wall distance methods were investigated for the aerodynamic analysis of a large-scale feeder ship at $\mathrm{Re}_\mathrm{L} = 5.0 \cdot 10^8$. A hybrid RANS Large-Eddy-Simulation (LES) approach, in line with the Improved Delayed Detached Eddy Simulation (IDDES) model, has been employed and demonstrated to be somewhat more sensitive to the choice of the wall distance model. In particular, the selected parametric of the Screened-Poisson approach resulted in errors of several percent in integral forces. The turbulence variables, whose balance variables depend directly on the wall distance, showed greater sensitivity to the wall distance model than further flow variables, such as the velocity vector or pressure field.

In summary, it can be concluded that the Eikonal approach based on a convective approximation --e.g., with an upwind-biased approximation and thus solid numerical viscosity—or a Hamilton-Jacobi method with a suitable, minor apparent viscosity-- with deferred correction provides a solid method with low numerical effort and, at the same time, fair approximation quality. The p-Poisson method is numerically too expensive, especially for p-values greater than two, the Laplace method is prone to singularities, and the Screened-Poisson method is generally too dependent on frequently unclear to define model parameters. Further studies should test the investigated strategies in further realistic applications, ideally with geometries moving relative to each other. Examples include propellers in free running and behind conditions, ideally equipped with cavitation and transition modeling.


\section{Declaration of Competing Interest}
The author declares that he has no known competing financial interests or personal relationships that could have appeared to influence the work reported in this paper.

\section{Acknowledgments}
The current work is part of the “Propulsion Optimization of Ships and Appendages” research project funded by the German Federal Ministry for Economics and Climate Action (Grant No. 03SX599C). The author gratefully acknowledges this support.
%
%
%

\begin{appendix}

\section{Governing Equations}
\label{app:governing_equations}

The conservation of mass and linear momentum governs the temporal ($t$) evolution of an incompressible fluid's (density $\rho$, dynamic viscosity $\mu$) velocity $v_i$ and pressure $p$, viz.
\begin{align}
	\frac{\partial v_\mathrm{k}}{\partial x_\mathrm{k}} &= 0 \label{equ:mass_conservation} \\
	\frac{\partial \rho v_i}{\partial t} + \frac{\partial}{\partial x_k} \left[ v_k \rho v_i + p^\mathrm{eff} \delta_{ik} - 2 \mu^\mathrm{eff} S_{ik} \right] &= 0 \label{equ:momentum_conservation}
\end{align}
where $\delta_{ik}$ denotes the Kronecker delta and $\mu^\mathrm{eff} = \mu + \mu^\mathrm{t}$ as well as $p^\mathrm{eff} = p + p^\mathrm{t}$ refer to effective viscosity and pressure, respectively, that contain turbulent contributions that depend on the underlying flow modeling approach. For this paper's studies, the employed numerical strategies differ in the determination of the eddy viscosity only, which is briefly described in the following.
Local material properties in Eqns. \eqref{equ:momentum_conservation} follow a concentration field $c$ in line with a Volume-of-Fluid method, i.e.,
\begin{align}
    \frac{\partial c}{\partial t} + \frac{\partial v_k c}{\partial x_k} &= 0 \label{equ:concentration_conservation} \, ,
\end{align}
that allows for the interpolation of the local fluid density and viscosity
\begin{align}
    \rho = \rho_\mathrm{a} c + (1-c) \rho_\mathrm{b}
    \qquad \qquad \mathrm{and} \qquad \qquad
    \mu = \mu_\mathrm{a} c + (1-c) \mu_\mathrm{b}
\end{align}
based on for- and background bulk properties ($\rho_\mathrm{a}, \mu_\mathrm{a}$) and ($\rho_\mathrm{b}, \mu_\mathrm{b}$), respectively.

\subsection{Reynolds-Averaged Navier-Stokes}
\label{app:rans}
In line with an incompressible Boussinesq Viscosity Model and the 2003 variant of Menter's shear stress transport model \cite{menter2003ten}, a two-equation Reynolds-Averaged Navier-Stokes turbulence closure is utilized. The transport equations for the turbulent kinetic energy $k$ as well its specific dissipation rate $\omega$ are implemented as follows
\begin{alignat}{3}
    &R^k &&= \frac{\partial \rho k}{\partial t} + \frac{\partial v_k \rho k}{\partial x_k} - \frac{\partial}{\partial x_k} \left[ \left( \mu + \sigma^\mathrm{k} \mu^\mathrm{t} \right) \frac{\partial k}{\partial x_k} \right] - P^\mathrm{k} + \beta^k \rho \omega k &&= 0 \label{equ:komegasst_k} \\
    &R^\omega &&= \frac{\partial \rho \omega}{\partial t} + \frac{\partial v_k \rho \omega}{\partial x_k} - \frac{\partial}{\partial x_k} \left[ \left( \mu + \sigma^\mathrm{\omega} \mu^\mathrm{t} \right) \frac{\partial \omega}{\partial x_k} \right] - \frac{\alpha \rho}{\mu^\mathrm{t}} P^\mathrm{k} + \beta^\omega \rho \omega^2 - 2 \left( 1 - F1 \right) \frac{ \rho \sigma_{\omega, 2} }{\omega} \frac{\partial k}{\partial x_k} \frac{\partial \omega}{\partial x_k} &&= 0 \, , \label{equ:komegasst_w}
\end{alignat}
where
\begin{align}
    F1 = \mathrm{tanh} \left( \mathrm{min} \left( \mathrm{max} \left( A, B \right) , C \right)^4 \right)
\end{align}
with
\begin{align}
    A = \frac{\mathrm{max} \left( \sqrt{k}, 0\right)}{\mathrm{max} \left( \beta^k \omega w, s \right)}, \quad
    B = \frac{500 \mu}{\mathrm{max} \left( \rho w^2 \omega, s \right)}, \quad 
    C = \frac{2 \rho \omega}{ \mathrm{max}(w^2 \frac{\partial k}{\partial x_k} \frac{\partial \omega}{\partial x_k} , 10^{-10} )} \, . \label{equ:w_in_eqn_0} 
\end{align}
The production term is assigned to $P^k = \mu^t S_{ik} \frac{\partial v_i}{\partial x_k}$ and realizability conditions referring to stagnation point corrections are considered, i.e.,
\begin{align}
    P^k = \mathrm{min}(P^k, C^l k C^\mu \omega \rho) \, ,
\end{align}
where the limiter's value reads $C^l= 15$ together with $C^\mu = 0.09$. Diffusive fluxes's relevant model data is linear interpolated based on the blending function, i.e.,
\begin{align}
    \sigma^k &= \sigma_{k,2} + (\sigma_{k,1} - \sigma_{k,2}) F1 \\
    \sigma^\omega &= \sigma_{\omega,2} + (\sigma_{\omega,1} - \sigma_{\omega,2}) F1 \\
    \beta^\omega &= \beta_{\omega,2} + (\beta_{\omega,1} - \beta_{\omega,2}) F1 \, .
\end{align}
Finally, the eddy viscosity follows
\begin{align}
    \mu^t = \frac{\rho k}{\mathrm{max} \left(\omega, F2 \sqrt{2 S_{ik} S_{ik}}/a_1\right)}
    \qquad \qquad \mathrm{with} \qquad \qquad
    F2 = \mathrm{tanh} \left( \left( 2A, B \right)^2 \right) \, .
\end{align}
Model constants are
\begin{alignat}{11}
    \sigma_{k,1} &= 0.85  \qquad
    &&\sigma_{k,2} &&= 1.0  \qquad
    &&\sigma_{\omega,1} &&= 0.5  \qquad
    &&\sigma_{\omega,2} &&= 0.856  \qquad
    &&\alpha_1 &&= 5/9 \qquad
    &&\alpha_2 = 0.44  \\
    a_1 &= 0.31  \qquad
    &&\beta_1 &&= 3/40  \qquad
    &&\beta_2 &&= 0.0828 \qquad
    &&\beta^k &&= 0.09 \qquad
    &&\kappa &&= 0.41 \qquad
\end{alignat}

\subsection{Detached Eddy Simulation}
\label{app:detached_eddy_simulation}

In line with the Improved Delayed Detached Eddy Simulation (IDDES) model of \cite{gritskevich2012development}, the turbulent viscosity is expressed as
\begin{align}
    \mu^\mathrm{t} &= \frac{\rho k}{\mathrm{max\left(\omega, f2 \frac{\sqrt{2 S_{ik} S_{ik}}}{a_{\mu}}\right)}} \, ,
\end{align}
where the required turbulent kinetic energy $k$ as well as its specific dissipation rate $\omega$ are obtained from the following transport equations
\begin{align}
	\frac{\partial \rho k}{\partial t} 
    + \frac{\partial}{\partial x_k} \left[ v_k \rho k 
    - \left( \mu + \mu^\mathrm{t} \, \sigma^\mathrm{k} \right) \frac{\partial k}{\partial x_k} \right] 
    - P^\mathrm{k} 
    + \rho \frac{k^{3/2}}{l^\mathrm{DES}} &= 0 \label{equ:tke_conservation} \\
    \frac{\partial \rho \omega}{\partial t} 
    + \frac{\partial}{\partial x_k} \left[ v_k \rho \omega
    - \left( \mu + \mu^\mathrm{t} \, \sigma^\mathrm{w} \right) \frac{\partial \omega}{\partial x_k} \right] 
    - 2 \sigma_\mathrm{\omega, 2} \frac{\rho}{\omega} \frac{\partial k}{\partial x_k} \frac{\partial \omega}{\partial x_k} \left( 1 - f1 \right)
    - \alpha P^\mathrm{k} \frac{\rho}{\mu^\mathrm{t}}
    + \beta \omega^2 \rho &= 0 \label{equ:omega_conservation}
\end{align}
where the TKE production is expressed as $P^\mathrm{k} = \mu^\mathrm{t} S_{ik} (\partial v_i / \partial x_k)$. Model coefficients
\begin{alignat}{3}
    \sigma^\mathrm{k} &=  \sigma_\mathrm{\omega,2}+\left( \sigma_\mathrm{\omega,1} - \sigma_\mathrm{\omega,2} \right) f1
    \qquad \qquad
    &&\sigma^\mathrm{\omega} &&= \sigma_\mathrm{\omega, 2}+\left( \sigma_\mathrm{\omega, 1} - \sigma_\mathrm{\omega, 2} \right) f1 \\
    \alpha &= \alpha_{\omega, 2} +\left( \alpha_{\omega, 1} - \alpha_{\omega, 2} \right) f1
    \qquad \qquad
    &&\beta &&= \beta_{\omega, 2} +\left( \beta_{\omega, 1} - \beta_{\omega, 2} \right) f1
\end{alignat}
as well as the denominator in the eddy viscosity are hyperbolically blended, viz.
\begin{align}
    f1 &= \mathrm{tanh} \left( \mathrm{min}\left(\mathrm{max}\left(
    \frac{ \sqrt{k}}{\beta^* \, \omega \, w},
    \frac{500 \mu}{\rho \, \omega \, w^2}
    \right),
    \frac{200 k \, \omega}{w^2 \, \frac{\partial k}{\partial x_k} \frac{\partial \omega}{\partial x_k}}
    \right)^4 \right) \label{equ:w_in_eqn_1} \\
    f2 &= \mathrm{tanh} \left( \mathrm{max}\left( 
    \frac{2 \sqrt{k}}{\beta^* \, \omega \, w},
    \frac{500 \mu}{\rho \, \omega \, w^2}
    \right)^2 \right) \, . \label{equ:w_in_eqn_2} 
\end{align}
The following Shear Stress Transport (SST, cf. \cite{menter1994two}) model coefficients are employed
\begin{alignat}{7}
    \sigma_{k,1} &= 0.85 \qquad
    &&\sigma_{k,2} &&= 1.0 \qquad
    &&\sigma_{\omega,1} &&= 0.5 \qquad
    &&\sigma_{\omega,2} &&= 0.856 \\
    \alpha_{\omega,1} &= 5/9 \qquad
    &&\alpha_{\omega,2} &&= 0.44 \qquad
    &&\beta_{\omega,1} &&= 3/40 \qquad
    &&\beta_{\omega,2} &&= 0.0828 \\
    a_{\mu} &= 0.31 \qquad
    &&\beta^* &&= 0.09 \, .
\end{alignat}
The DES length scale determination follows \cite{gritskevich2012development}, i.e.,
\begin{align}
    l^\mathrm{DES} &= \tilde{f}^\mathrm{d} \, l^\mathrm{RANS} + (1 - \tilde{f}^\mathrm{d}) \, l^\mathrm{LES}
    \qquad \qquad \mathrm{with} \qquad \qquad
    l^\mathrm{RANS} = \frac{\sqrt{k}}{\omega \beta^*}
    \qquad \mathrm{and} \qquad
    l^\mathrm{LES} = C^\mathrm{DES} \Delta \, , \label{equ:length_scales_des}
\end{align}
where $\Delta = \mathrm{min}(C^\mathrm{d} \, \mathrm{max}(w, h), h)$ accounts for a local grid spacing that either refers to maximum edge length $h := h(\Delta V)$ per computational cell in the underlying Finite-Volume framework or a scaled ($C^\mathrm{d} = 0.15$) distance to the nearest wall. The LES's length scale coefficient is linear interpolated
\begin{align}
    C^\mathrm{DES} = C^\mathrm{DES, 1} f1 + C^\mathrm{DES, 2} (1 - f1) \, ,
\end{align}
and the blending between RANS and LES is performed via
\begin{align}
    \tilde{f}^\mathrm{d} = \mathrm{max}\left(\left(1 - f^\mathrm{d}\right),f^\mathrm{b}\right)
    \qquad \qquad \mathrm{with} \qquad \qquad
    f^\mathrm{d} = 1 - \mathrm{tanh}\left( \left(C^\mathrm{d, 1} \, r^\mathrm{d} \right)^{C^\mathrm{d, 2}}\right) \, ,
    f^\mathrm{b} = \mathrm{min}\left( 2 e^{-9 \alpha^\mathrm{fb}}, 1 \right) \, . \label{equ:w_in_eqn_3} 
\end{align}
The required exponent reads
\begin{align}
    r^\mathrm{d} = \frac{\mu^\mathrm{t}}{\rho (\kappa w)^2 \sqrt{0.5} S \, W }  \label{equ:w_in_eqn_4} 
\end{align}
and accounts for strain $S = \sqrt{2 S_{ik} S_{ik}}$ and vorticity $W = \sqrt{2 W_{ik} W_{ik}}$ rate magnitude, respectively, where $W_{ik} = 0.5 (\partial v_i / \partial x_k - \partial v_k / \partial x_i)$. The employed model constants are assigned to
\begin{align}
    C^\mathrm{DES, 1} = 0.78 \qquad \qquad
    C^\mathrm{DES, 2} = 0.61 \qquad \qquad
    C^\mathrm{d, 1} = 20 \qquad \qquad
    C^\mathrm{d, 2} = 3 \qquad \qquad
    \kappa = 0.41 \, .
\end{align}

\end{appendix}


\end{document}

%% file: tex/settings.tex
\usepackage[utf8]{inputenc}
\usepackage{amsmath, amsthm, amssymb, amsfonts}     
\usepackage{mathrsfs}                               
\usepackage{accents}                                
\usepackage{natbib}                                 
\usepackage{hyperref}                               
\usepackage{doi}                                    
\usepackage{authblk}                                
\usepackage{siunitx}                                
\usepackage{booktabs}                               
\usepackage{ulem}                                   
\usepackage{nomencl}                                
\makenomenclature                                   
\usepackage[linesnumbered,ruled,vlined]{algorithm2e}
\SetCommentSty{mycommfont}
\usepackage{subfigure}

\usepackage{bm}                     


%% file: tex/plotting.tex
\usepackage{graphicx}
\usepackage{tikz}
\usepackage{pgfplots}

\usetikzlibrary{patterns}

\usetikzlibrary{decorations.pathmorphing,arrows,intersections,arrows.meta,angles,quotes}

\pgfplotsset{
	kurze Legende/.style={
		legend image code/.code={
			\draw[##1,mark repeat=2,line width=0.6pt]
			plot coordinates {
				(0cm,0cm)
				(0.3cm,0cm)
			};
		}
	}
}

\pgfplotsset{
	compat = newest,
	scale only axis, 
	max space between ticks = 50pt,
	ticklabel style = {font=\footnotesize},
	legend style =  {font=\footnotesize},
	grid = major,
	grid style = {dotted},
	legend columns=1, 
	xtick pos=left,
	ytick pos=left
}

\pgfplotsset{select coords between index/.style 2 args={
		x filter/.code={
			\ifnum\coordindex<#1\fi
			\ifnum\coordindex>#2\fi
		}
}}

\definecolor{color1}{HTML}{0060AD} 
\definecolor{color2}{HTML}{FF4500} 
\definecolor{color3}{HTML}{FFA500} 
\definecolor{color4}{HTML}{006400} 
\definecolor{color5}{HTML}{9400D3} 
\definecolor{color6}{HTML}{800000} 
\definecolor{color7}{HTML}{000000} 
\definecolor{color8}{HTML}{0000FF} 
\definecolor{color9}{HTML}{FF0000} 
\definecolor{mycolor_blue}{RGB}{66,124,161}
\definecolor{mycolor_grey}{RGB}{198,198,198} 

\definecolor{bblue}{HTML}{4F81BD}
\definecolor{rred}{HTML}{C0504D}
\definecolor{ggreen}{HTML}{9BBB59}
\definecolor{ppurple}{HTML}{9F4C7C}

\tikzstyle{line1} = [color=color7,semithick] 
\tikzstyle{line2} = [color=color2,densely dotted,semithick]
\tikzstyle{line3} = [color=color1,densely dashed,semithick]
\tikzstyle{line4} = [color=color5,dash dot,semithick]
\tikzstyle{line5} = [color=color4,dash dot dot,semithick]
\tikzstyle{line6} = [color=color6,semithick]
\tikzstyle{line7} = [color=color3,densely dotted,semithick]
\tikzstyle{line8} = [color=color8,densely dashed,semithick]

\tikzstyle{mark1} = [color=color7,mark=x,mark size=2pt,mark options=solid,semithick] 
\tikzstyle{mark2} = [color=color2,mark=o,mark size=2pt,mark options=solid,semithick]
\tikzstyle{mark3} = [color=color1,mark=*,mark size=2pt,mark options=solid,semithick]
\tikzstyle{mark4} = [color=color5,mark=triangle,mark size=2pt,mark options=solid,semithick]
\tikzstyle{mark5} = [color=color4,mark=square,mark size=2pt,mark options=solid,semithick]
\tikzstyle{mark6} = [color=color7,mark=o,mark size=2pt,mark options=solid,semithick]
\tikzstyle{mark7} = [color=color3,mark=*,mark size=2pt,mark options=solid,semithick]
\tikzstyle{mark8} = [color=color8,mark=triangle,mark size=2pt,mark options=solid,semithick]